%% file: paper.tex
\documentclass[aps, pre, superscriptaddress, twocolumn, longbibliography, nofootinbib]{revtex4-1}

\usepackage{amsmath,amssymb}
\usepackage{graphics}
\usepackage{graphicx}
\DeclareMathOperator\erf{erf}
\usepackage{hyperref}
\usepackage{color}

\begin{document}

\title{Link persistence and conditional distances in multiplex networks}
\author{Fragkiskos Papadopoulos}
\email{f.papadopoulos@cut.ac.cy}
\affiliation{Department of Electrical Engineering, Computer Engineering and Informatics, Cyprus University of Technology, 33 Saripolou Street, 3036 Limassol, Cyprus}
\author{Kaj-Kolja Kleineberg}
\affiliation{Computational Social Science, ETH Zurich, Clausiusstrasse 50, 8092, Zurich, Switzerland}

\date{\today}

\begin{abstract}
Recent progress towards unraveling the hidden geometric organization of real multiplexes revealed significant correlations across the hyperbolic node coordinates in different network layers, which facilitated applications like trans-layer link prediction and mutual navigation. But are geometric correlations alone sufficient to explain the topological relation between the layers of real systems? Here we provide the negative answer to this question. We show that connections in real systems tend to persist from one layer to another irrespectively of their hyperbolic distances. This suggests that in addition to purely geometric aspects the explicit link formation process in one layer impacts the topology of other layers. Based on this finding, we present a simple modification to the recently developed Geometric Multiplex Model to account for this effect, and show that the extended model can reproduce the behavior observed in real systems. We also find that link persistence is significant in all considered multiplexes and can explain their layers' high edge overlap, which cannot be explained by coordinate correlations alone. Furthermore, by taking both link persistence and hyperbolic distance correlations into account we can improve trans-layer link prediction. These findings guide the development of multiplex embedding methods, suggesting that such methods should be accounting for both coordinate correlations and link persistence across layers.
\end{abstract}

\maketitle

\section{Introduction}
\label{sec:intro}

It has been shown that random geometric graphs in hyperbolic spaces are adequate models for complex networks, as they naturally and simultaneously possess many of their common structural and dynamical characteristics, including heterogeneous distributions of node degrees, strong clustering, and preferential attachment, cf.~\cite{Krioukov2009, Krioukov2010, Papadopoulos2012, Panagiotou2012, cliques2015, diameter2015, Zuev2015, fountoulakis2017, fountoulakis2018}. Specifically, the $\mathbb{H}^{2}$ model~\cite{Krioukov2009, Krioukov2010} constructs networks by randomly distributing nodes on a hyperbolic disc of radius $R$, such that each node $i$ has the polar coordinates, or hidden variables, $r_i, \theta_i$, and connecting each pair of nodes with a probability that decreases with their hyperbolic distance.

Given the ability of the $\mathbb{H}^{2}$ model to construct synthetic networks that resemble real networks, it has been shown that one can meaningfully map (embed) real networks into the hyperbolic plane, in a way congruent with the model~\cite{Boguna2010}. Mapped networks include the Autonomous Systems Internet~\cite{Boguna2010}, biological networks~\cite{Serrano2011, alanis2018, allard2018}, social networks~\cite{Papadopoulos2012, kleineberg2016np}, and the international trade system~\cite{perez2016}. Using the constructed hyperbolic maps one can facilitate important applications, which include identifying node communities on a geometric basis~\cite{Papadopoulos2012, Boguna2010, Serrano2011, blasius16}; predicting missing and future links~\cite{Serrano2011, frag:hypermap, frag:hypermap_cn}; and performing efficient network navigation or search~\cite{Boguna2010, frag:hypermap, frag:hypermap_cn, ndn2016, colnav, Ortiz2017, allard2018}. Model-free mapping methods have also been developed~\cite{carlo1,carlo2}.

The work in~\cite{Papadopoulos2012} explained the emergence of hyperbolic geometry by extending the static approach of \cite{Krioukov2010} to growing networks. It has been shown that the radial coordinate of a node abstracts its \emph{popularity}. The smaller this coordinate, the more popular the node is, and the more likely it attracts connections. The angular distance between two nodes, $i, j$,  $\Delta \theta_{ij}$, abstracts their \emph{similarity}. The smaller this distance, the more similar the two nodes are, and the more likely they are connected. The hyperbolic distance between two nodes, very well approximated by $x_{ij}=r_i+r_j+2 \ln{\sin{(\Delta \theta_{ij}/2)}}$~\cite{Krioukov2010}, is then a single-metric representation of a combination of the two attractiveness attributes, radial popularity and angular similarity.

The above framework has mainly focused on individual complex networks. In this view, given the connection probability function, the node coordinates alone determine the network's observed topology, and vice versa, the network's observed topology alone is used to infer its node coordinates. However, there are cases where nodes from one network may also exist in other networks. 
This gives rise to multiplex systems~\cite{kivel2014, Boccaletti2014}, which are collections of networks (called layers) that share common nodes. Examples include the different social networks that a person may belong to~\cite{multi2010, mucha2010}; the Internet's IPv4 and IPv6 topologies~\cite{ark2009}; structural and functional brain networks~\cite{Bullmore2009}; and networks of different types of genetic interactions~\cite{manlio2015}.  This observation calls for extending the network geometry paradigm to the multiplex domain, where the different layers are treated simultaneously and not independently. This is because the coordinates and connections of nodes in one layer can, in principle, provide information about the coordinates and connections of the same nodes in other layers. 

A first step towards this direction is the finding that if the layers comprising real multiplexes are independently embedded into hyperbolic spaces, their coordinates exhibit significant correlations~\cite{kleineberg2016np}. This finding motivated new applications, like multidimensional community detection and trans-layer link prediction on a geometric basis, as well as multilayer greedy routing~\cite{kleineberg2016np}. Furthermore, it was shown that the discovered correlations play an important role in the robustness of multiplexes against targeted attacks to high degree nodes~\cite{koljaprl2017} and to the outcome of evolutionary dynamics~\cite{amato17,kleineberg18}. Yet, despite these advances, it is still not fully understood to what extend can coordinate correlations alone explain the topological relation between the layers of real systems. In particular, are coordinate correlations alone sufficient to explain the high edge overlap observed among the layers of real systems~\cite{Boccaletti2014, Battiston2014, Ginestra2013}? This is an important metric as it has been shown to significantly affect the outcome of dynamical processes~\cite{Ginestra_overlap2013, Shlomo2013, Salehi2015}.

Here we provide the negative answer to the above question. We first provide empirical evidence from real multiplexes suggesting that connections can persist from one layer to another irrespectively of the hyperbolic distances that they span. We then consider a simple modification to the Geometric Multiplex Model (GMM)~\cite{kleineberg2016np} to account for this effect, and show that the extended model can reproduce the behavior observed in real systems. We also estimate the link persistence probability in the considered systems and find that it is significant in all cases, explaining their layers' high edge overlap, which cannot be explained by coordinate correlations alone. Furthermore, we show that by taking link persistence into account one can improve trans-layer link prediction. 

The rest of the paper is organized as follows. In Section~\ref{sec:preliminaries} we review the $\mathbb{H}^{2}$ model, the GMM and the HyperMap embedding method~\cite{frag:hypermap, frag:hypermap_cn}. In Section~\ref{sec:evidence} we provide empirical evidence of link persistence in real multiplexes. In Section~\ref{sec:model} we present the modified GMM and show that it reproduces the behavior observed in the real data. In Section~\ref{sec:overlap} we show that link persistence explains the high edge overlap in real systems.  In Section~\ref{sec:analysis} we analyze the model and explain how one can estimate the link persistence probability from the layers' hyperbolic embeddings. In Section~\ref{sec:link_prediction} we show that the accuracy of trans-layer link prediction can be improved by taking link persistence into account. Finally, in Section~\ref{sec:conclusion} we discuss open problems and conclude the paper.

\section{Preliminaries}
\label{sec:preliminaries}

In this section we review the $\mathbb{H}^{2}$ model~\cite{Krioukov2010}, the GMM~\cite{kleineberg2016np} and HyperMap~\cite{frag:hypermap, frag:hypermap_cn}. We limit ourselves only to the basic details that we will need in the rest of the paper. 
Throughout the paper, symbol `$\approx$' means \emph{approximately equal}. Symbol `$\propto$' means \emph{proportional to}, i.e., $f(t) \propto g(t)$ means $f(t)=c g(t)$, where $c$ is a constant, $0 < c < \infty$. Sometimes there are additive terms so that $f(t) \propto g(t)$ can also mean $f(t)=c g(t)+d$.

\subsection{$\mathbb{H}^{2}$ model}
\label{sec:H2}

In the $\mathbb{H}^{2}$ model each node $i$ has radial (popularity) and angular (similarity) coordinates $r_i, \theta_i$. To construct a network that has size $N$, average node degree $\bar{k}$, a power law degree distribution with exponent $\gamma > 2$, and temperature $T \in [0,1)$, we perform the following steps:
\begin{enumerate}
\item[(1)] coordinate assignment: we sample the angular coordinates of nodes $\theta_i$, $i=1,2,\ldots,N$, uniformly at random from $[0, 2\pi]$, and their radial coordinates $r_{i}$, $i=1,2,\ldots,N$, from the probability density function (PDF):
\begin{align}
\rho(r)=\frac{1}{2\beta} \frac{\sinh{\frac{r}{2 \beta}}}{\cosh{\frac{R}{2\beta}-1}} \approx \frac{1}{2\beta} e^{\frac{1}{2\beta}(r-R)},
\end{align}
where $\beta=\frac{1}{\gamma-1}$, while $R=2 \ln{\frac{N}{c}}$ is the radius of the hyperbolic disc where nodes reside, and $c=\bar{k} \frac{\sin{T \pi}}{2T}\left(\frac{\gamma -2}{\gamma-1}\right)^2$;
\item[(2)] creation of edges: we connect every pair of nodes $i, j$ with the Fermi-Dirac connection probability:
\begin{equation}
\label{eq:p_x_ij}
p(x_{ij})=\frac{1}{1+e^{\frac{1}{2T}(x_{ij}-R)}}.
\end{equation}
\end{enumerate}
In the last expression, $x_{ij} \leq 2 R$ is the hyperbolic distance between nodes $i$ and $j$~\cite{Bonahon09-book}:
\begin{align}
\label{eq:hyperbolic_distance}
\nonumber x_{ij}&=\mathrm{arccosh}\left(\cosh{r_i}\cosh{r_j}-\sinh{r_i} \sinh {r_j} \cos{\Delta\theta_{ij}}\right)\\
&\approx r_i+r_j+2 \ln{\sin{\frac{\Delta \theta_{ij}}{2}}} \approx r_i+r_j+2 \ln{\frac{\Delta \theta_{ij}}{2}},
\end{align}
where $\Delta \theta_{ij}=\pi-|\pi-|\theta_i-\theta_j||$ is the angular  distance between the nodes. The approximations in Eq.~(\ref{eq:hyperbolic_distance}) hold for sufficiently large $r_i$, $r_j$, and $\Delta\theta_{ij}>2\sqrt{e^{-2r_i}+e^{-2r_j}}$~\cite{Krioukov2010}. Temperature $T$ controls the average clustering~\cite{Dorogovtsev10-book} in the network, which is maximized at $T = 0$, and nearly linearly decreases to zero with $T \in [0, 1)$. At $T \to 0$ the connection probability in Eq.~(\ref{eq:p_x_ij}) becomes the step-function $p(x_{ij}) \to 1$ if $x_{ij} \leq R$, and $p(x_{ij}) \to 0$ if $x_{ij}  > R$. 

We also recall that the average degree of a node at radial coordinate $r$, $\bar{k}(r)$, is:
\begin{align}
\label{eq:bar_k_r_H2}
\bar{k}(r) \approx \bar{k}_0 e^{\frac{1}{2}(R-r)} \propto e^{-\frac{1}{2}r},
\end{align}
where $\bar{k}_0 \equiv \bar{k}(\gamma-2)/(\gamma-1)$ is the expected minimum degree in the network. Therefore, $r(k) \propto 2\ln(1/k)$. This combined with the fact that the density of radial coordinates increases exponentially, $\rho(r) \propto e^{\frac{1}{2\beta}r}$, gives the power law degree  distribution, $P(k) \approx \rho(r(k))|r'(k)| \propto k^{-\gamma}$, $\gamma=1+1/\beta > 2$. Without loss of generality, we assume here a hyperbolic plane of curvature $K=-1$. See~\cite{Krioukov2010} for further details. 

\subsection{Geometric Multiplex Model (GMM)}
\label{sec:GMM}

In the GMM each single layer is constructed according to the $\mathbb{H}^{2}$ model, while accounting for correlations among the radial and angular coordinates of nodes in the different layers, whose strength can be tuned. We overview here the GMM for two-layer systems, where each node $i \leq N$ exists in both layers. See~\cite{kleineberg2016np} for the extension of the model to more than two layers and to layers with different sizes.

In a nutshell, we perform the following steps to construct a two-layer system:
\begin{enumerate}
\item[(1)] assignment of coordinates $r_{1, i}, \theta_{1, i}$ to each node $i$ in layer~1, as in the $\mathbb{H}^{2}$ model (Eqs.~(\ref{eq:eq_rho_1}), (\ref{eq:eq_theta}) below);
\item[(2)] assignment of coordinates $r_{2, i}, \theta_{2, i}$ to each node $i$ in layer~2, depending on the node's coordinates in layer~1---the assignment here is done such that the marginal (unconditional) distribution of $r_{2, i}, \theta_{2, i}$ is still the one in the $\mathbb{H}^{2}$ model (Eqs.~(\ref{eq:r_cond_text}), (\ref{eq:assign_theta_text}));
\item[(3)] creation of edges, by connecting node pairs in each layer with the corresponding $\mathbb{H}^{2}$ connection probability, which depends exclusively on the node coordinates in each layer (Eqs.~(\ref{eq:c_prob_1}), (\ref{eq:c_prob_2})).
\end{enumerate}
Below, we explain these steps in more detail. To proceed, let:
\begin{equation}
\label{eq:not}
\beta_i=\frac{1}{\gamma_i-1}, R_i=2 \ln{\frac{N}{c_i}}, c_i=\bar{k}_i \frac{\sin{T_i \pi}}{2T_i}\left(\frac{\gamma_i-2}{\gamma_i-1}\right)^2,
\end{equation}
where $\bar{k}_i$, $\gamma_i$ and $T_i$ are respectively the target average degree, power law degree distribution exponent, and temperature in layer~$i=1,2$.

\emph{(1) Assignment of coordinates in layer~1.} 
For each node  $i=1, 2, \ldots, N$ in layer~1 we sample  its radial coordinate $r_{1, i}$ from the PDF:
\begin{equation}
\label{eq:eq_rho_1}
\rho_1(r_1)=\frac{1}{2\beta_1}e^{\frac{1}{2\beta_1}(r_1-R_1)},
\end{equation}
while its angular coordinate $\theta_{1, i}$ is sampled from the uniform PDF:
\begin{equation}
\label{eq:eq_theta}
f(\theta)=\frac{1}{2\pi},~\theta\in [0, 2\pi].
\end{equation}

\emph{(2) Assignment of coordinates in layer~2.} The radial coordinate $r_{2, i}$ of each node $i=1,2, \dots, N$ in layer~2 is sampled from the conditional PDF:
\begin{align}
\label{eq:r_cond_text}
\nonumber \rho_2(r_2 | r_1=r_{1, i}, \eta) = \frac{1}{2\beta_2} e^{\phi_1-(\phi_1^\eta+\phi_2^\eta)^\frac{1}{\eta}}(\phi_1 \phi_2)^{\eta-1} \\
 \times (\phi_1^\eta+\phi_2^\eta)^{\frac{1}{\eta}-2} \left((\phi_1^\eta+\phi_2^\eta)^{\frac{1}{\eta}}+\eta-1\right),\\
\nonumber \phi_i \equiv \frac{R_i-r_i}{2\beta_i}, i=1,2,\text{~~} \eta \equiv \frac{1}{1-\nu} \in [1, \infty),
\end{align}
where $\nu \in [0, 1)$ is the radial correlation strength parameter. The higher the value of $\nu$ the stronger is the correlation between $r_{2, i}$ and $r_{1, i}$. At $\nu \to 1$, $r_{1, i}, r_{2, i}$ are maximally correlated, while at $\nu=0$,  $r_{1, i}, r_{2, i}$ are uncorrelated. We note that $r_{1, i}=r_{2, i}$ at $\nu \to 1$ only if $R_1=R_2$ and $\beta_1=\beta_2$. To derive Eq.~(\ref{eq:r_cond_text}) we use the bivariate Gumbel-Hougaard copula~\cite{kleineberg2016np}, see Appendix~\ref{sec:conditional_radial_pdf}. The copula ensures that no matter the value of $\nu$ the marginal PDF of $r_{2, i}$ is the same as in the $\mathbb{H}^{2}$ model:
\begin{align}
\label{eq:eq_rho_2}
\rho_2(r_2)=\frac{1}{2\beta_2}e^{\frac{1}{2\beta_2}(r_2-R_2)}.
\end{align}

The angular coordinate $\theta_{2, i}$ of each node $i=1,2, \ldots, N$ in layer~2 is obtained by:
\begin{equation}
\theta_{2, i} = \mod\left[\theta_{1, i} + \frac{2 \pi l_i}{N},2\pi\right],
\label{eq:assign_theta_text}  
\end{equation}
where $l_i$ is a directed arc length on the circle of radius $R=N/(2\pi)$, sampled from the zero-mean truncated Gaussian PDF:
\begin{align}
\label{eq:truncated_normal}
f_g(l) =\frac{\phi\left(\frac{l}{\sigma}\right) }{ \sigma \erf(\frac{N}{2 \sqrt{2} \sigma})},\\
\nonumber -\frac{N}{2} \leq l \leq \frac{N}{2}, \text{~~} \sigma \equiv  \sigma_0 \left(\frac{1}{g}-1\right),
\end{align}
where $\sigma \in [0,\infty)$ is the standard deviation of the PDF, while $g \in (0,1]$ is the angular correlation strength parameter. Furthermore, $\sigma_0 = \min[100,N/(4\pi)]$ denotes the standard deviation for $g=0.5$, $\phi(x) = \frac{1}{\sqrt{2 \pi}} e^{-\frac{1}{2} x^2}$, and $\erf(x) = \frac{2}{\sqrt{\pi}} \int_0^x e^{-t^2} \mathrm{d}t$ is the Gauss error function. 

The higher the value of $g$ the stronger is the correlation between $\theta_{2,i}$ and $\theta_{1,i}$. At $g \to 0$, $\sigma \to \infty$, $f_g(l)$ becomes the uniform PDF, and $\theta_{2, i}$, $\theta_{1, i}$ are uncorrelated. At $g=1$, $\sigma=0$, and $l_i=0$, meaning that the angles of each node are the same in the two layers. The marginal PDF of $\theta_{2, i}$ is still the uniform PDF (Eq.~(\ref{eq:eq_theta})).

\emph{(3) Creation of edges.} Once all node coordinates are assigned, we connect each node pair $i, j$ in layers~1 and 2 with the corresponding $\mathbb{H}^{2}$  connection probabilities given in Eqs.~(\ref{eq:c_prob_1}), (\ref{eq:c_prob_2}) below:
\begin{eqnarray}
\label{eq:c_prob_1}
p_1(x_1^{ij})=\frac{1}{1+e^{\frac{1}{2 T_1}(x_1^{ij}-R_1)}},\\
\label{eq:c_prob_2}
p_2(x_2^{ij})=\frac{1}{1+e^{\frac{1}{2 T_2}(x_2^{ij}-R_2)}},
\end{eqnarray}
where $x_1^{ij} \leq 2 R_1, x_2^{ij} \leq 2 R_2$ are the hyperbolic distances between nodes $i, j$ in layers~1 and 2.

\subsection{HyperMap}
\label{sec:HyperMap} 

Finally, given a real network, the HyperMap method~\cite{frag:hypermap, frag:hypermap_cn} can be used to infer the popularity and similarity coordinates of its nodes. The method is based on maximum likelihood estimation. On its input it takes the network adjacency matrix  $\alpha_{ij}$ ($\alpha_{ij}=\alpha_{ji}=1$ if there is a link between nodes $i$ and $j$, and $\alpha_{ij}=\alpha_{ji}=0$ otherwise), and computes radial and angular coordinates $r_i, \theta_i$, for all nodes $i \leq N$. The radial coordinates are related to the observed node degrees $k_i$:
\begin{equation}
r_i \propto R-2\ln{k_i},
\end{equation}
while the angular coordinates are found by maximizing the likelihood:
\begin{equation}
\label{eq:likelihood}
\mathcal L=\prod_{1 \leq j < i \leq N} p(x_{ij})^{\alpha_{ij}}\left[1-p(x_{ij})\right]^{1-\alpha_{ij}}.
\end{equation}
The product in the above relation goes over all node pairs $i, j$ in the network, $x_{ij}$ is the hyperbolic distance between pair $i, j$ and $p(x_{ij})$ is the connection probability in Eq.~(\ref{eq:p_x_ij}). HyperMap was used in~\cite{kleineberg2016np} to independently map the layers of different real multiplexes into hyperbolic spaces. Its implementation is available at~\cite{hypermap_code}.

\section{Evidence of link persistence in real multiplexes}
\label{sec:evidence}

\begin{table*}
\centering{
\begin{tabular}{lllll}
\textbf{Name} &  \textbf{Type} & \textbf{Nodes} & \textbf{layer~1, layer~2} &~~~$\nu$,~$g$\\\hline
Internet & Technological & Autonomous Systems & IPv4 AS topology, IPv6 AS topology &0.40, 0.40 \\\hline
Drosophila & Biological & Proteins & Suppressive genetic interaction, additive genetic interaction &0.47, 0.82 \\\hline
C. Elegans & Biological & Neurons & Electric, chemical monadic synaptic junctions & 0.22, 0.48 \\\hline
Human Brain & Biological & Brain regions & Structural network, functional network & 0.18, 0.42 \\\hline
arXiv & Collaboration & Authors &  cond-mat.disnn and physics.bioph categories  & 0.45, 0.93 \\\hline
Physicians & Social & Physicians & Discussion, advise relations & 0.47, 0.84\\\hline
\end{tabular}
\caption{Overview of the considered real-world multiplex network data. 
\label{tab_datasets}}}
\end{table*}

We now provide empirical evidence from real multiplexes suggesting that connections can persist from one layer to another irrespectively of the hyperbolic distances that they span. We call such links \emph{persistent links}. To this end, we consider layer pairs of different real multiplexes and their hyperbolic embeddings from~\cite{kleineberg2016np}. See Table~\ref{tab_datasets} for an overview of the data and Appendix~\ref{sec:real_data} for further details. The correlation strengths $\nu, g$ in Table~\ref{tab_datasets} were estimated in~\cite{kleineberg2016np}. The considered multiplexes are paradigmatic systems from different domains: technological (Internet), biological (Drosophila, C. Elegans, Human Brain), scientific collaboration (arXiv), and society (Physicians). We have verified that similar results hold in the other multiplexes and layer pairs considered in~\cite{kleineberg2016np}.

Let us classify all pairs of nodes in layer~1 that also exist in layer~2 into three sets. First, $S_\text{c}$ contains the pairs connected in layer~1. Second, $S_\text{d}$ contains the pairs disconnected in layer~1. Finally, $S_{\text{all}}$ simply contains all pairs in $S_\text{c}$ and $S_\text{d}$. For each set, we compute the empirical trans-layer connection probability, which is the probability that a pair in the set is connected in layer~2 given its hyperbolic distance $x_1$ in layer~1. These probabilities are respectively denoted by $p_{\text{trans}}^{\text{c}}(x_1)$, $p_{\text{trans}}^{\text{d}}(x_1)$, $p_{\text{trans}}^{\text{all}}(x_1)$, and computed as follows. For each set, we compute the hyperbolic distances among its pairs in layer~1. We then bin the range of hyperbolic distances from zero to the maximum distance into small bins. For each bin we find all the pairs located at the hyperbolic distances falling within the bin. The percentage of pairs in this set of pairs that are connected in layer~2 is the value of the corresponding trans-layer connection probability at the bin.  

For each of $S_\text{c}, S_\text{d}, S_{\text{all}}$, we also compute the empirical connection probability in layer~2, $p_2^{\text{c}}(x_2)$, $p_2^{\text{d}}(x_2)$, $p_2^{\text{all}}(x_2)$. These probabilities are computed following the same binning procedure as above, except that we consider the hyperbolic distances $x_2$ among the pairs in layer~2 instead of layer~1.  For each of $S_\text{c}, S_\text{d}, S_{\text{all}}$, we further compute the average hyperbolic distance among its pairs in layer~2 conditioned on their distance in layer~1, $E^{\text{c}}[x_2|x_1]$, $E^{\text{d}}[x_2|x_1]$, $E^{\text{all}}[x_2|x_1]$. This metric is used to illustrate the hyperbolic distance correlations for each set of nodes in the two layers.

\begin{figure*}
\includegraphics[width=2.32in, height=1.35in]{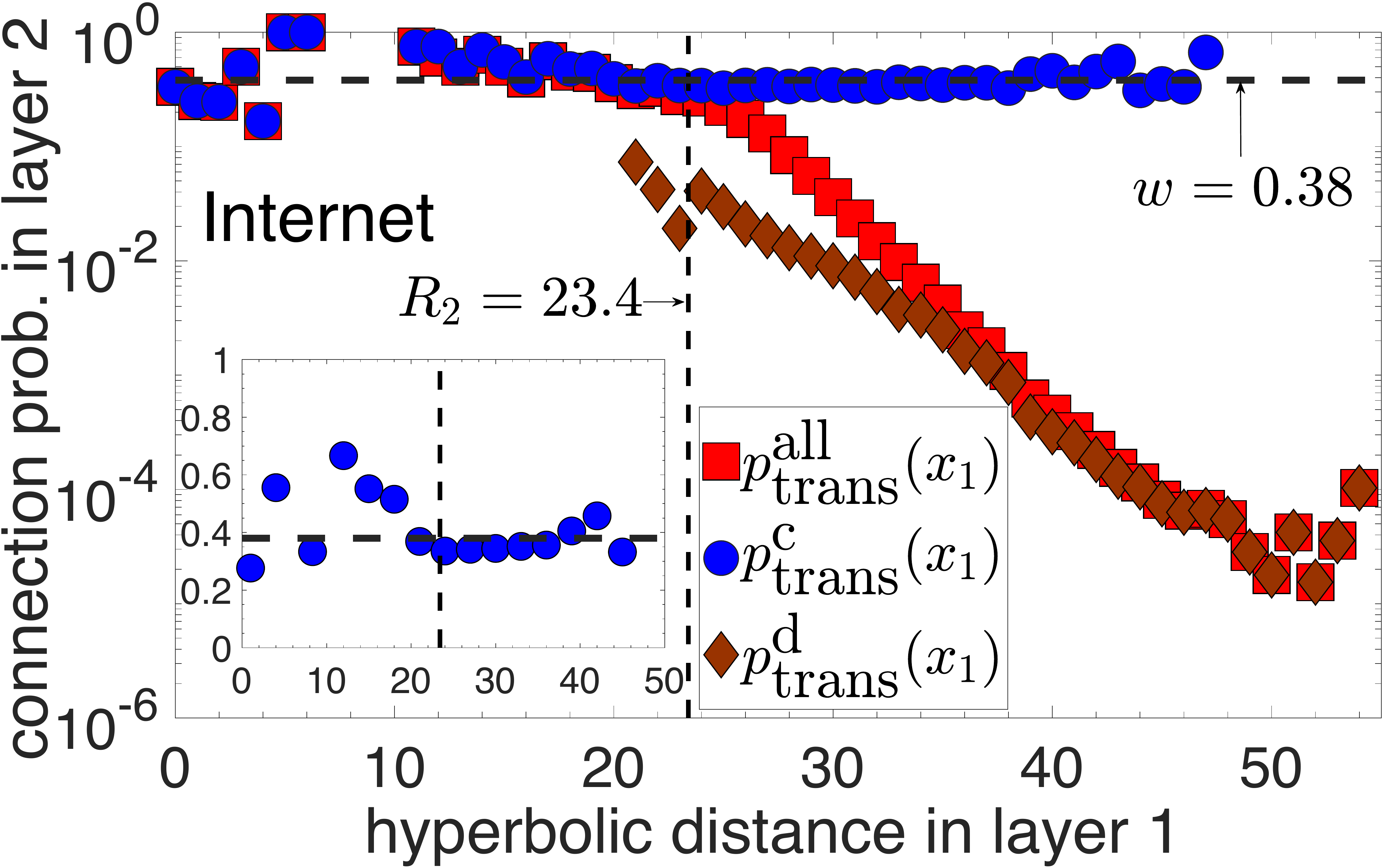}
\includegraphics[width=2.32in, height=1.35in]{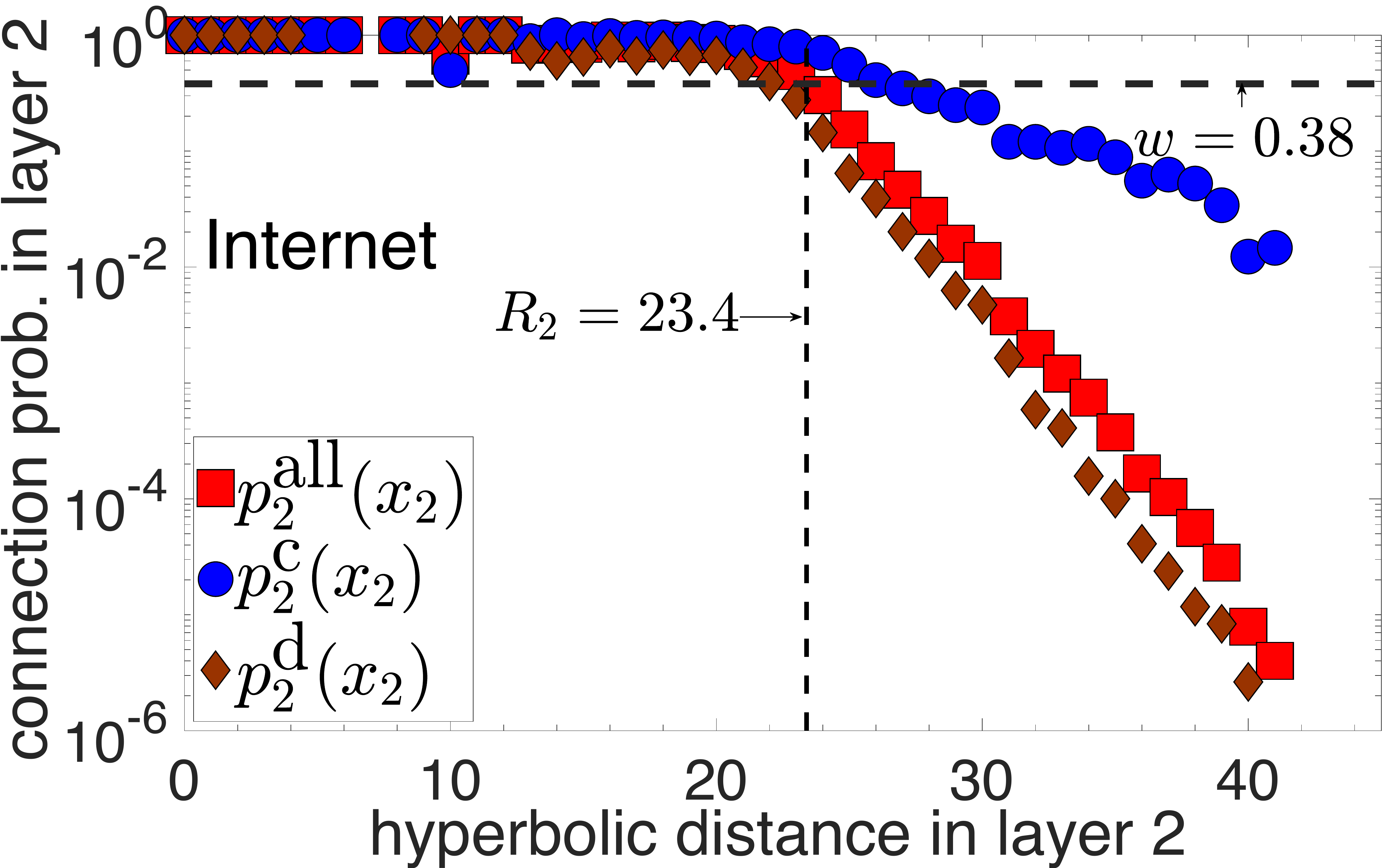}
\includegraphics[width=2.30in, height=1.33in]{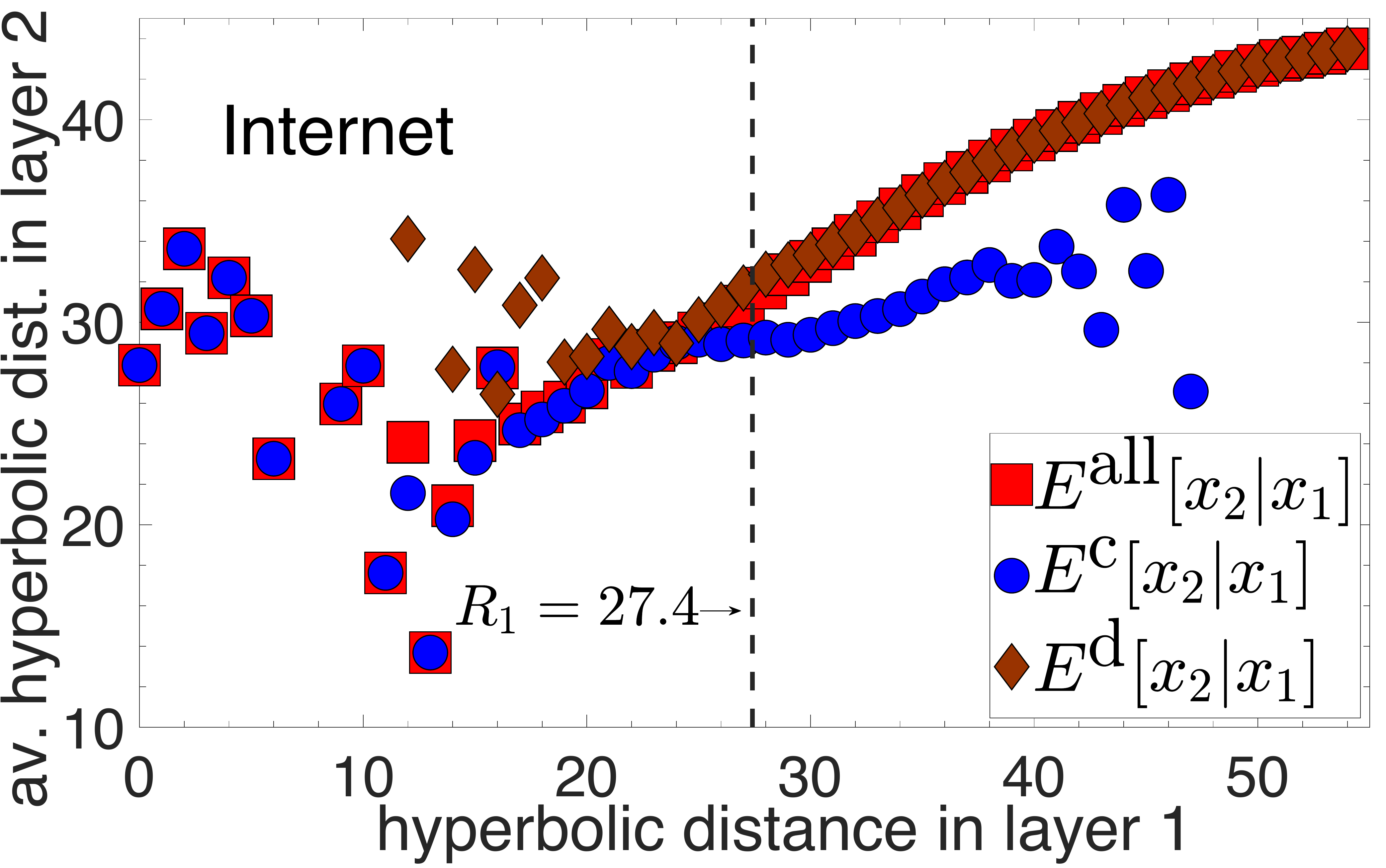}\\
\includegraphics[width=2.32in, height=1.35in]{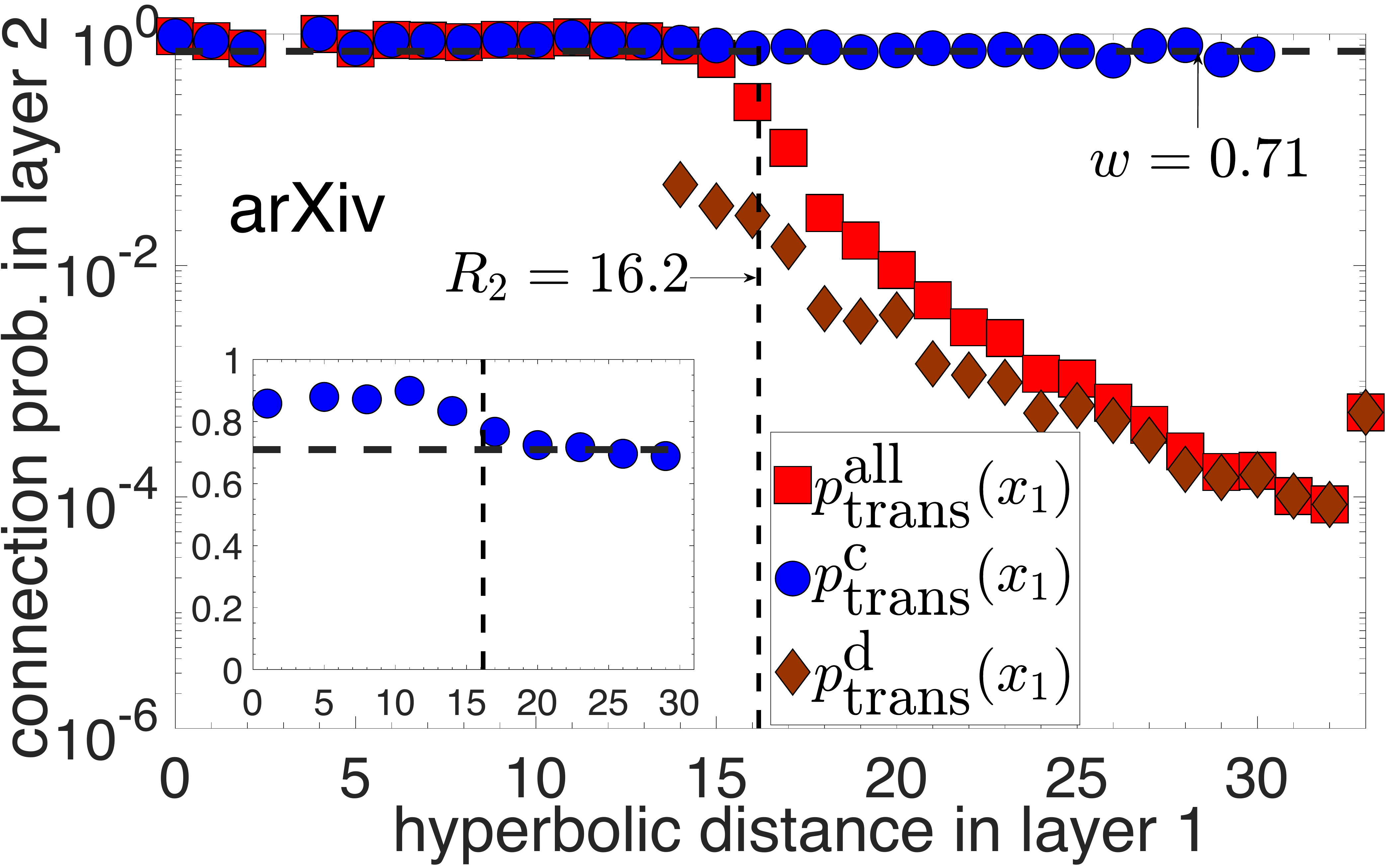}
\includegraphics[width=2.32in, height=1.35in]{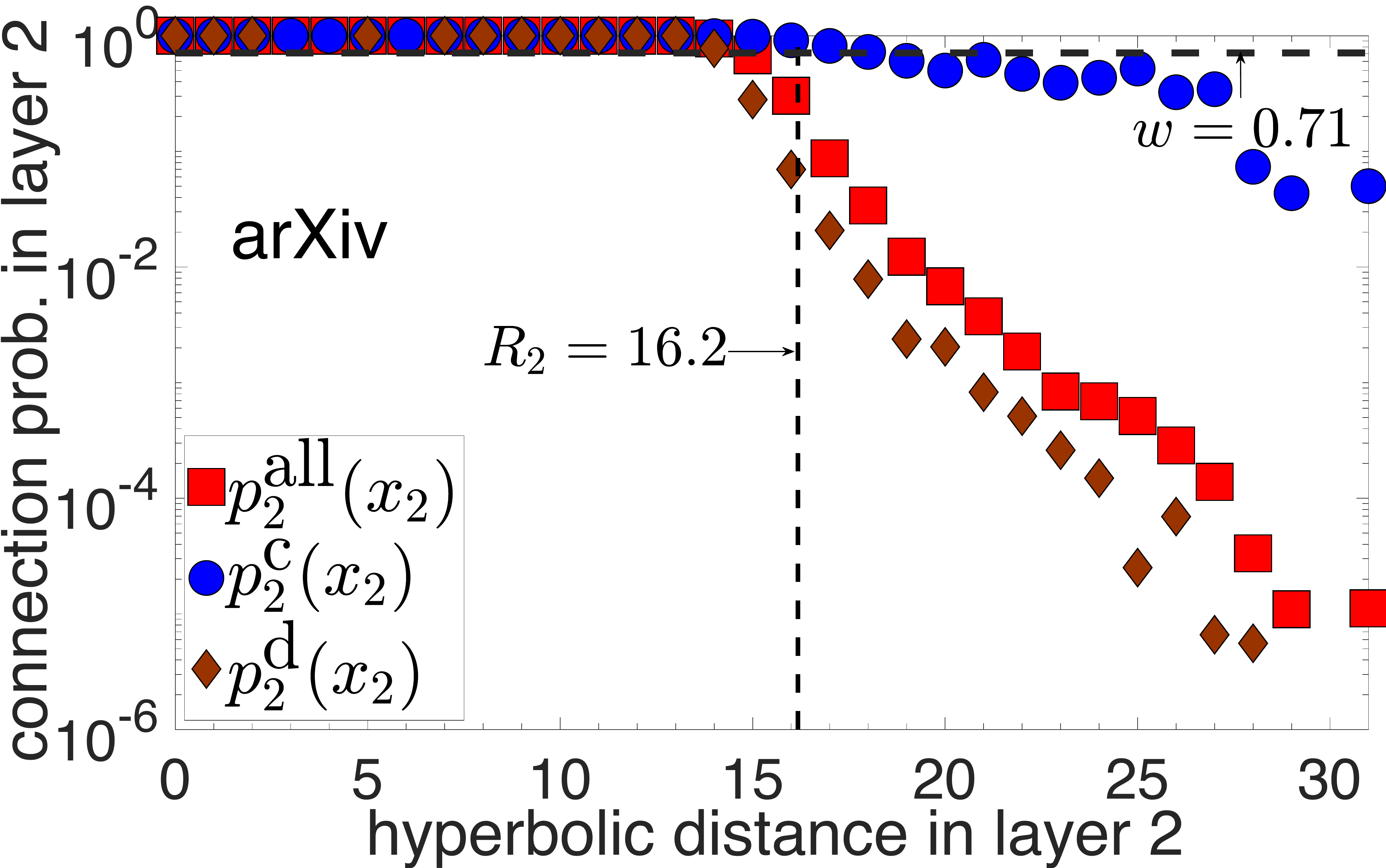}
\includegraphics[width=2.32in, height=1.35in]{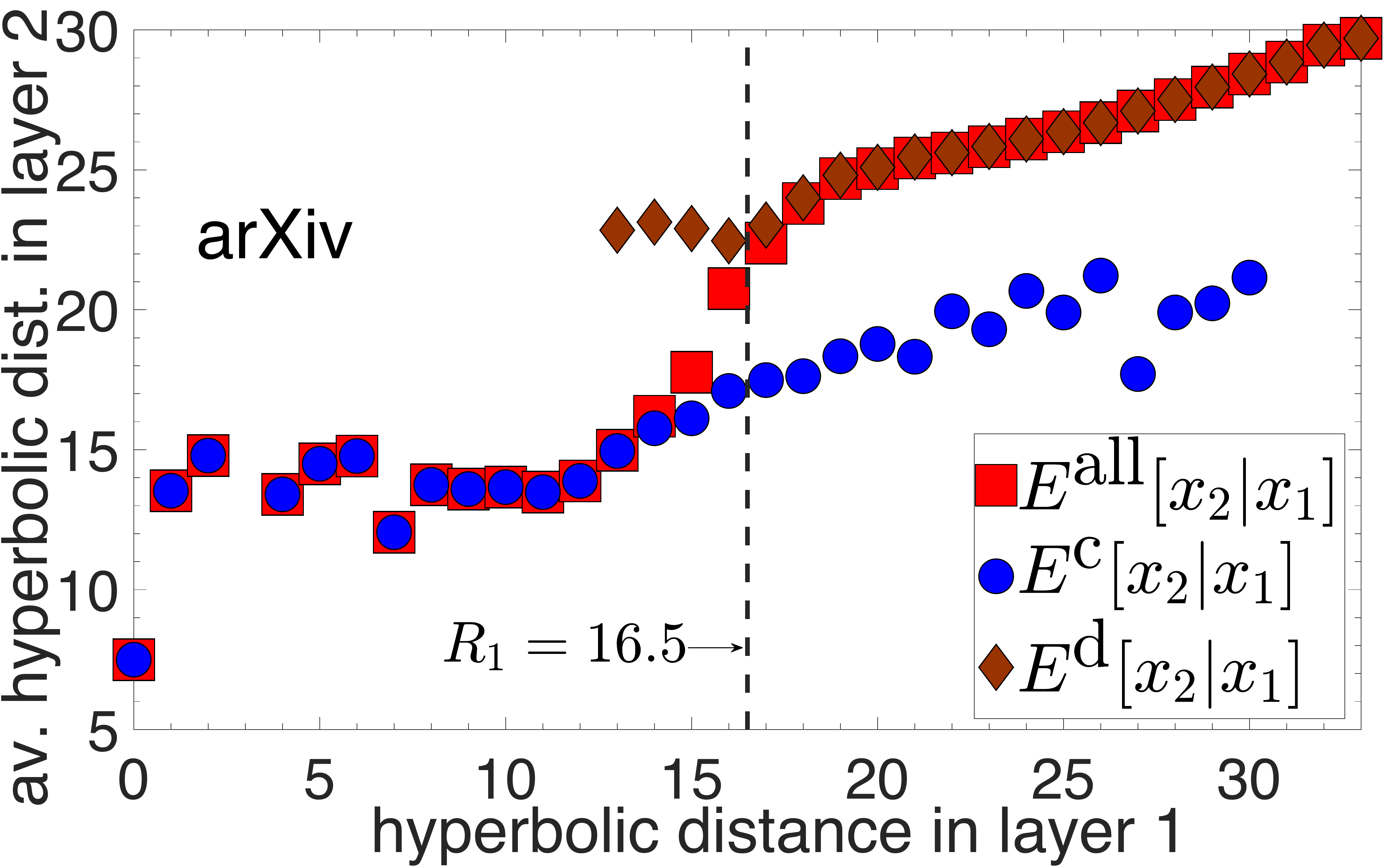}\\
\includegraphics[width=2.32in, height=1.35in]{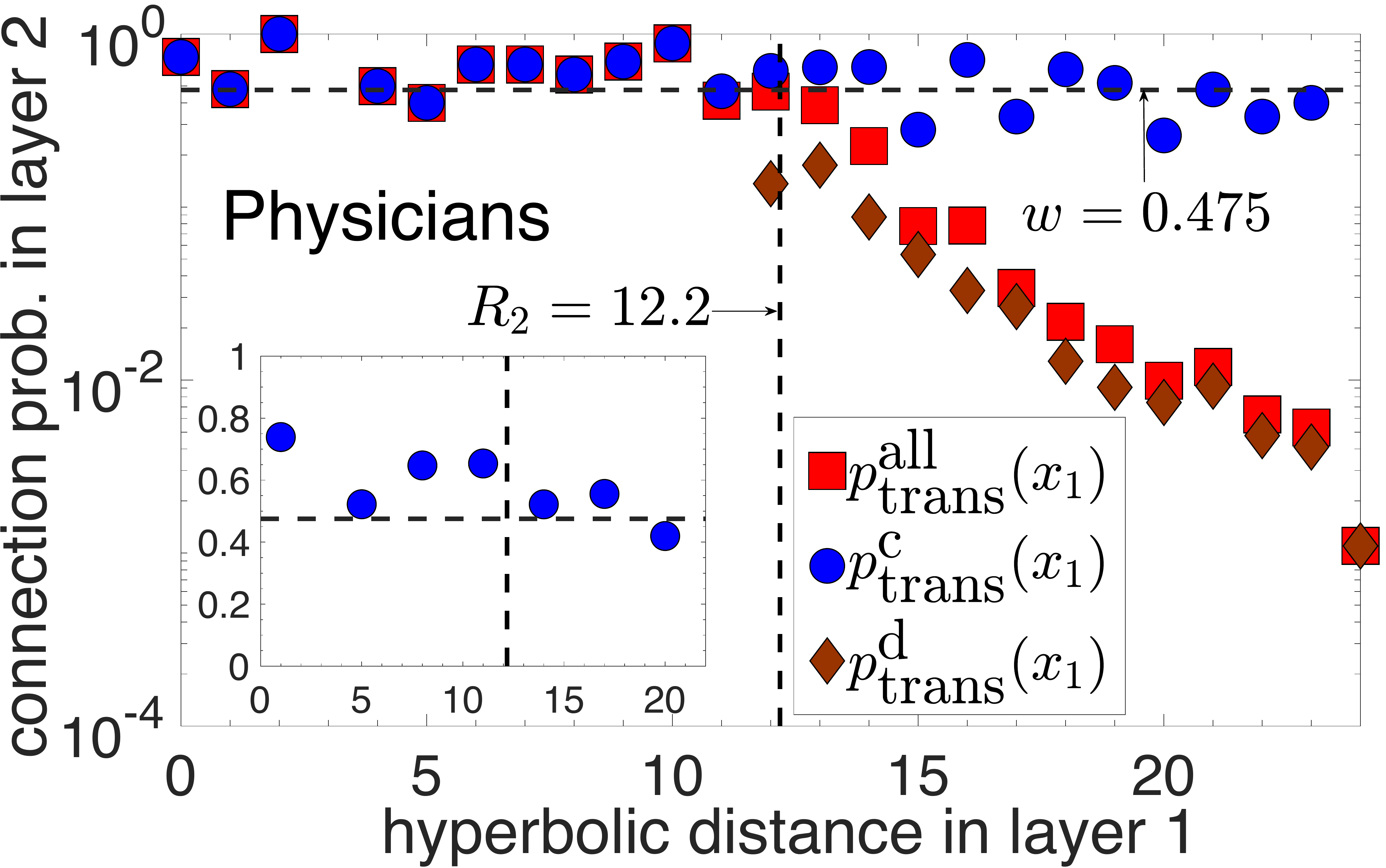}
\includegraphics[width=2.32in, height=1.35in]{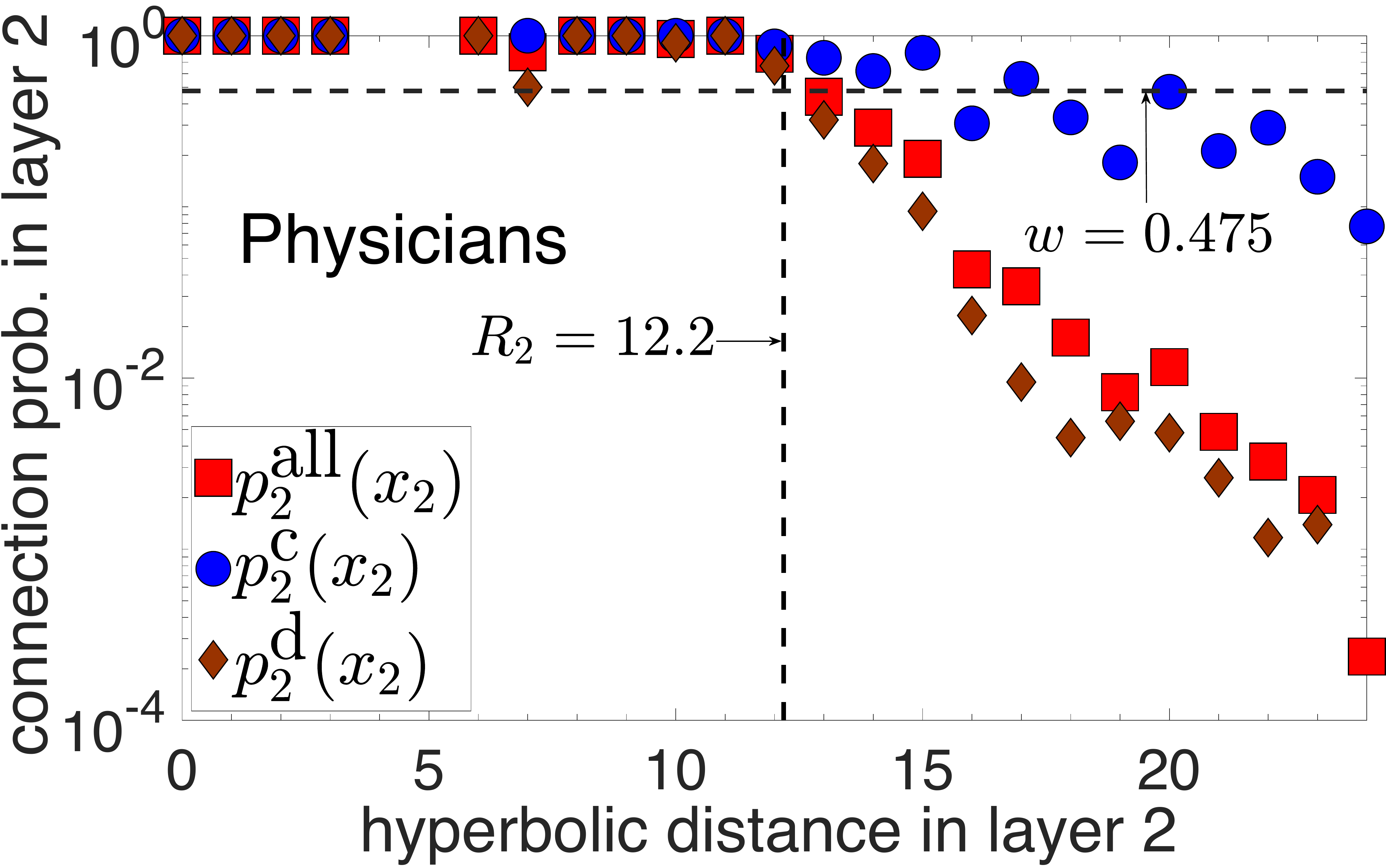}
\includegraphics[width=2.32in, height=1.35in]{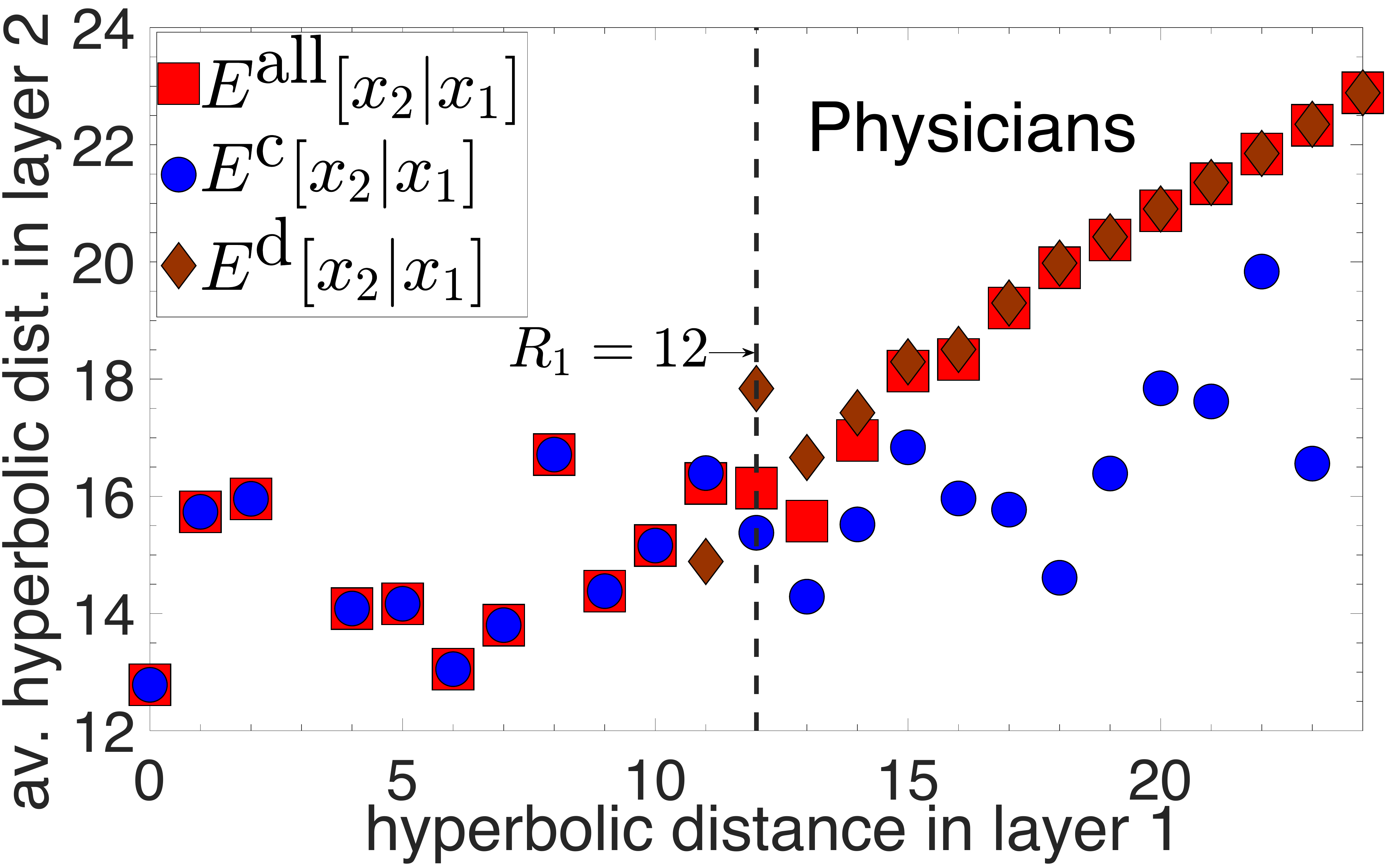}\\
\includegraphics[width=2.32in, height=1.35in]{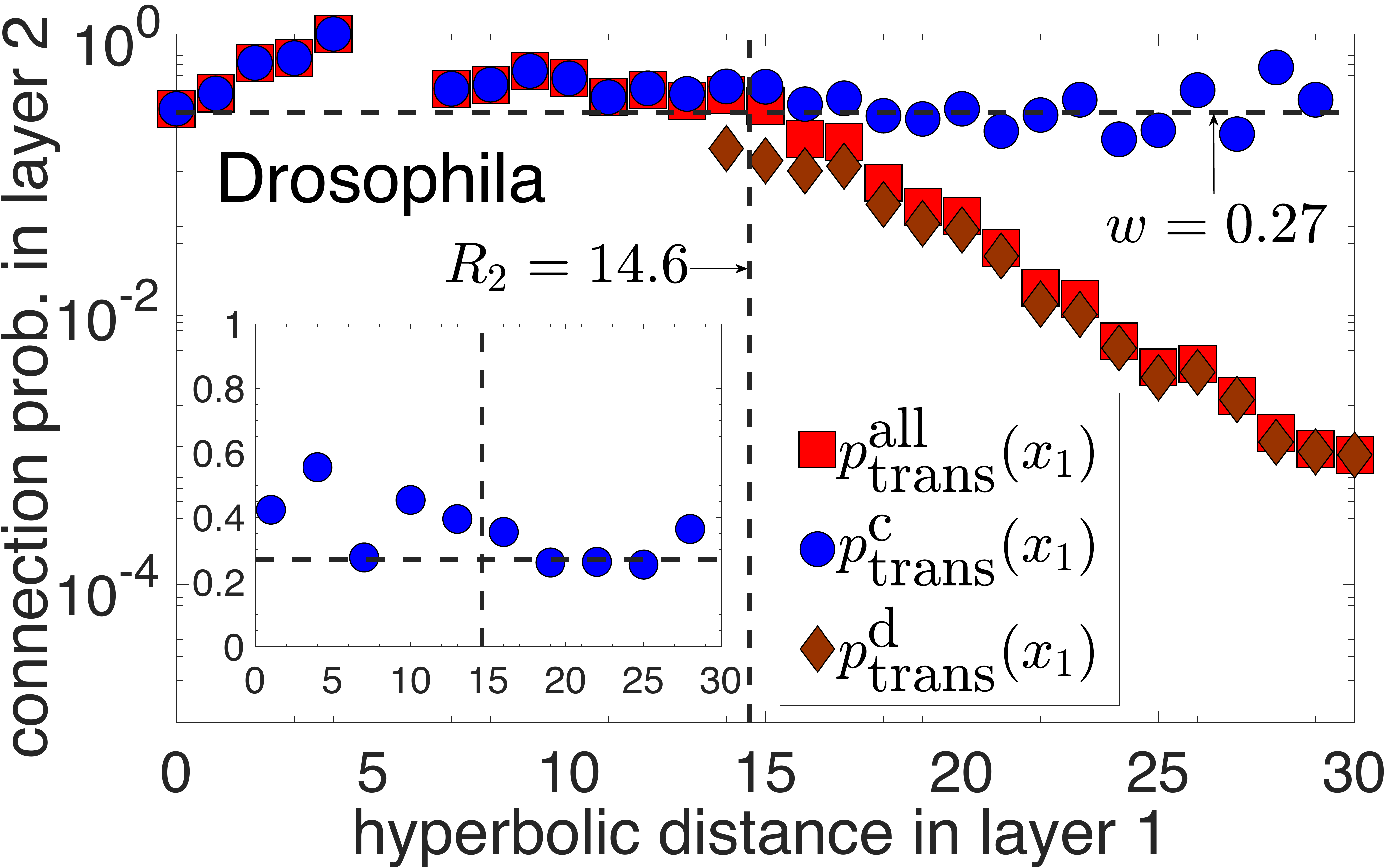}
\includegraphics[width=2.32in, height=1.35in]{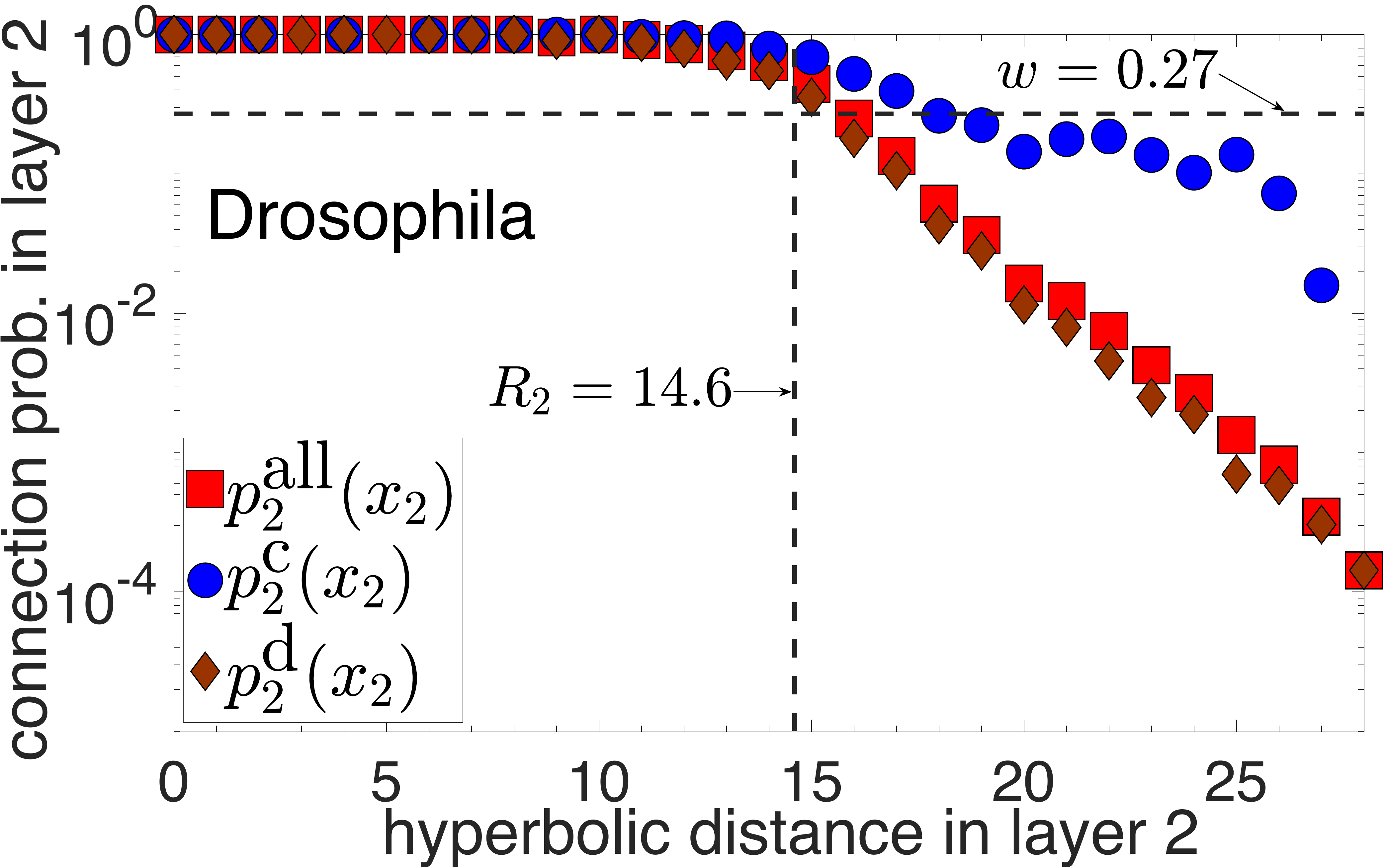}
\includegraphics[width=2.32in, height=1.35in]{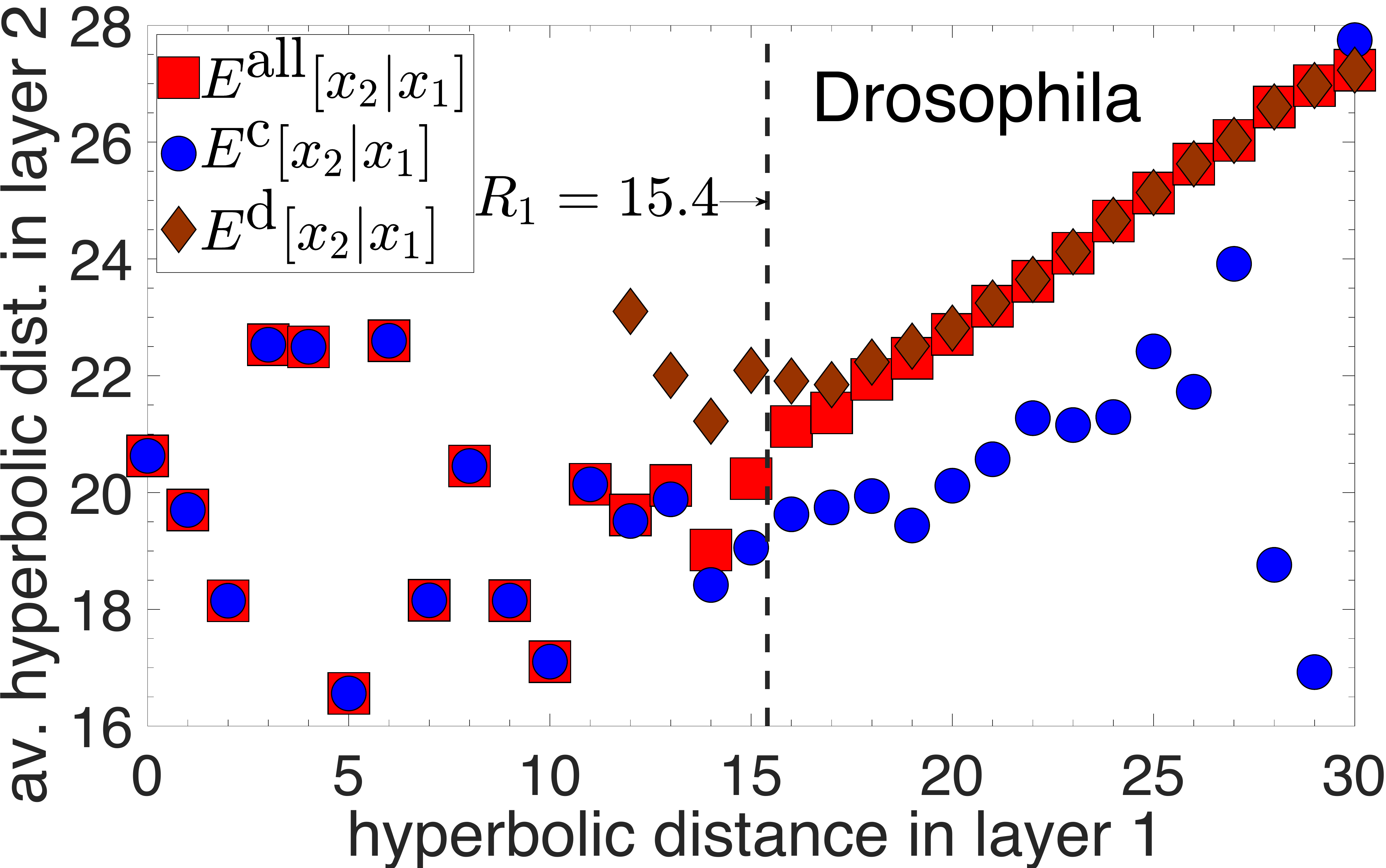}\\
\includegraphics[width=2.33in, height=1.35in]{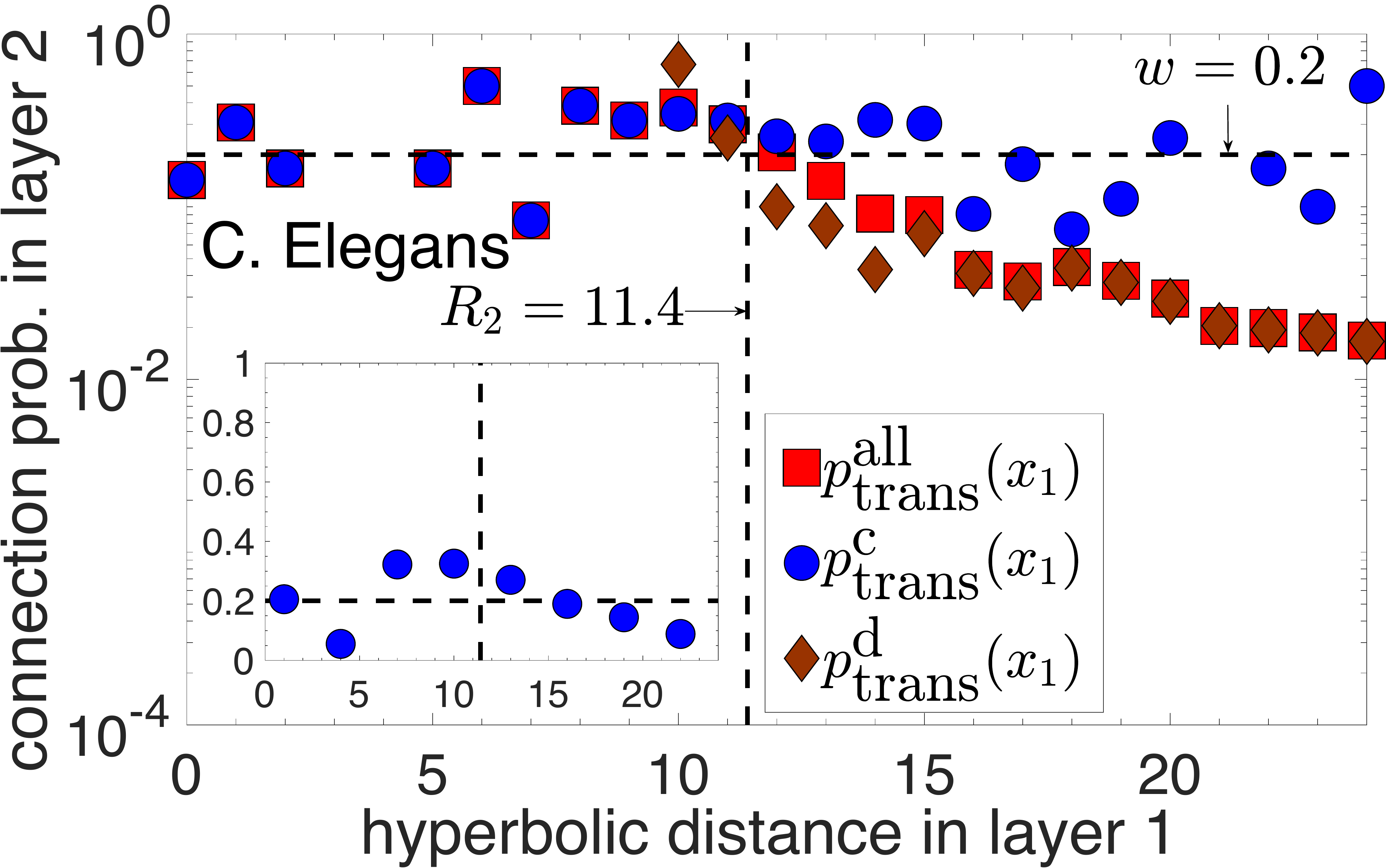}
\includegraphics[width=2.32in, height=1.35in]{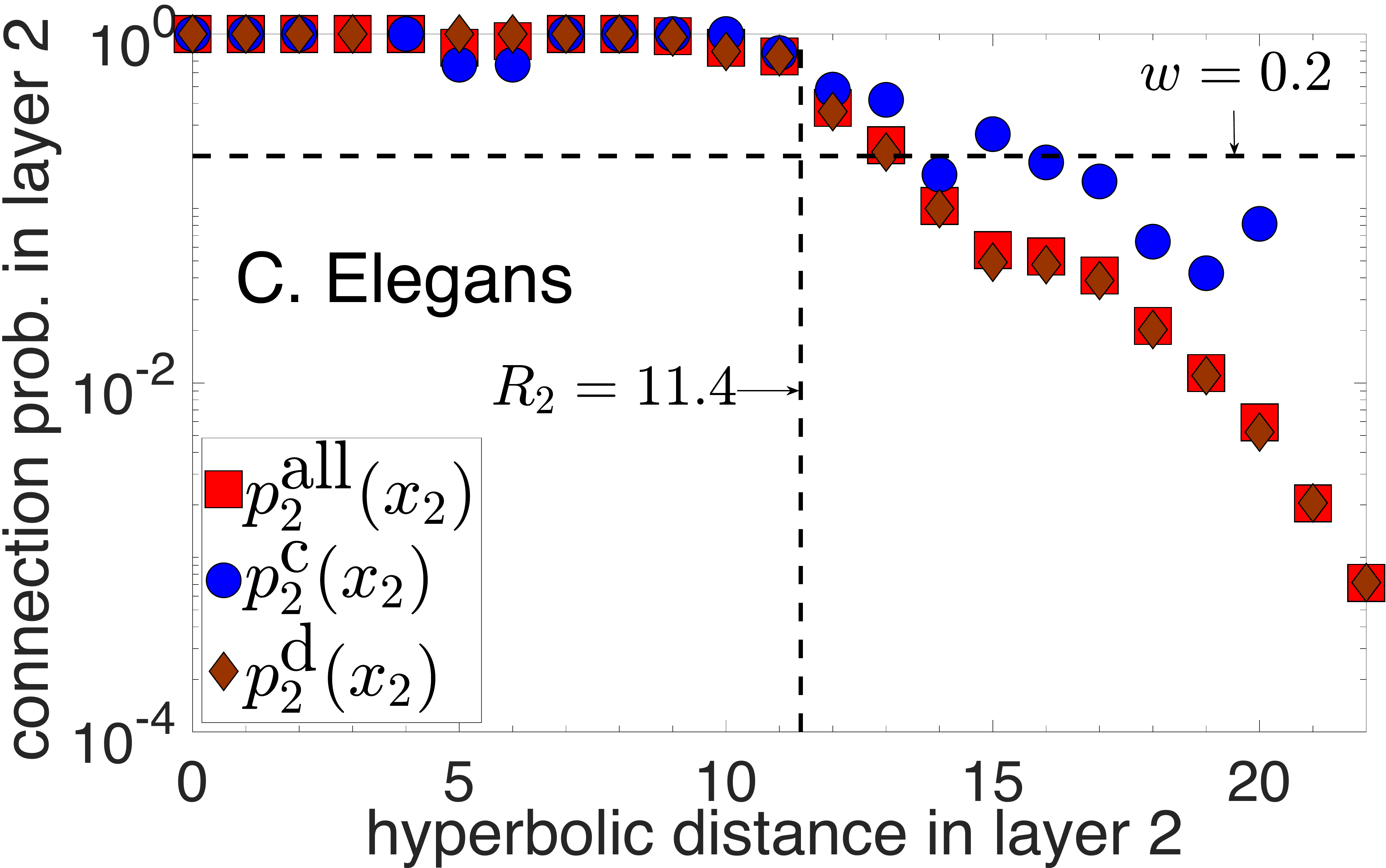}
\includegraphics[width=2.32in, height=1.35in]{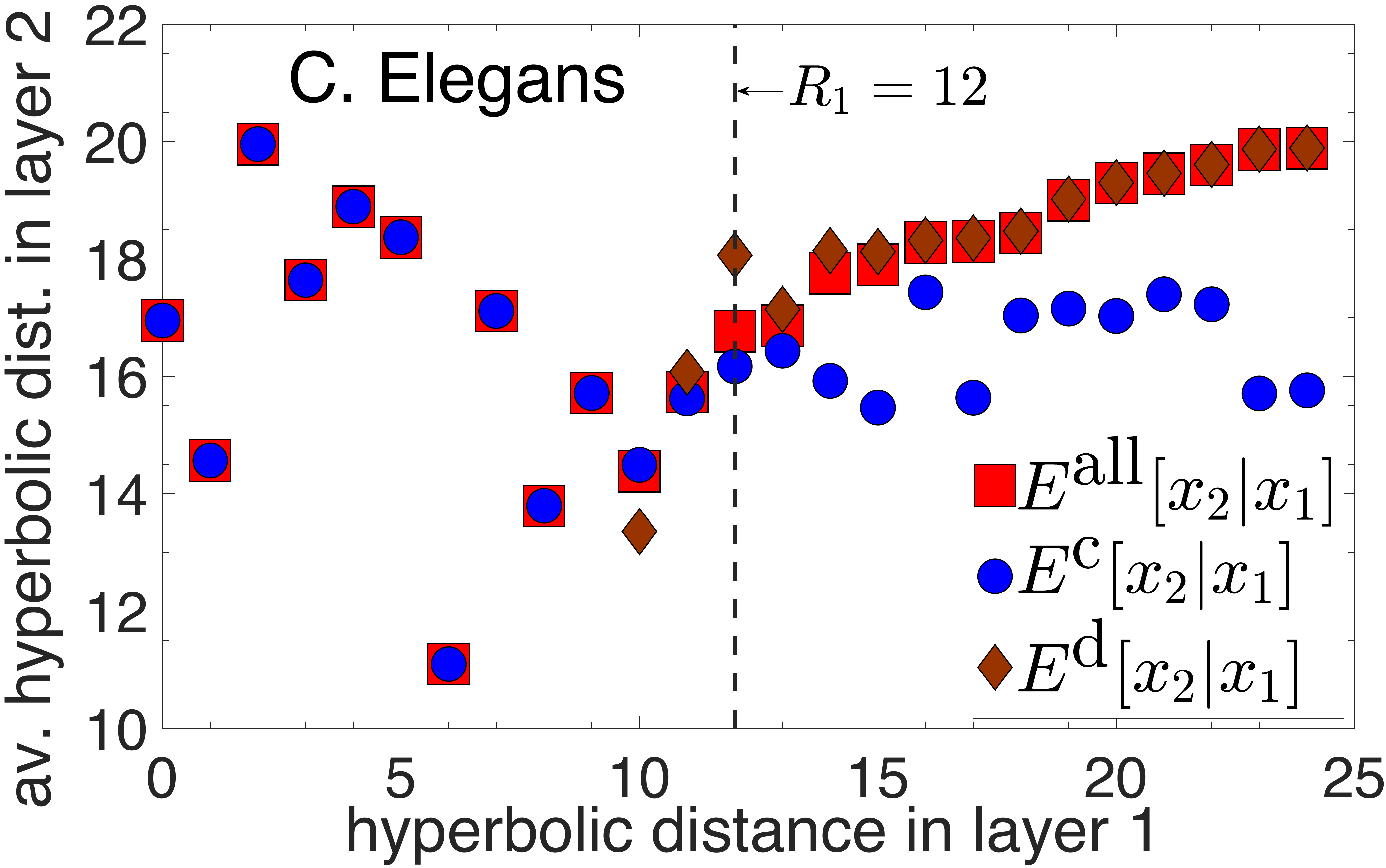}\\
\includegraphics[width=2.31in, height=1.35in]{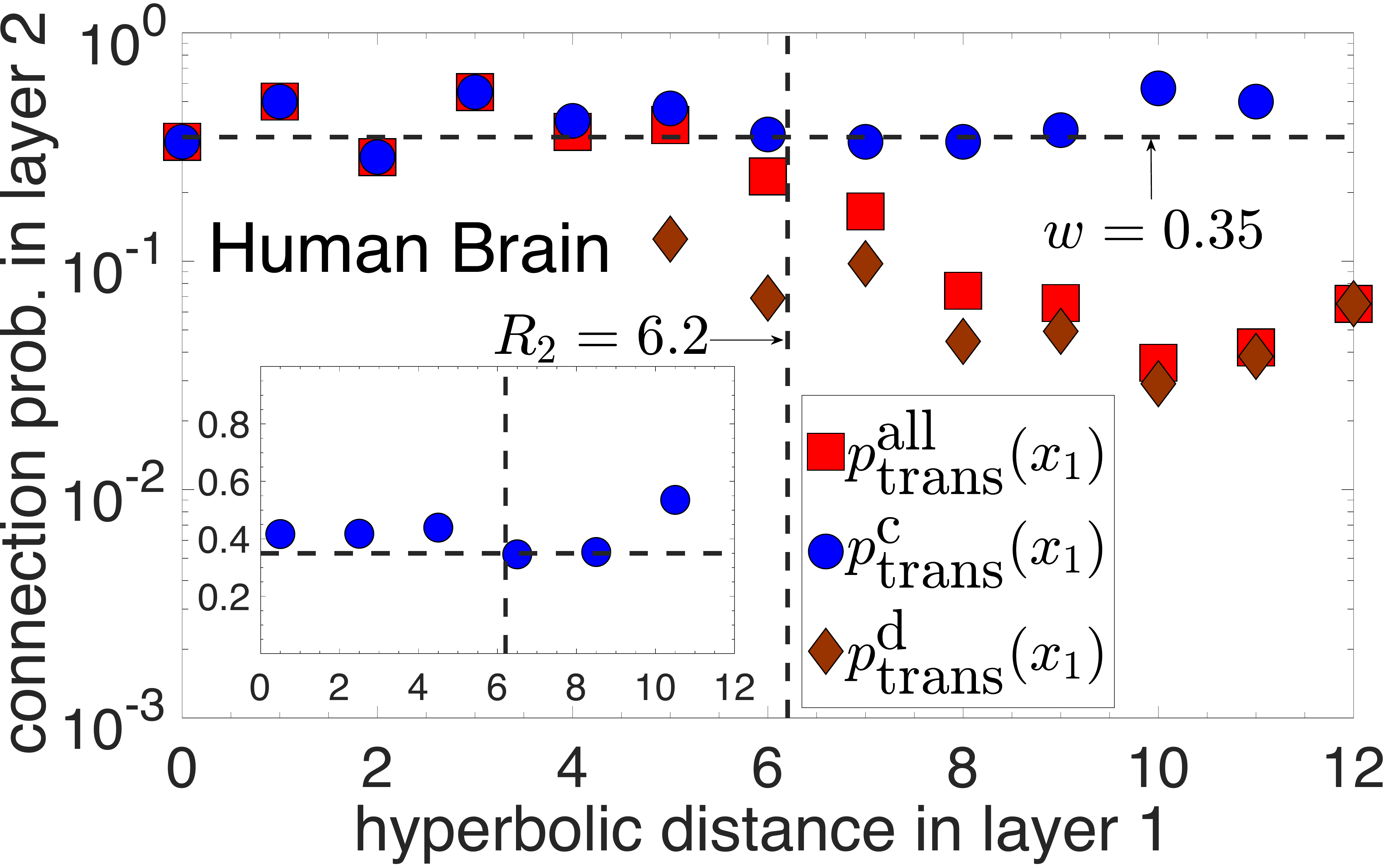}
\includegraphics[width=2.31in, height=1.35in]{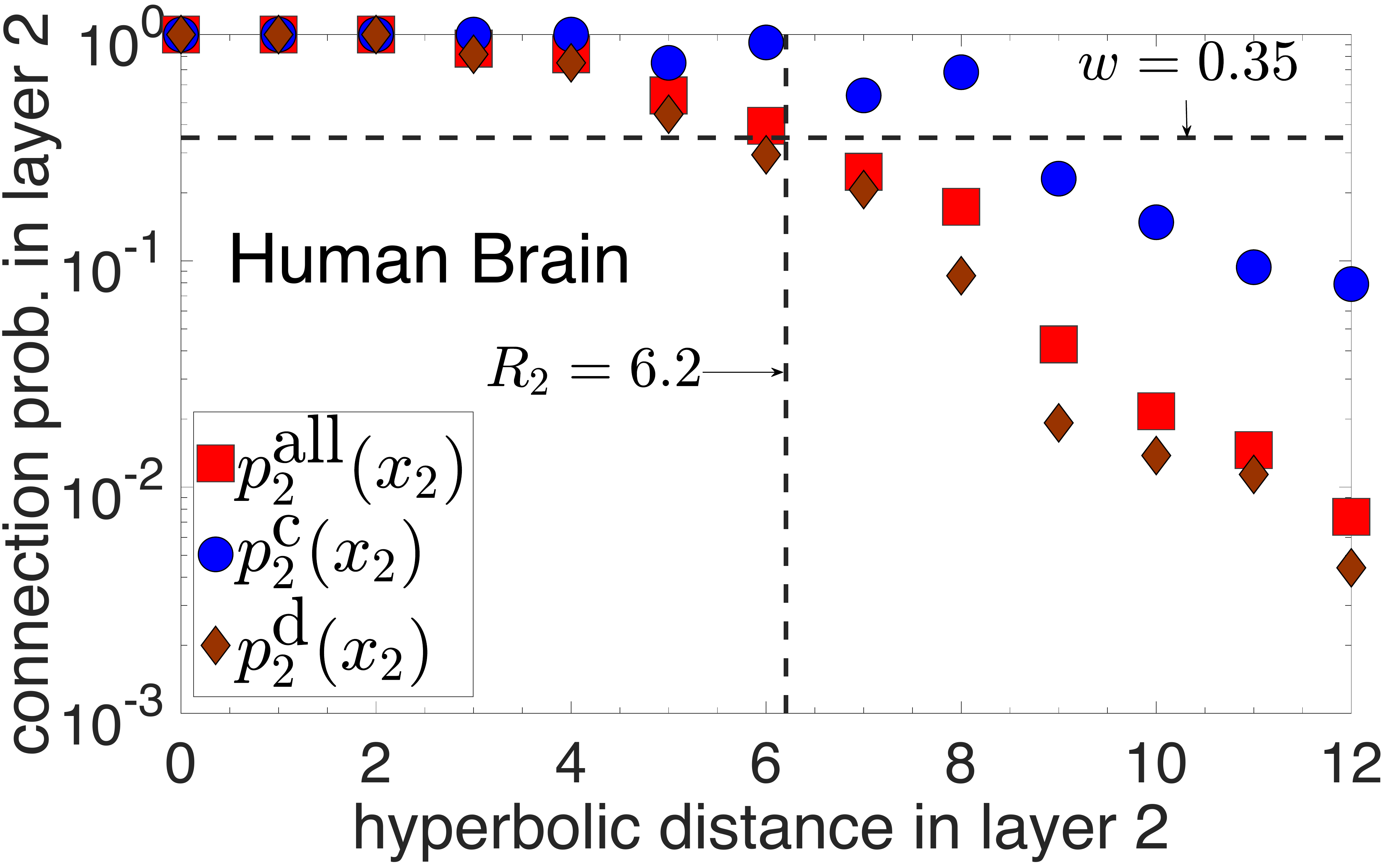}
\includegraphics[width=2.35in, height=1.38in]{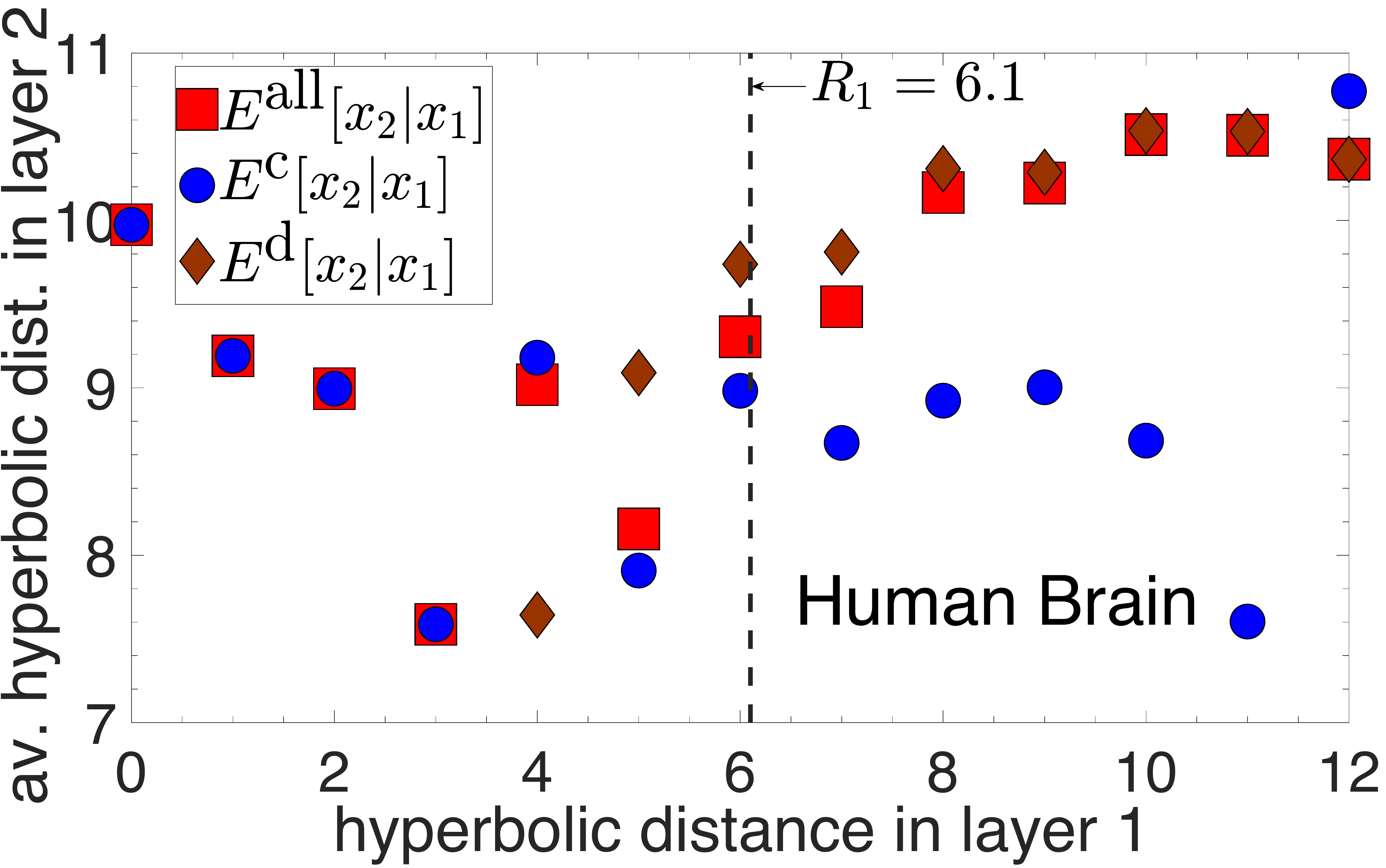}
\caption{Link persistence and hyperbolic distance correlations in the IPv4/IPv6 Internet, arXiv, Physicians,  Drosophila, C.~Elegans and Human Brain multiplexes. The plots show the trans-layer connection probabilities $p_{\text{trans}}^{\text{all}}(x_1)$, $p_{\text{trans}}^{\text{c}}(x_1)$, $p_{\text{trans}}^{\text{d}}(x_1)$, the connection probabilities $p_2^{\text{all}}(x_2)$, $p_2^{\text{c}}(x_2)$, $p_2^{\text{d}}(x_2)$, and the conditional average hyperbolic distances $E^{\text{all}}[x_2|x_1]$, $E^{\text{c}}[x_2|x_1]$, $E^{\text{d}}[x_2|x_1]$. The $y$-axes in the connection probability plots are in log-scale. The insets show $p_{\text{trans}}^{\text{c}}(x_1)$ in linear scale. The bins in the $x$-axes have size $1$ except from the insets where larger bins have been used to reduce fluctuations. The horizontal dashed lines indicate the value of the estimated link persistence probability $w$ (Sec.~\ref{sec:w_infer}). The vertical dashed lines indicate the hyperbolic disc radii $R_1, R_2$ of layers~1 and 2.
\label{fig:empirical_results}}
\end{figure*}

The results are shown in Fig.~\ref{fig:empirical_results}, where we make similar observations in all considered multiplexes.\footnote{Similar results hold if we consider the layers in the direction $2$ to $1$ instead of $1$ to $2$ (see Sec.~\ref{sec:w_infer} for the estimated link persistence probabilities in the direction $2$ to $1$).} First, we see that $p_{\text{trans}}^{\text{all}}(x_1)$ decreases with $x_1$. This means that nodes at smaller hyperbolic distances in layer~1 have higher chances of being connected in layer~2, as already observed~\cite{kleineberg2016np}. However, here we also observe that $p_{\text{trans}}^{\text{c}}(x_1)$ is significantly larger than $p_{\text{trans}}^{\text{d}}(x_1)$, and virtually independent of $x_1$ at  $x_1 > R_2$. Furthermore, even though it may have an increasing trend at $x_1 < R_2$, it does not drastically change. These observations suggest that a significant percentage of connected layer~1 pairs remain connected in layer~2 irrespectively of their distances $x_1$; and consequently, irrespectively of their distances $x_2$. We also see that at smaller $x_1$, $p_{\text{trans}}^{\text{c}}(x_1)=p_{\text{trans}}^{\text{all}}(x_1)$. This is because there are no disconnected pairs at those distances, i.e., $S_c=S_{\text{all}}, S_d=\emptyset$ for those distances. At larger $x_1$, where $S_d \neq \emptyset$, we see that $p_{\text{trans}}^{\text{d}}(x_1)$ decreases with $x_1$, similarly to $p_{\text{trans}}^{\text{all}}(x_1)$. 

The empirical connection probabilities $p_2^{\text{c}}(x_2)$, $p_2^{\text{d}}(x_2)$,  $p_2^{\text{all}}(x_2)$ also show that at large $x_2 > R_2$ the connected layer~1 pairs have significantly higher chances of being connected in layer~2, compared to the disconnected layer~1 pairs. On the other hand, at smaller $x_2 < R_2$ both sets of pairs have similarly high chances of being connected.

Based on the above observations in the next section we consider a simple modification to the GMM, where a percentage of connected pairs in layer 1 remain connected in layer 2 irrespectively of their distances, and show that the modified GMM reproduces the behavior observed in Fig.~\ref{fig:empirical_results}.

\section{GMM with link persistence (GMM-LP)}
\label{sec:model}

To construct a two-layer synthetic multiplex with the modified GMM (GMM-LP) we follow exactly the same steps as in the GMM (Sec.~\ref{sec:GMM}), except that nodes in layer~2 connect according to the following procedure: 
\begin{itemize}
\item[(1)] if the nodes are connected in layer~1, they remain connected in layer~2 with probability $w$ or connect according to Eq.~(\ref{eq:c_prob_2});
\item[(2)] if the nodes are disconnected in layer~1, or exist only in layer~2, they connect according to Eq.~(\ref{eq:c_prob_2}).
\end{itemize}
We call $w \in [0, 1]$ \emph{link persistence probability}. In other words, the connection probability among connected layer~1 pairs in layer~2 is:
\begin{align}
\label{eq:p_c}
\nonumber p_2^{\text{c}}(x_2)&=w+(1-w)p_2(x_2)\\
&=p_2(x_2)+(1-p_2(x_2))w,
\end{align}
while for the disconnected layer~1 pairs:
\begin{equation}
\label{eq:p_d}
p_2^{\text{d}}(x_2)=p_2(x_2).
\end{equation}
We can see from Eqs.~(\ref{eq:p_c}),~(\ref{eq:p_d}) that the considered modification is equivalent to the following simpler procedure that we implement in the GMM-LP\footnote{The code implementing GMM-LP can be found online at~\cite{gmm_lp}.}:
\begin{itemize}
\item[(1)] construct layers~1 and~2 using the GMM;
\item[(2)] select at random a percentage $w$ of connected pairs in layer~1 that also exist in layer~2;
\item[(3)] connect the selected pairs in layer~2 if they are not already connected.
\end{itemize}

\begin{figure*}
\centering{
\includegraphics[width=1.73in, height=1.1in]{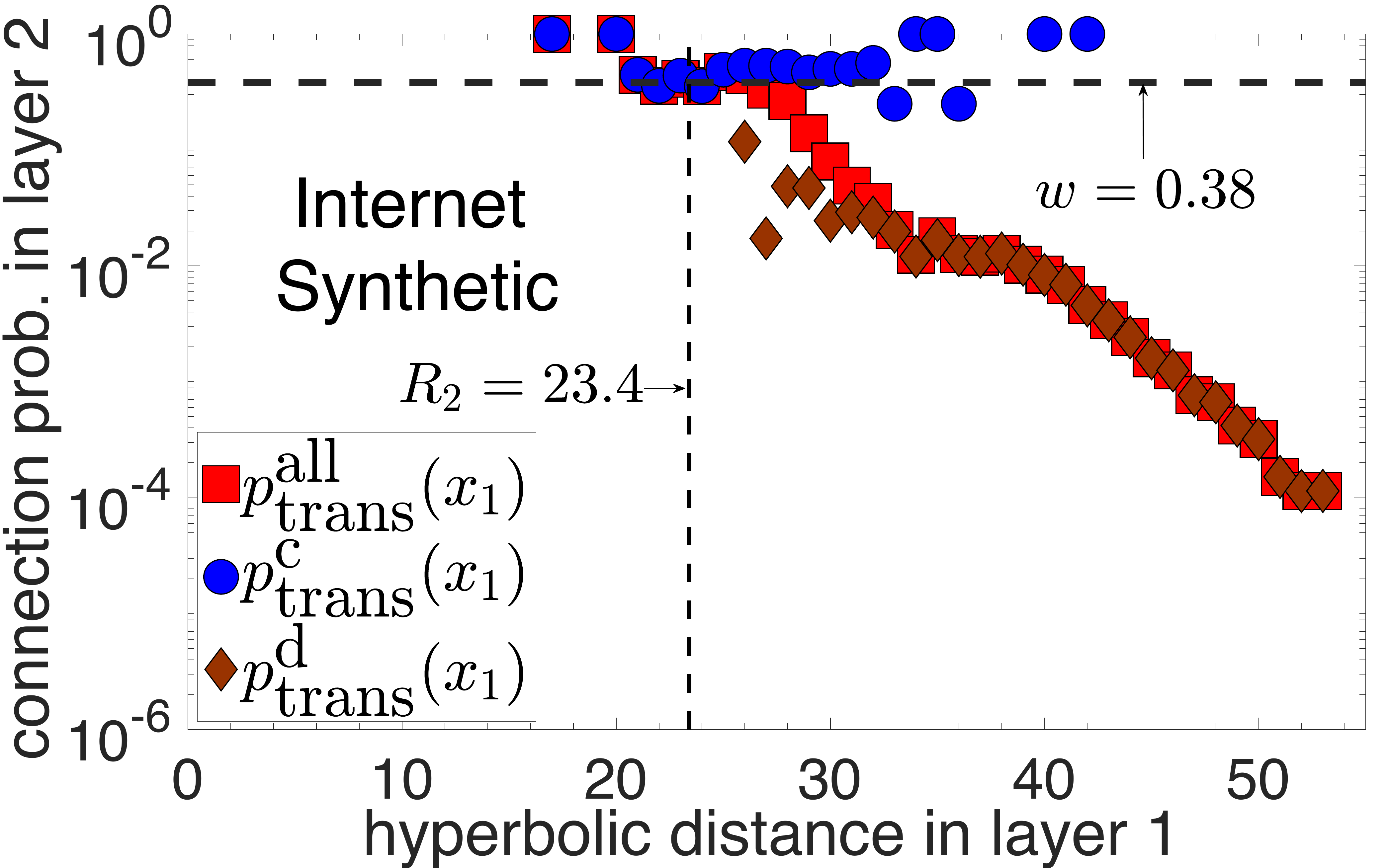}
\includegraphics[width=1.73in, height=1.1in]{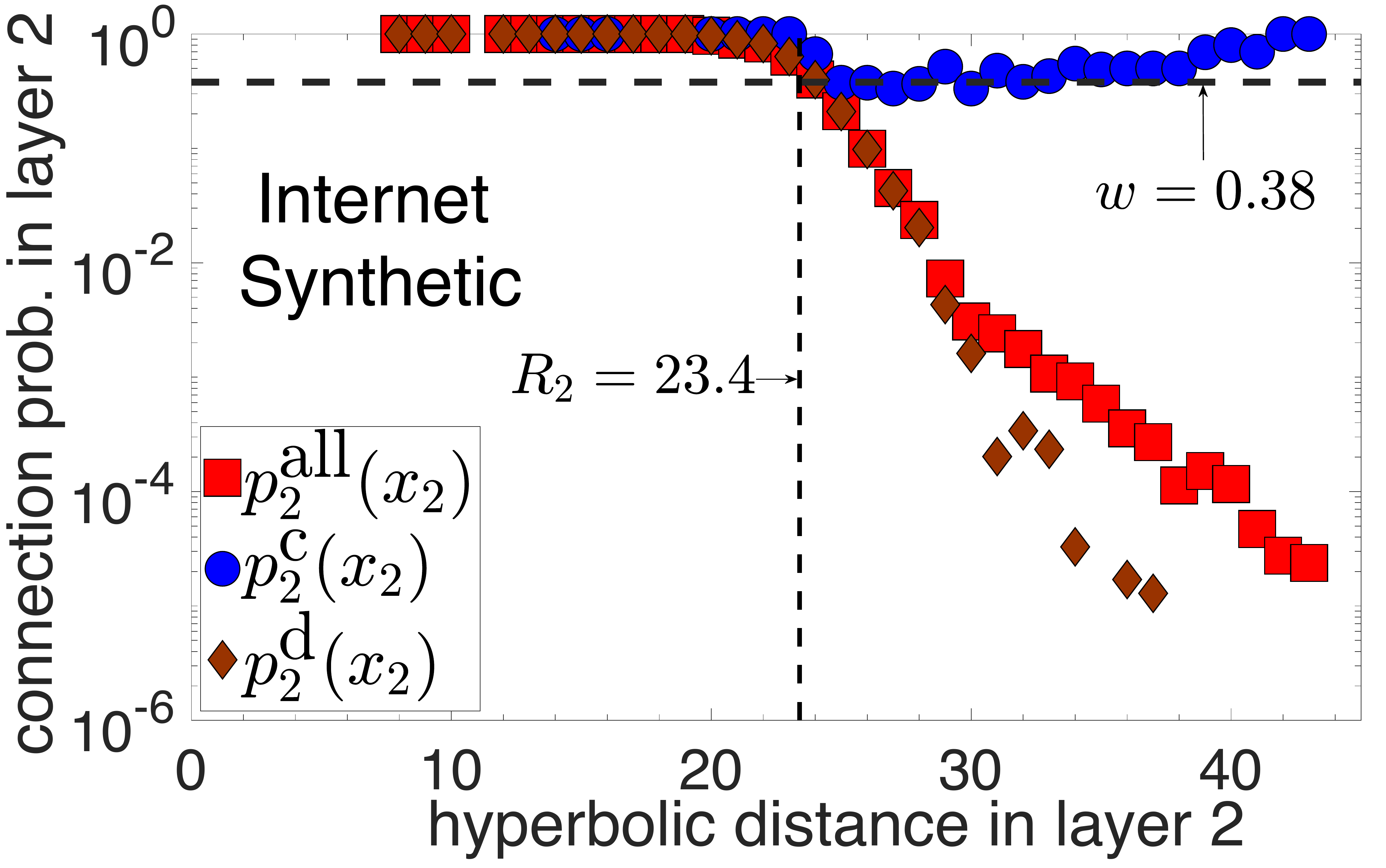}
\includegraphics[width=1.73in, height=1.1in]{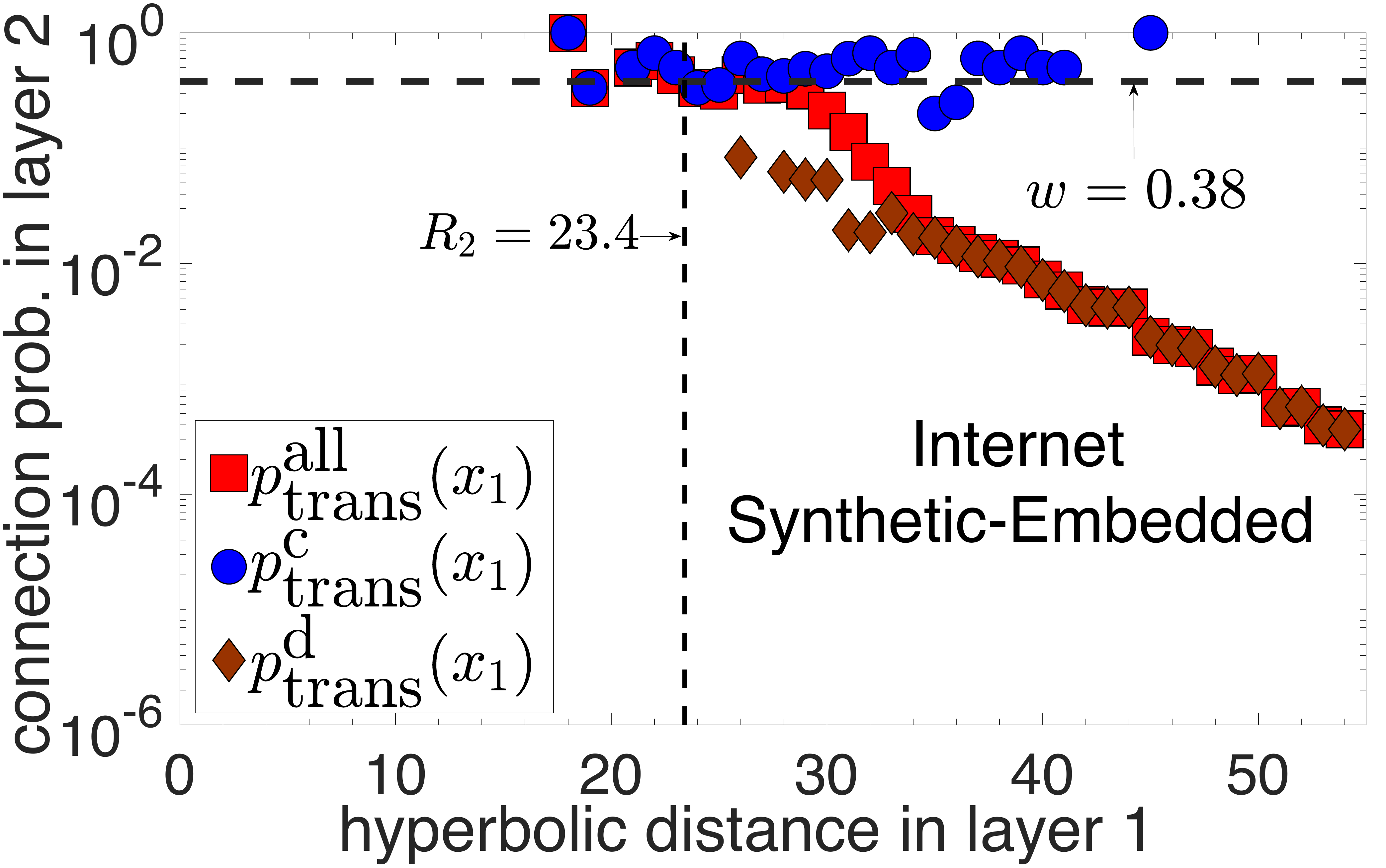}
\includegraphics[width=1.73in, height=1.1in]{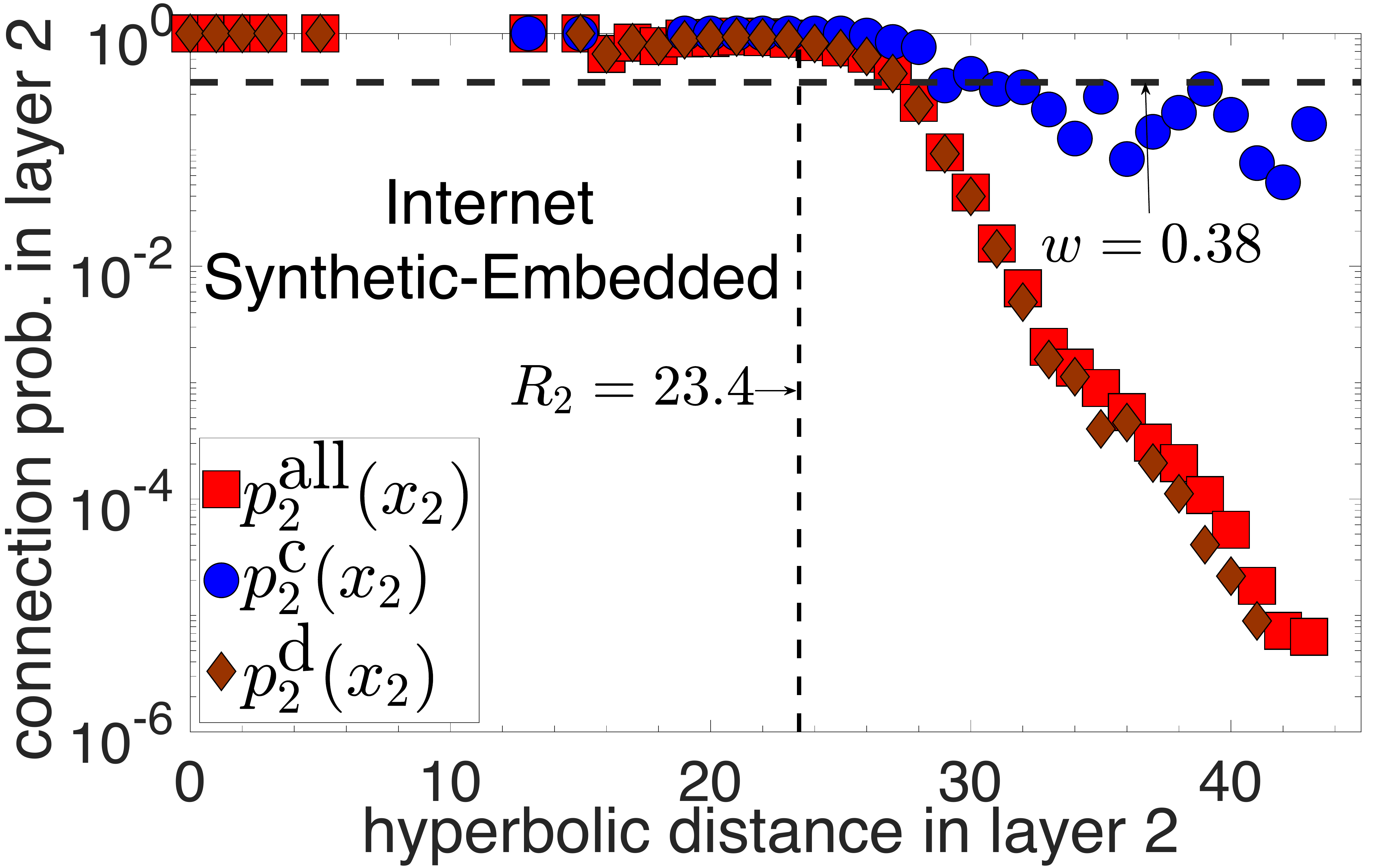}\\
\includegraphics[width=1.73in, height=1.1in]{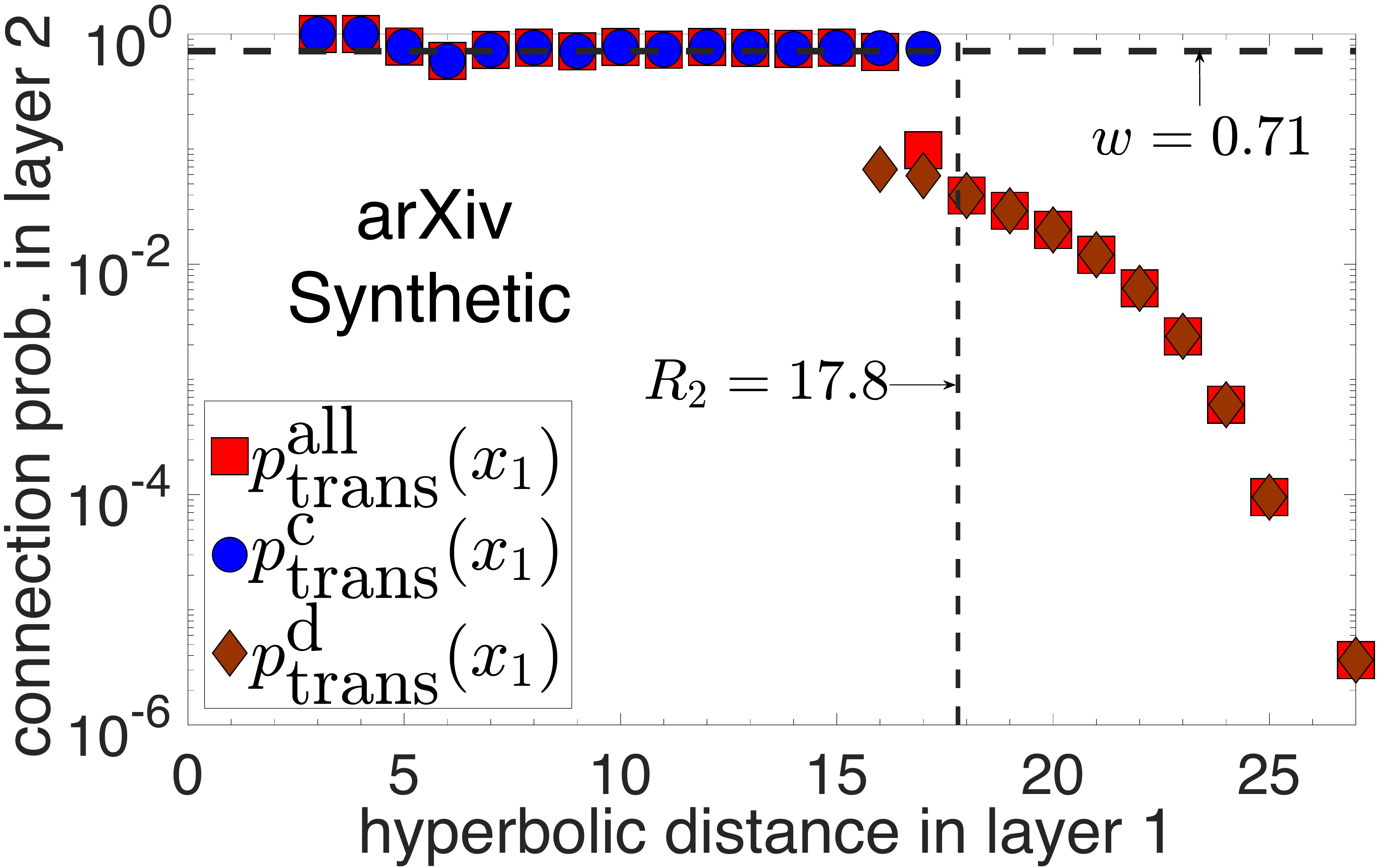}
\includegraphics[width=1.73in, height=1.1in]{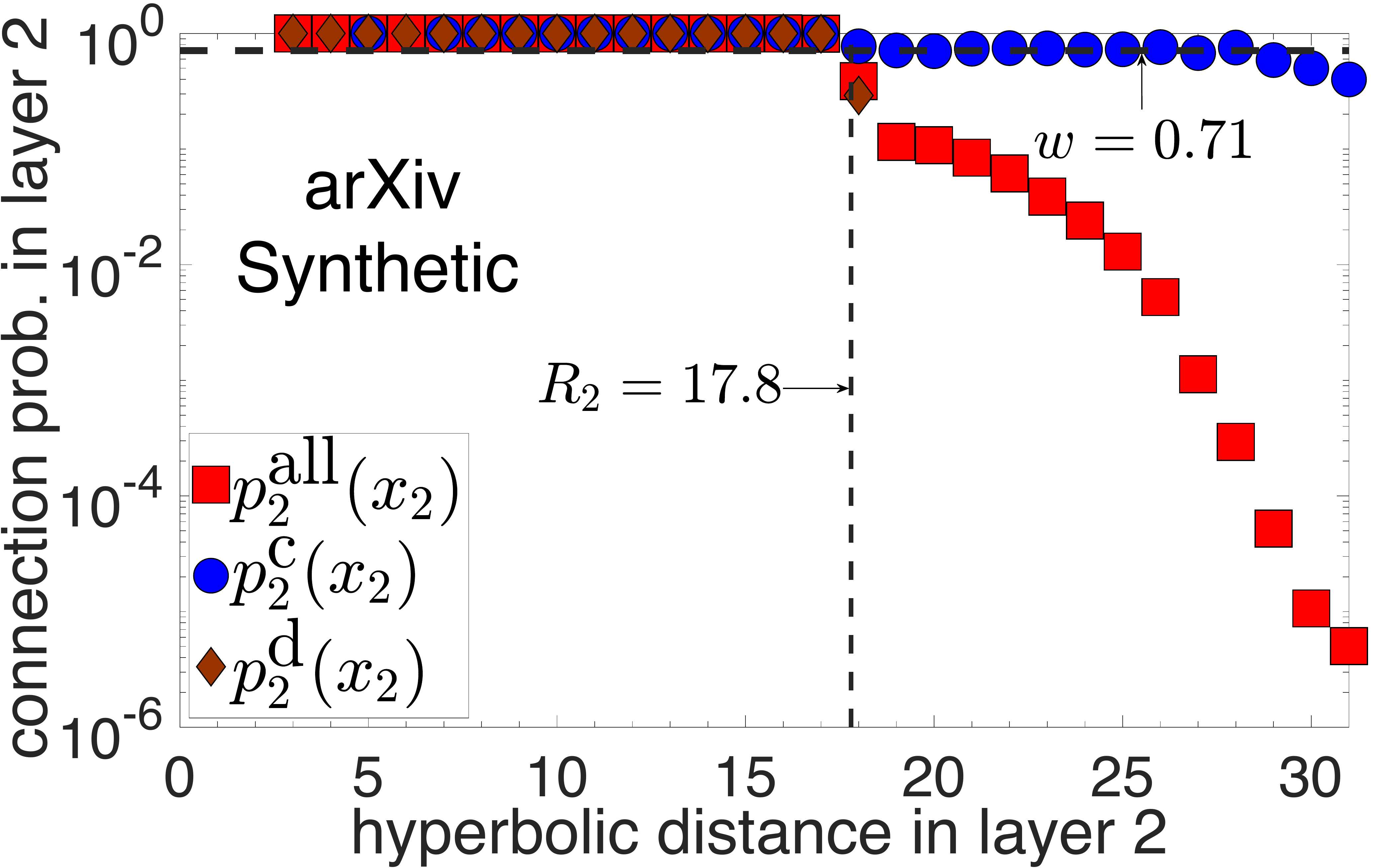}
\includegraphics[width=1.73in, height=1.1in]{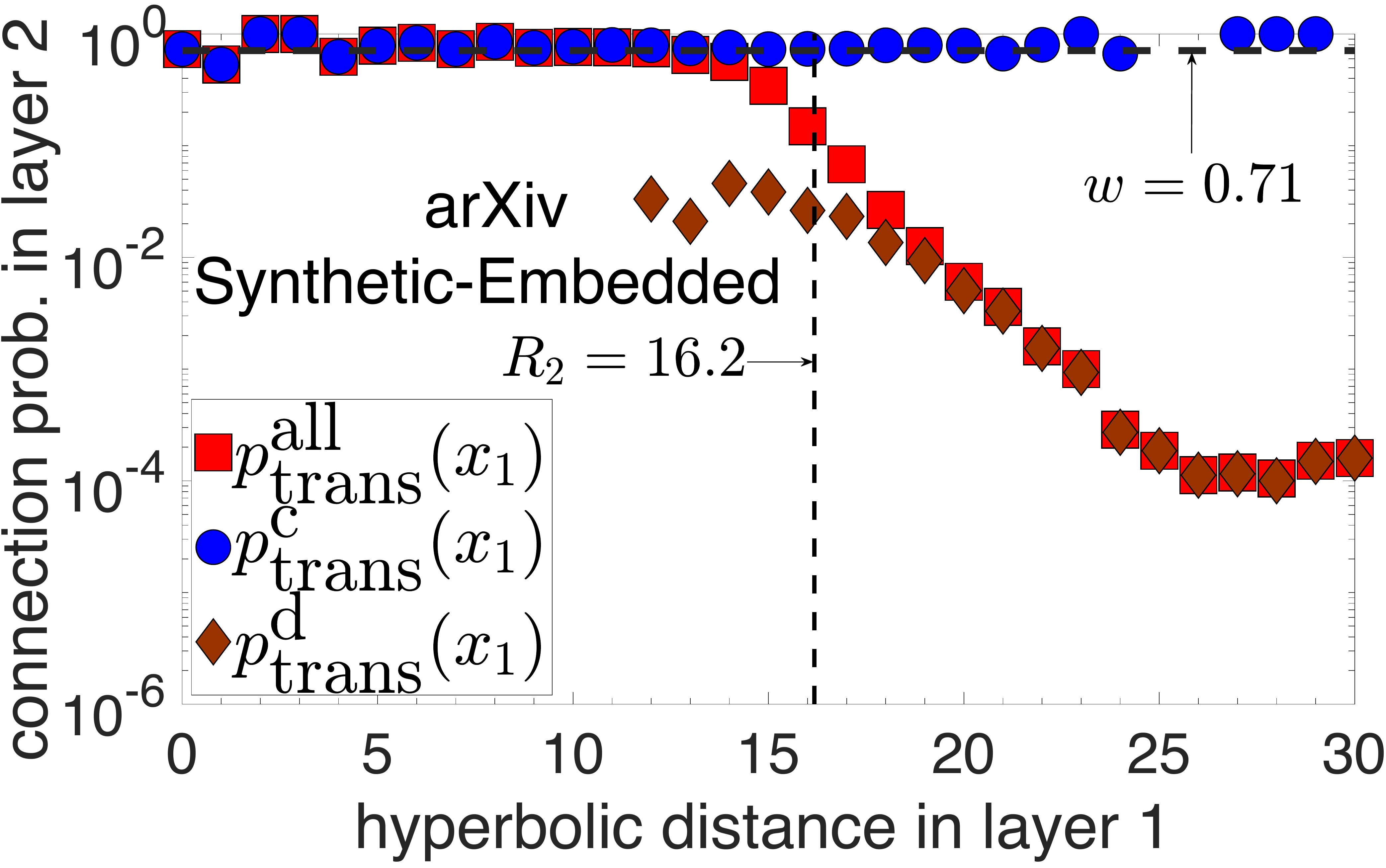}
\includegraphics[width=1.73in, height=1.1in]{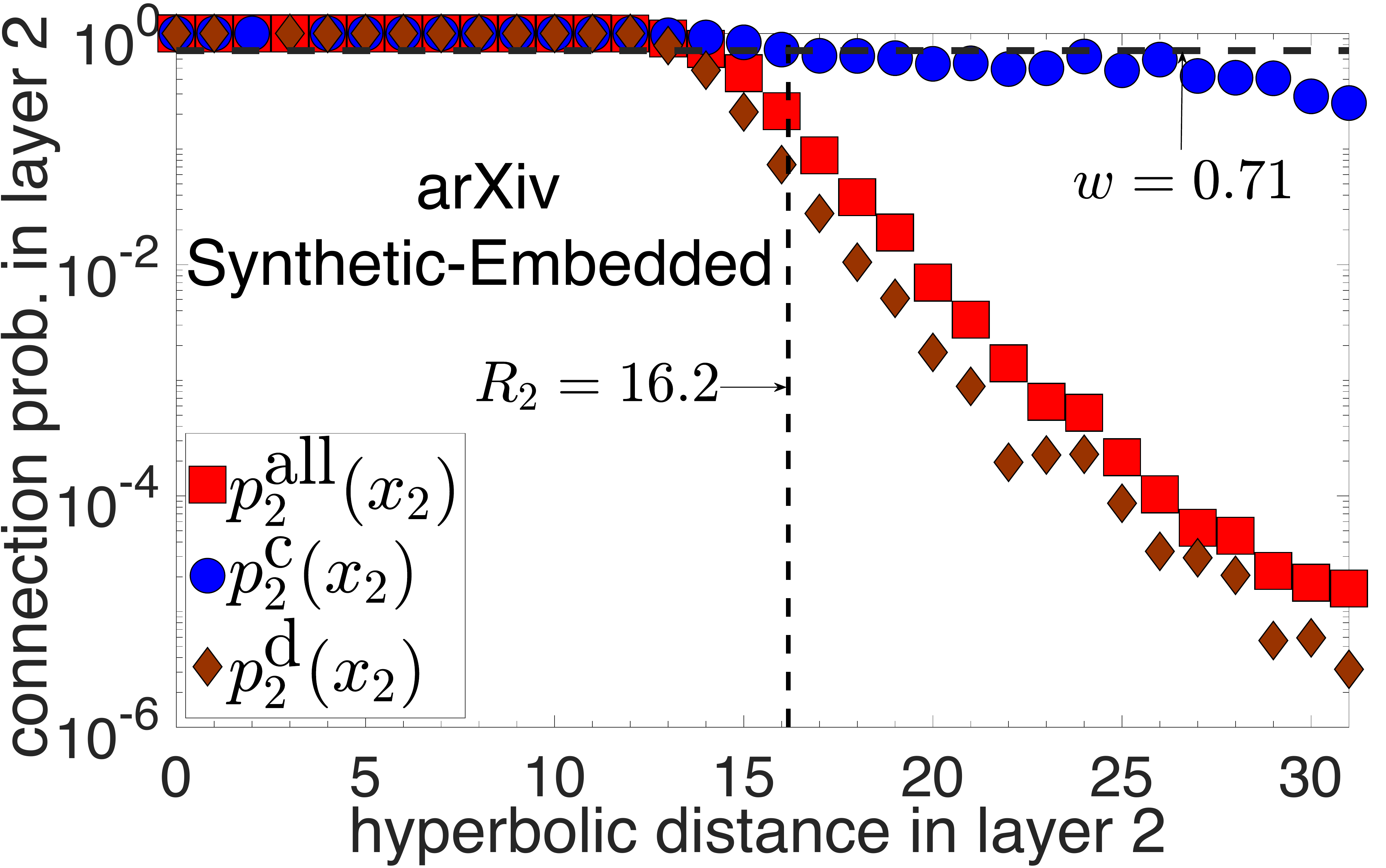}\\
\includegraphics[width=1.73in, height=1.1in]{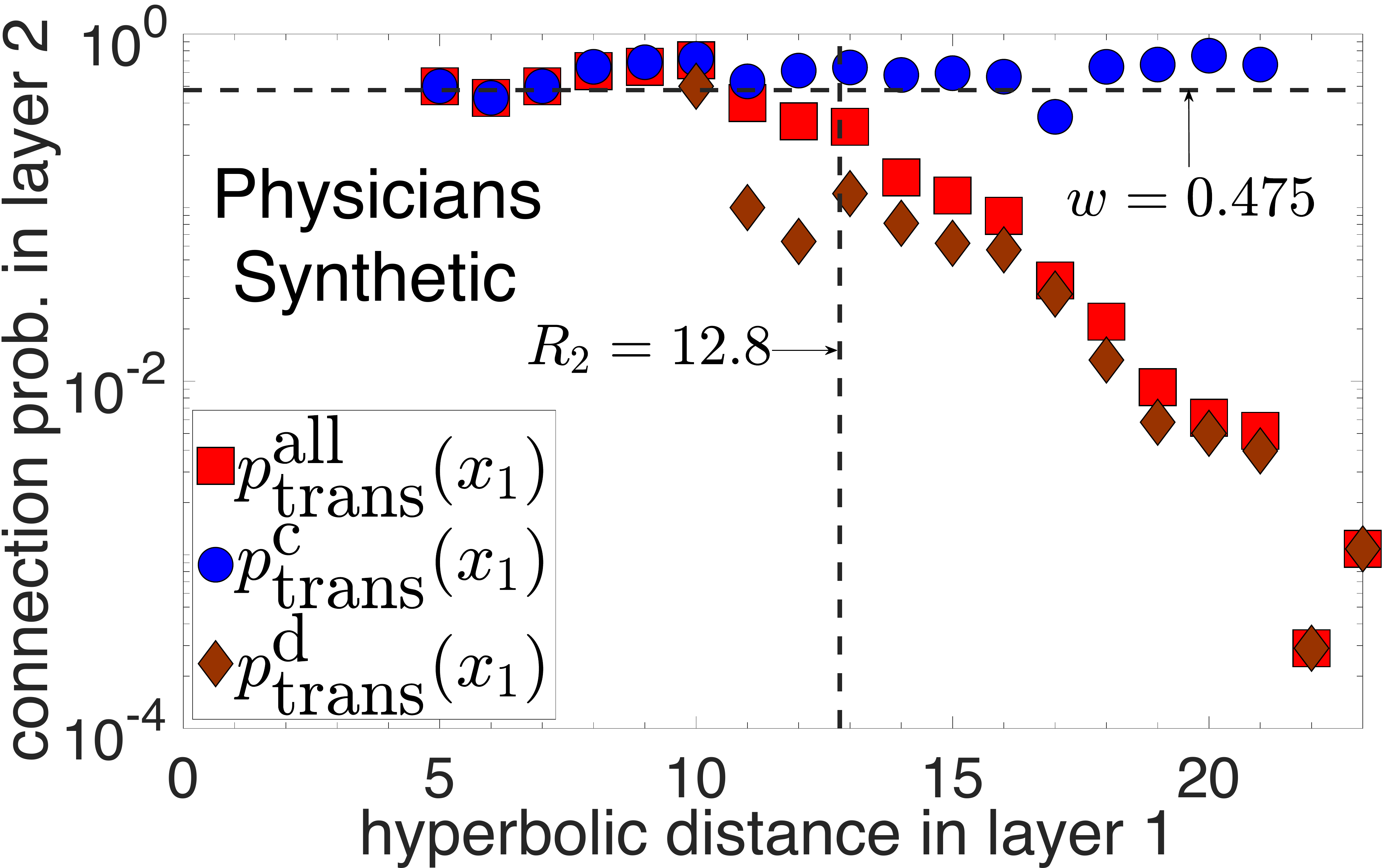}
\includegraphics[width=1.73in, height=1.1in]{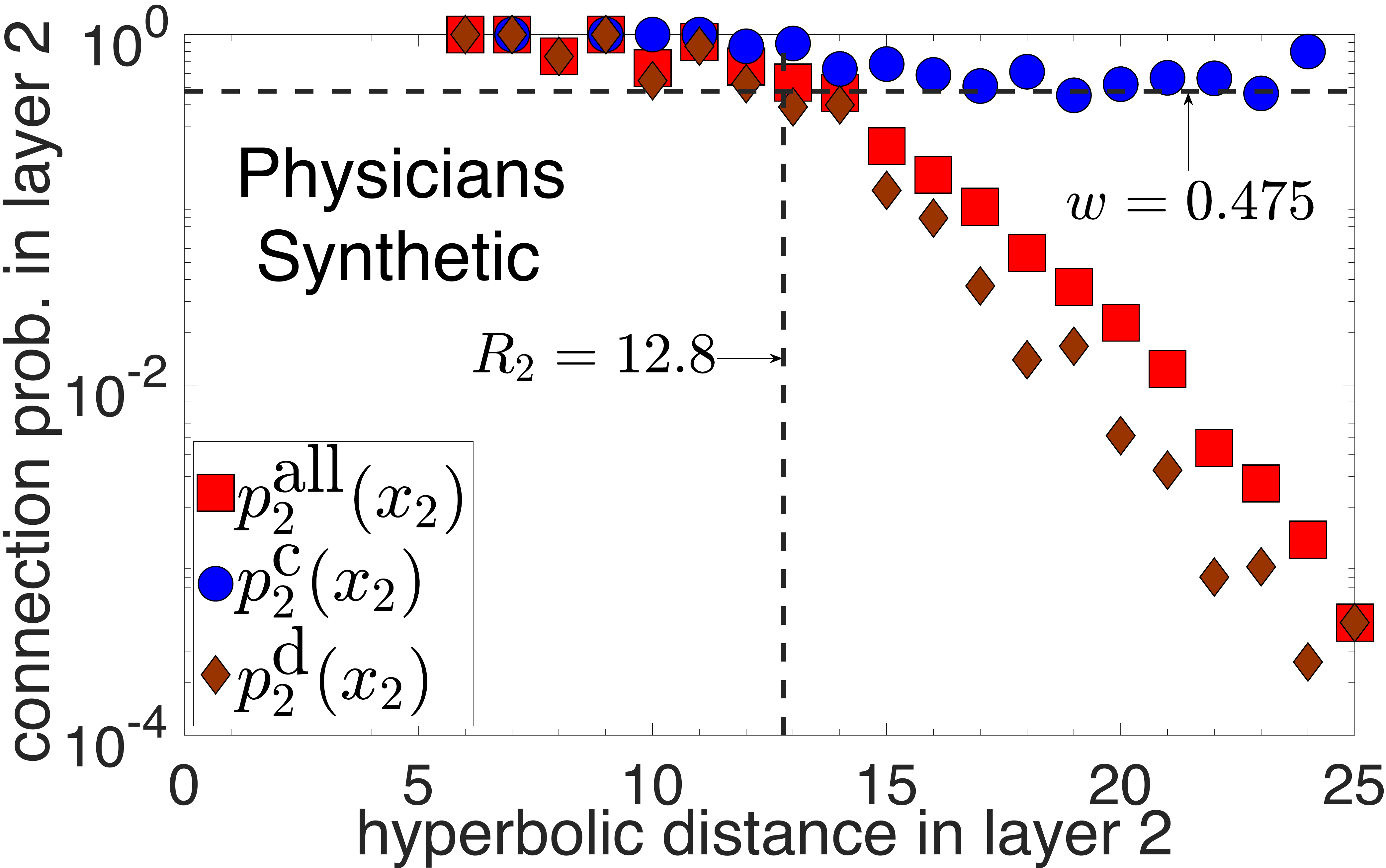}
\includegraphics[width=1.73in, height=1.1in]{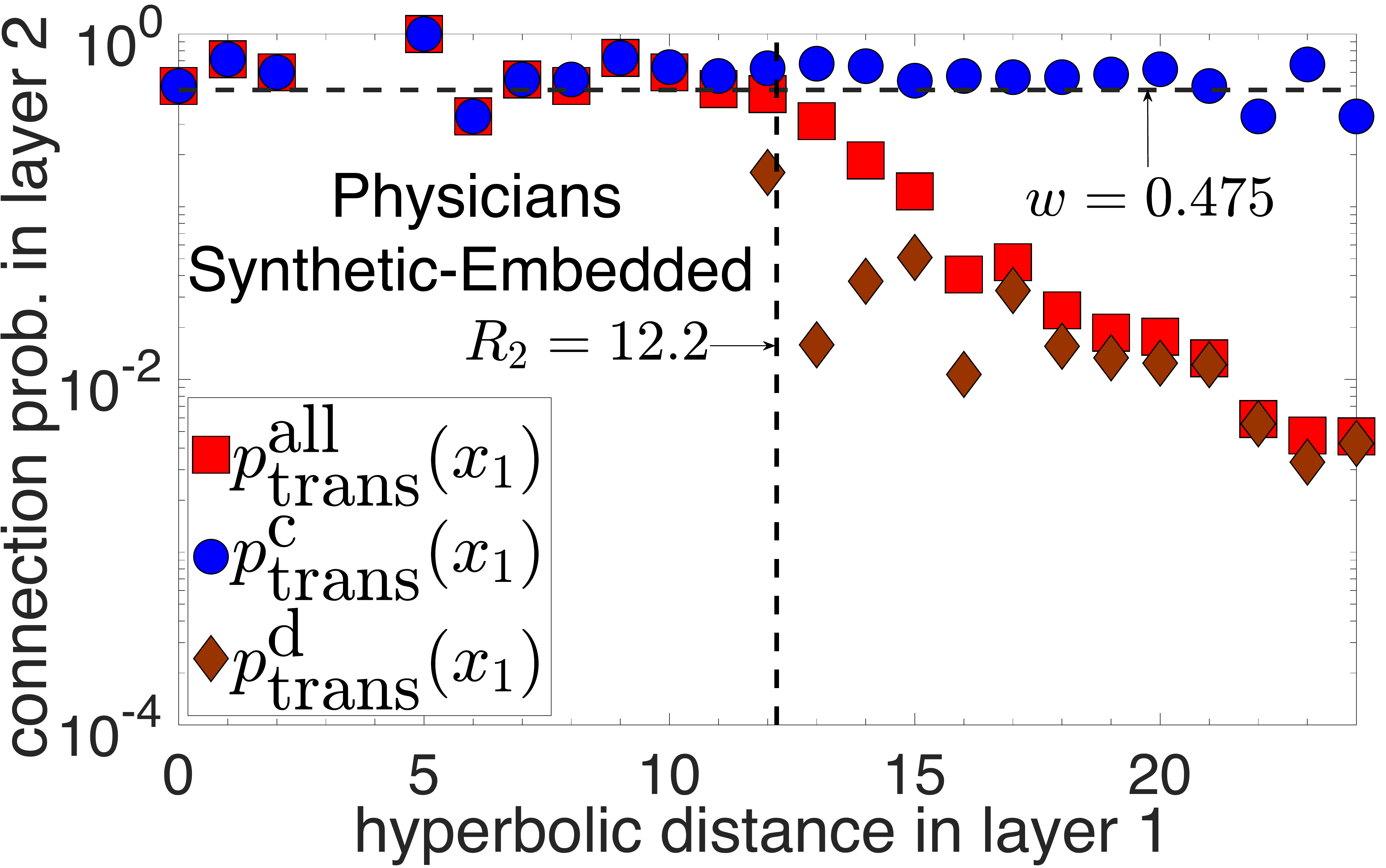}
\includegraphics[width=1.73in, height=1.1in]{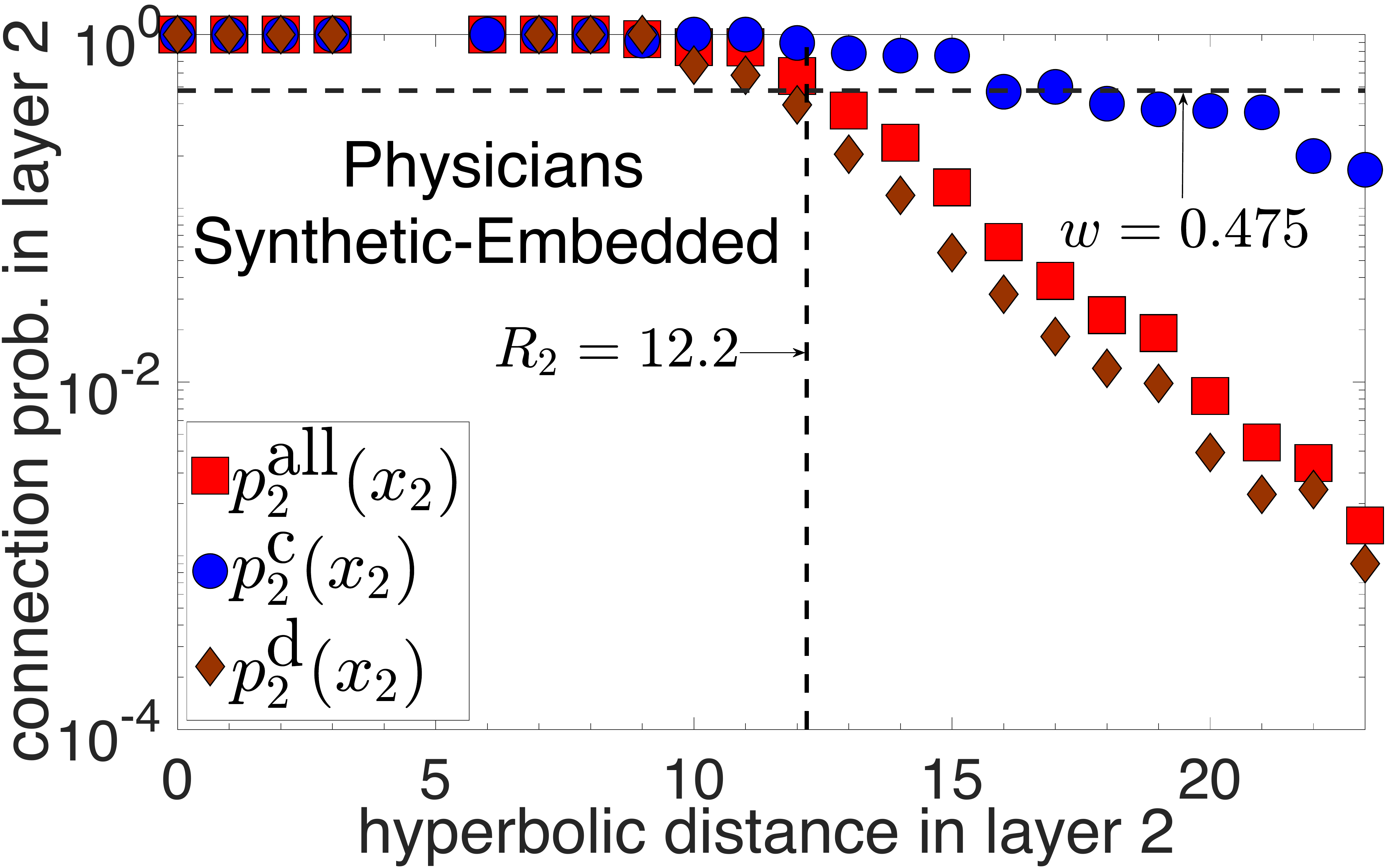}\\
\includegraphics[width=1.73in, height=1.1in]{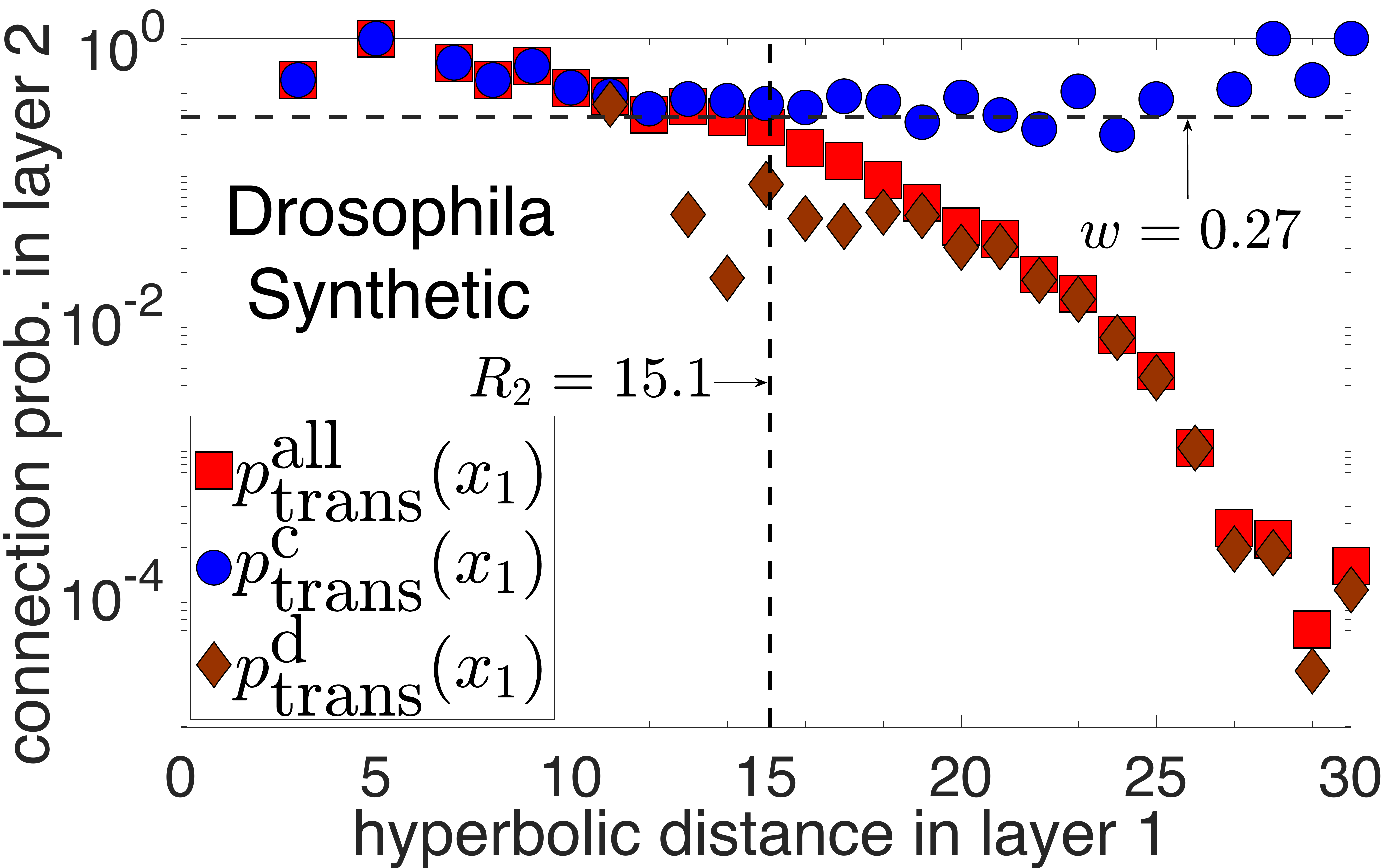}
\includegraphics[width=1.73in, height=1.1in]{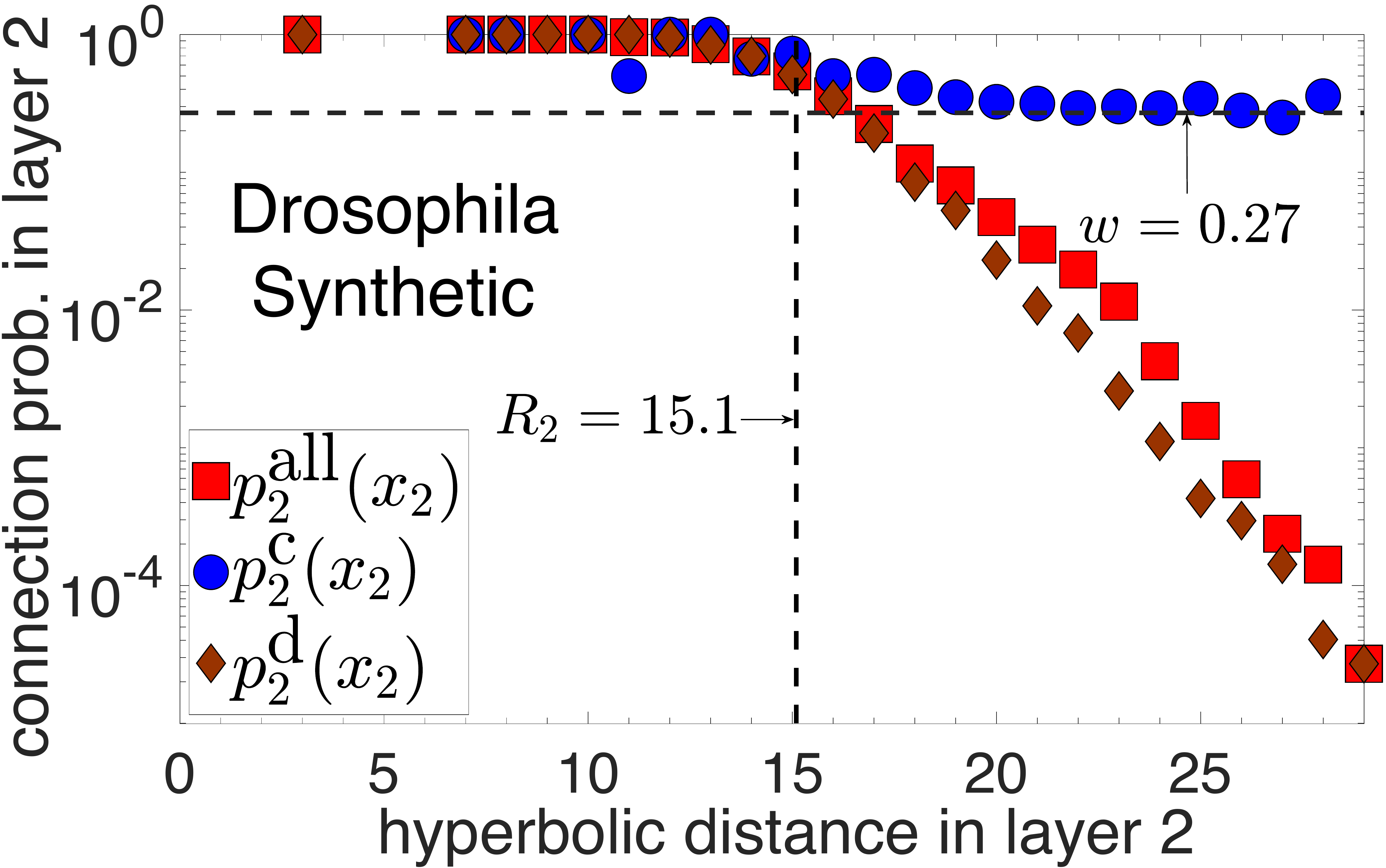}
\includegraphics[width=1.73in, height=1.1in]{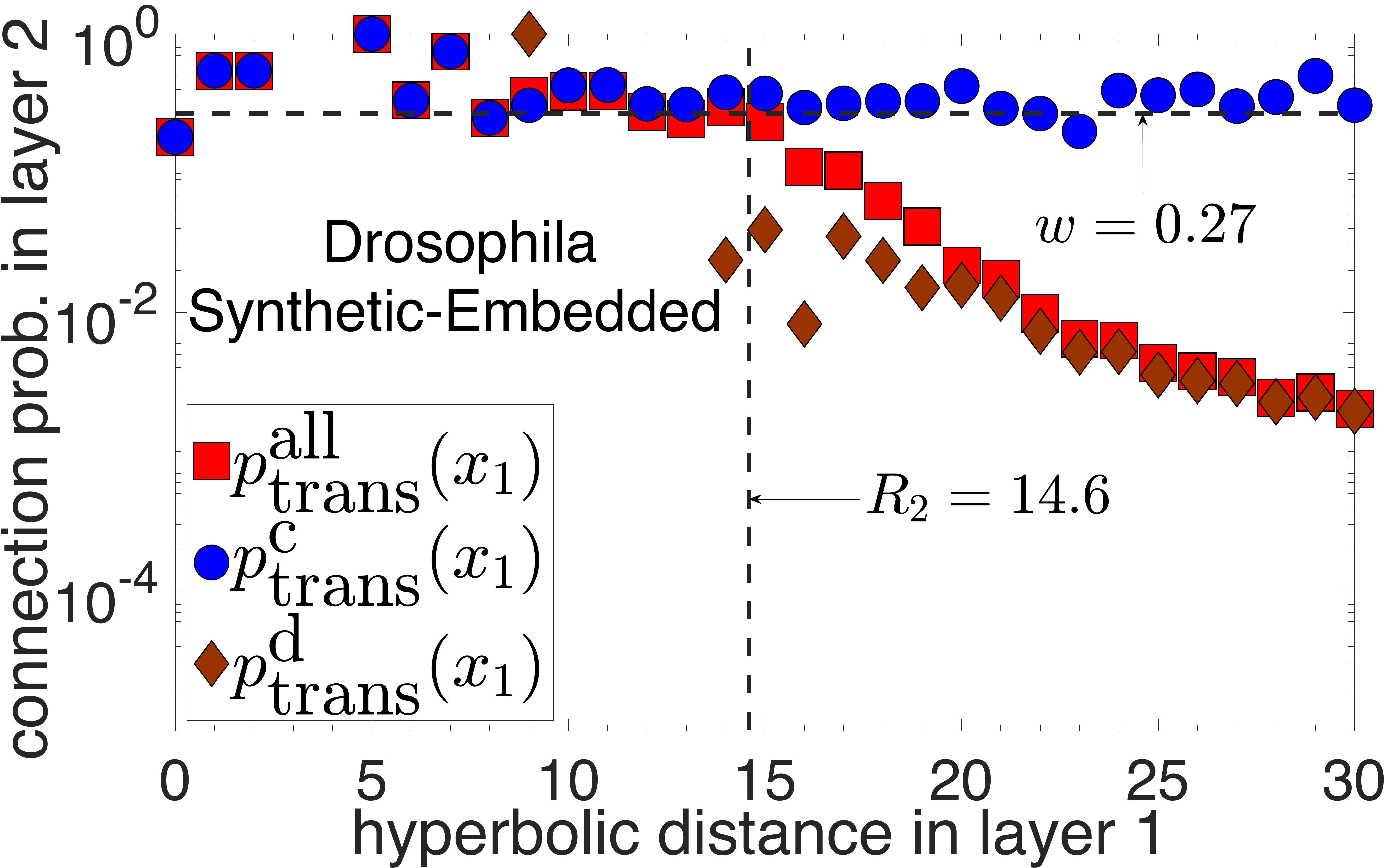}
\includegraphics[width=1.73in, height=1.1in]{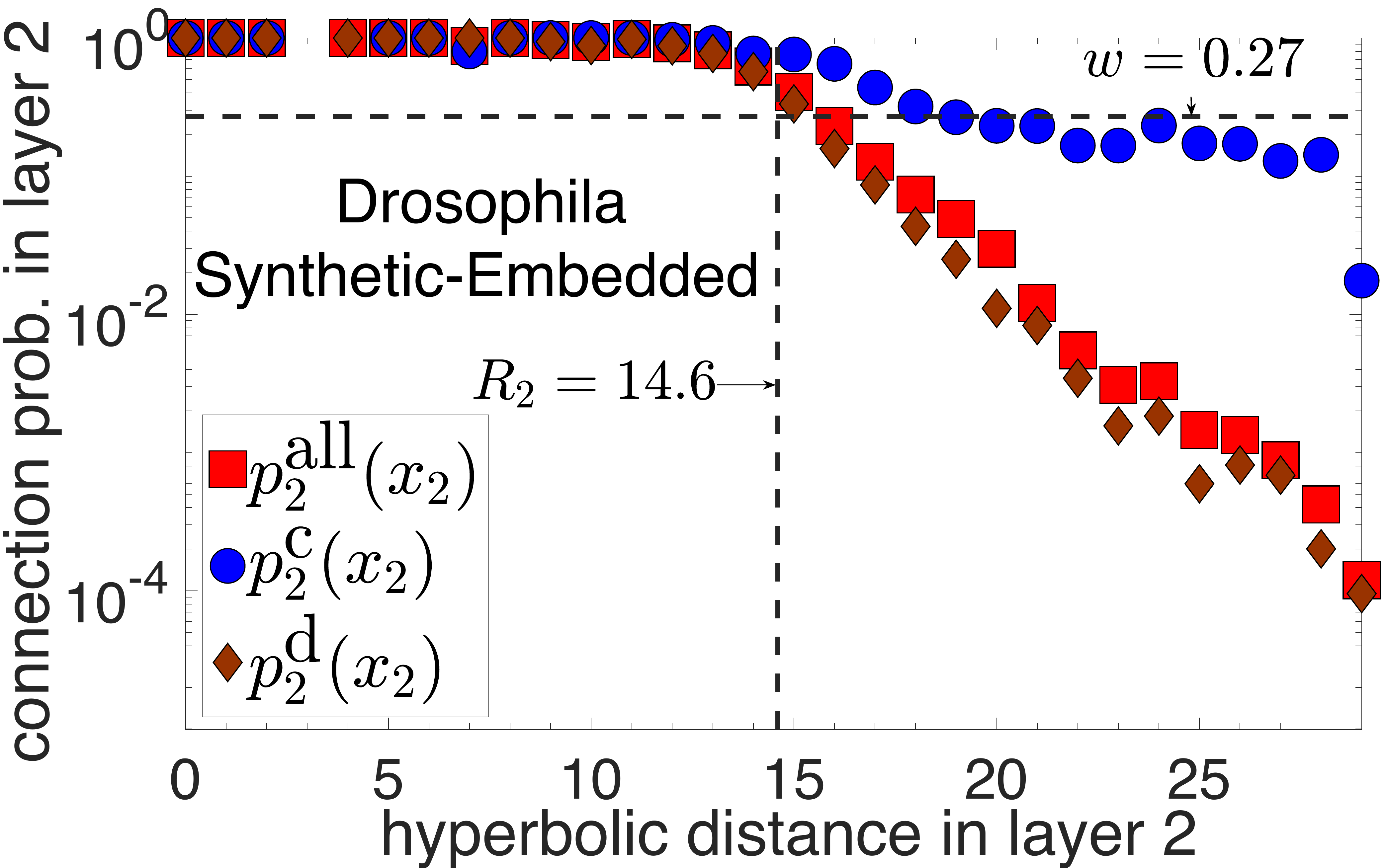}\\
\includegraphics[width=1.73in, height=1.1in]{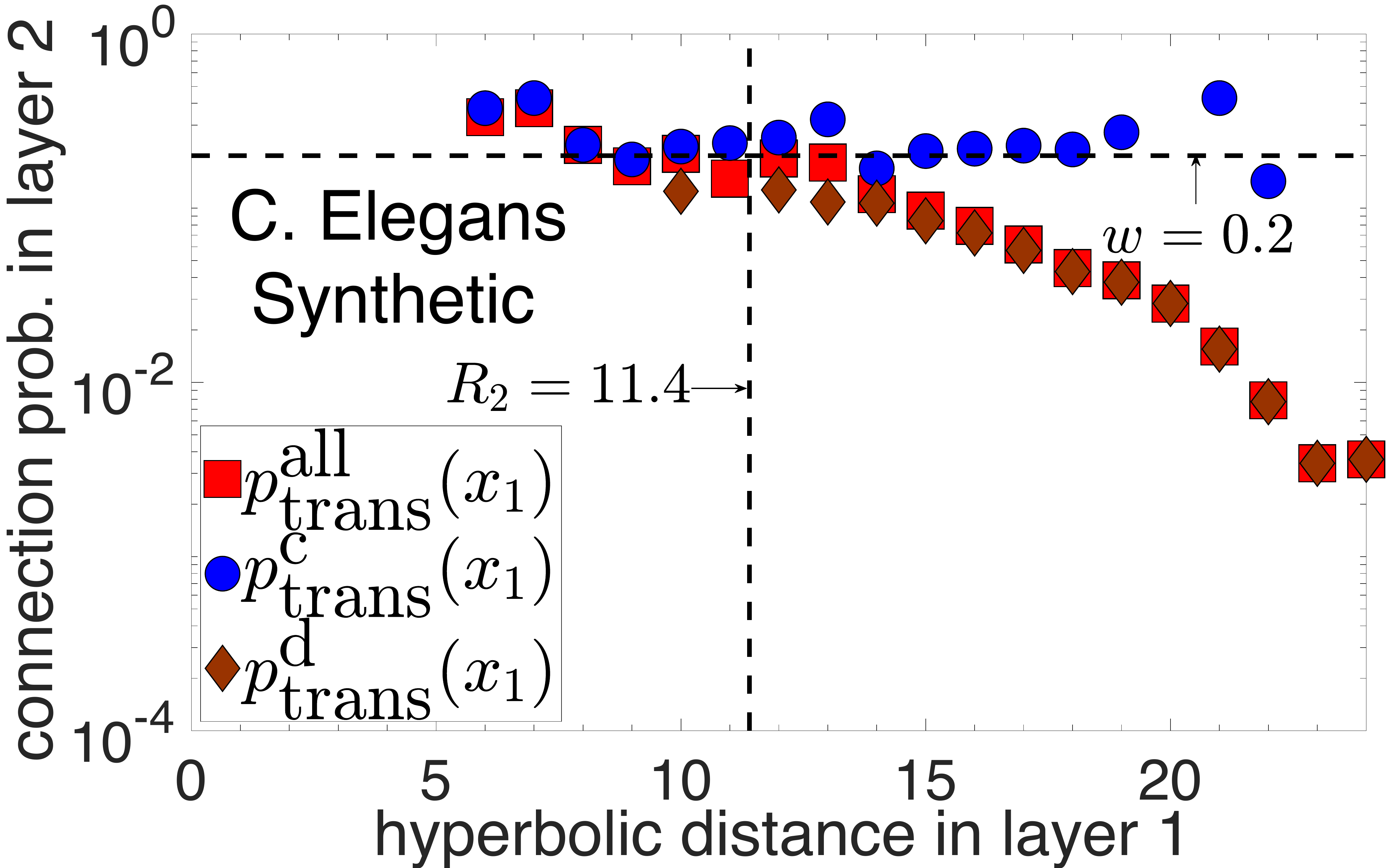}
\includegraphics[width=1.73in, height=1.1in]{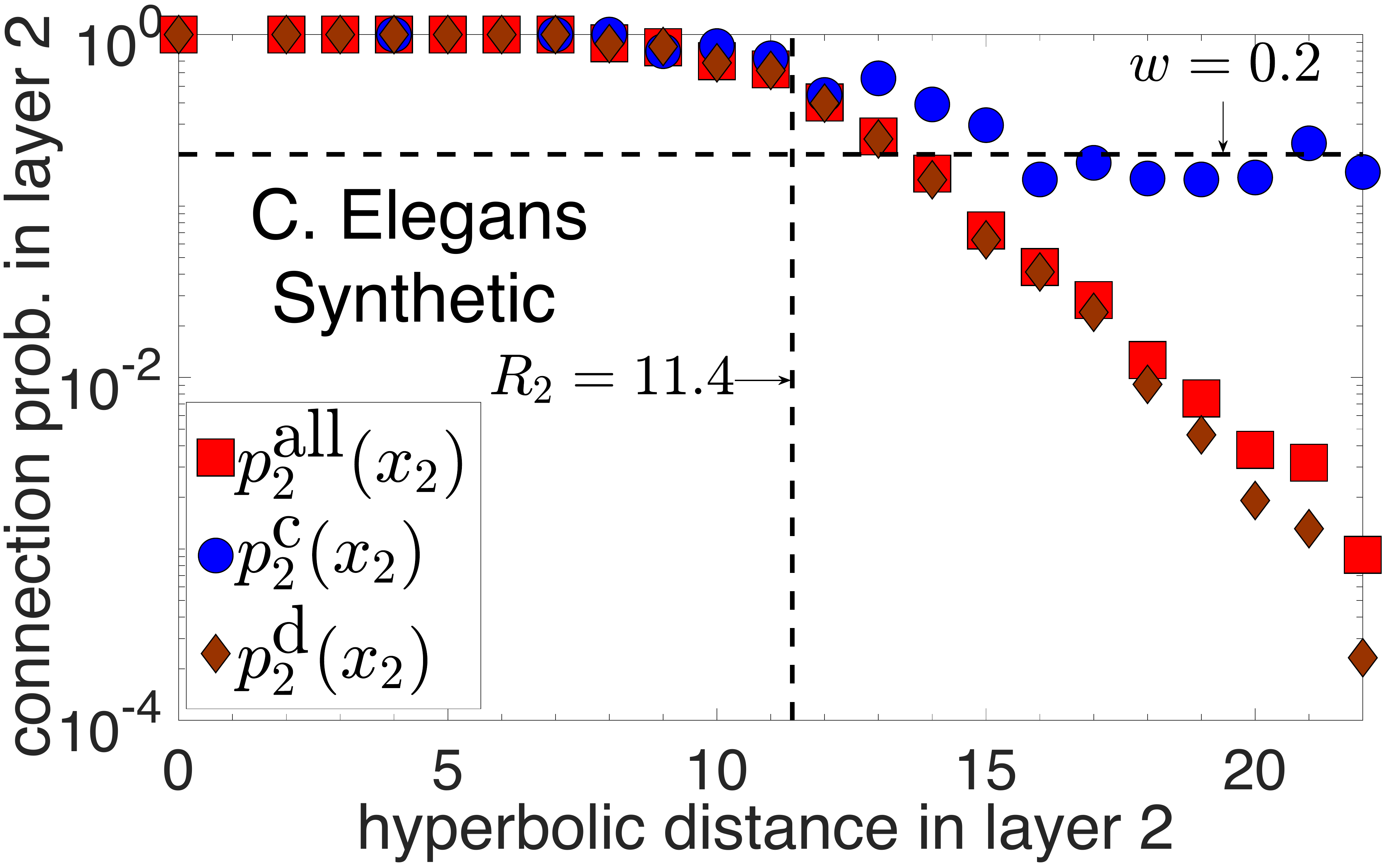}
\includegraphics[width=1.73in, height=1.1in]{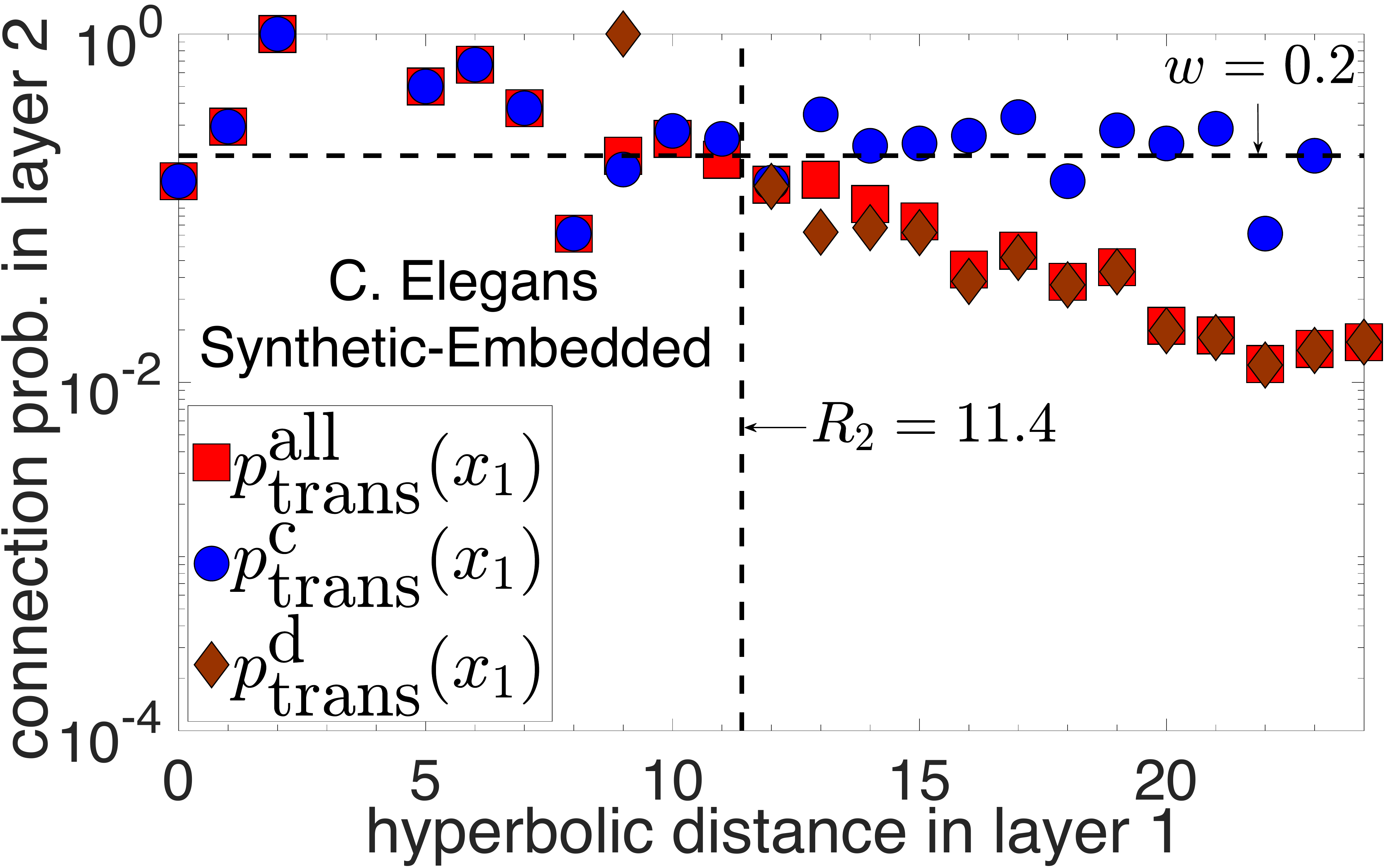}
\includegraphics[width=1.73in, height=1.1in]{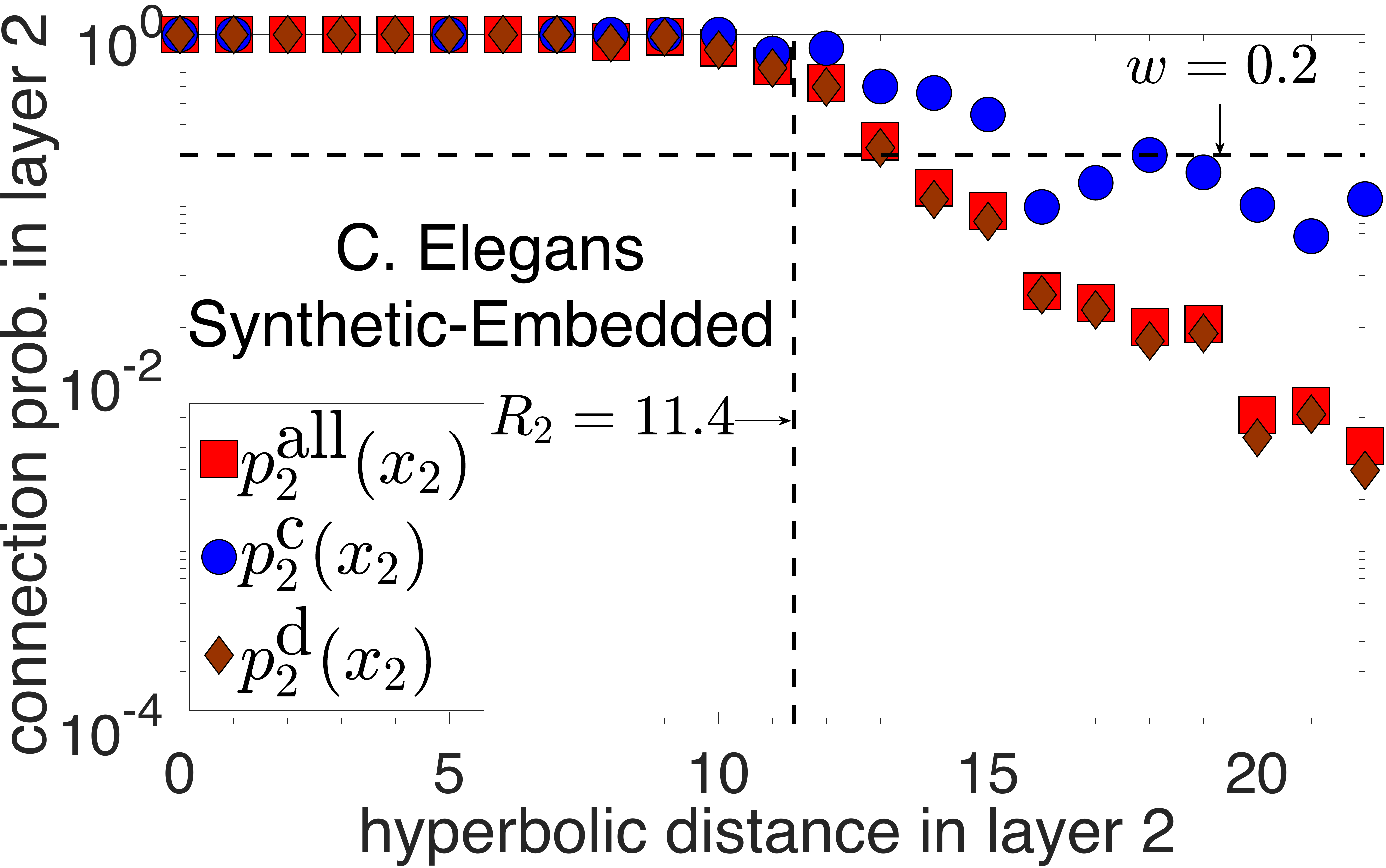}\\
\includegraphics[width=1.73in, height=1.1in]{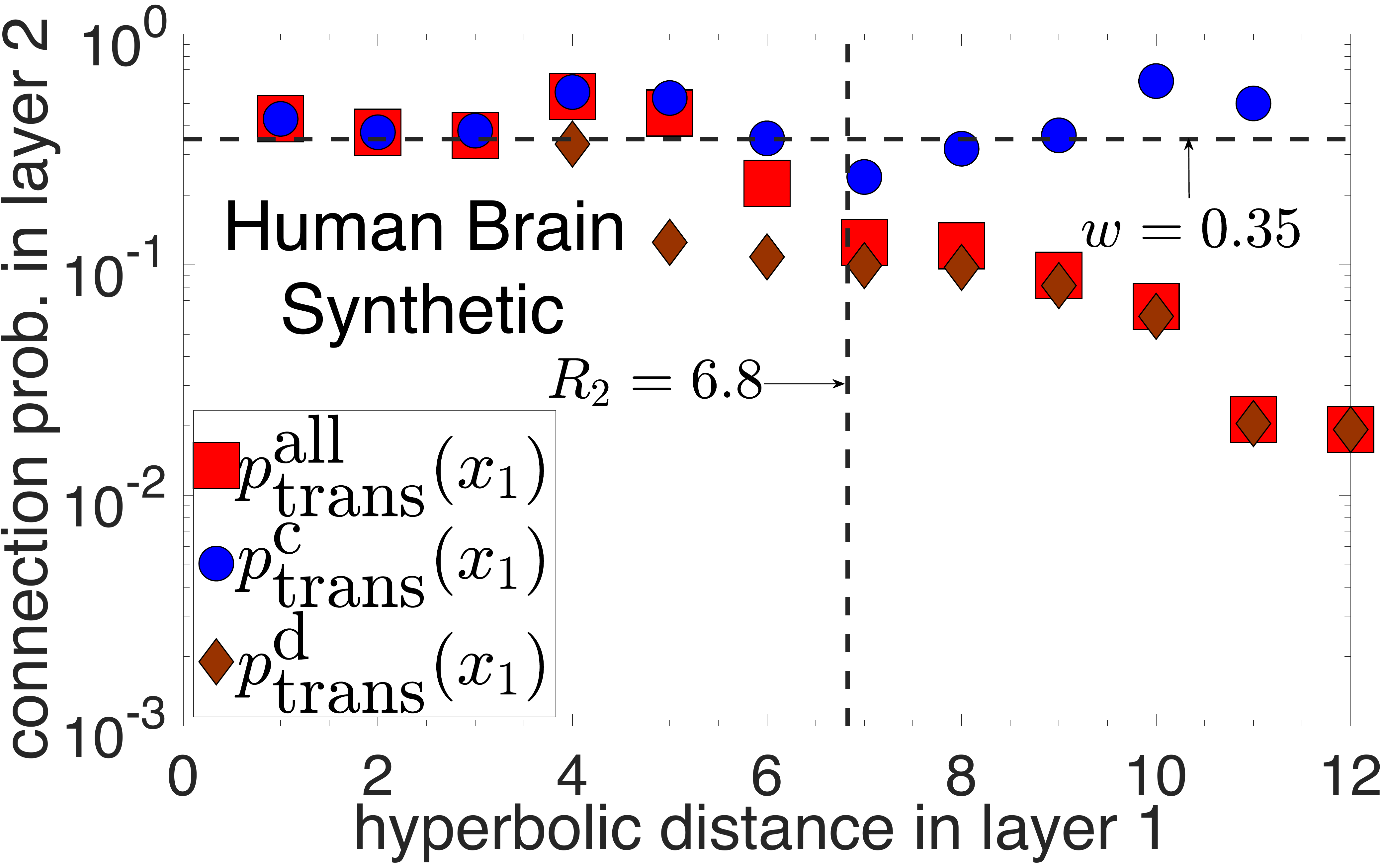}
\includegraphics[width=1.73in, height=1.1in]{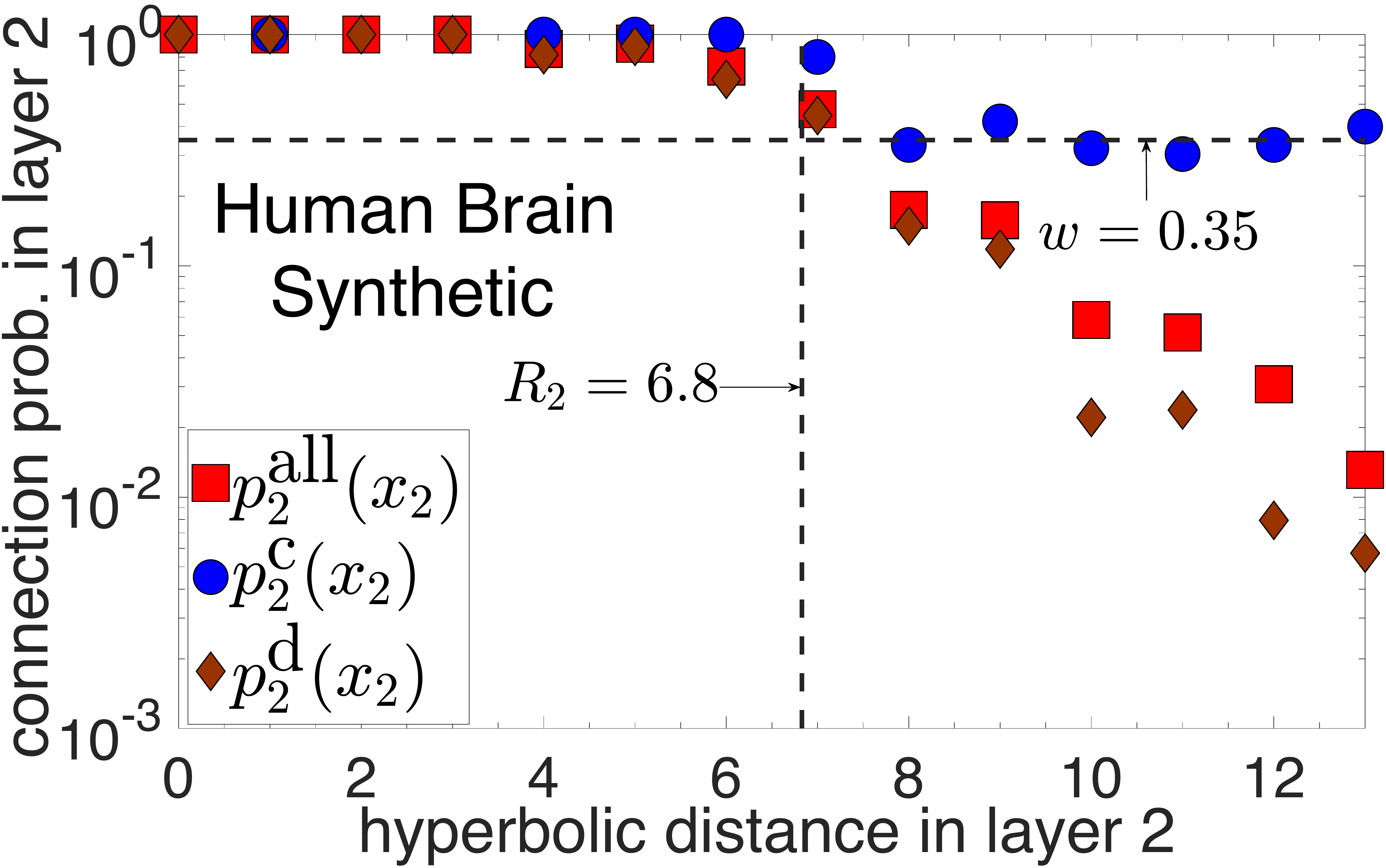}
\includegraphics[width=1.73in, height=1.1in]{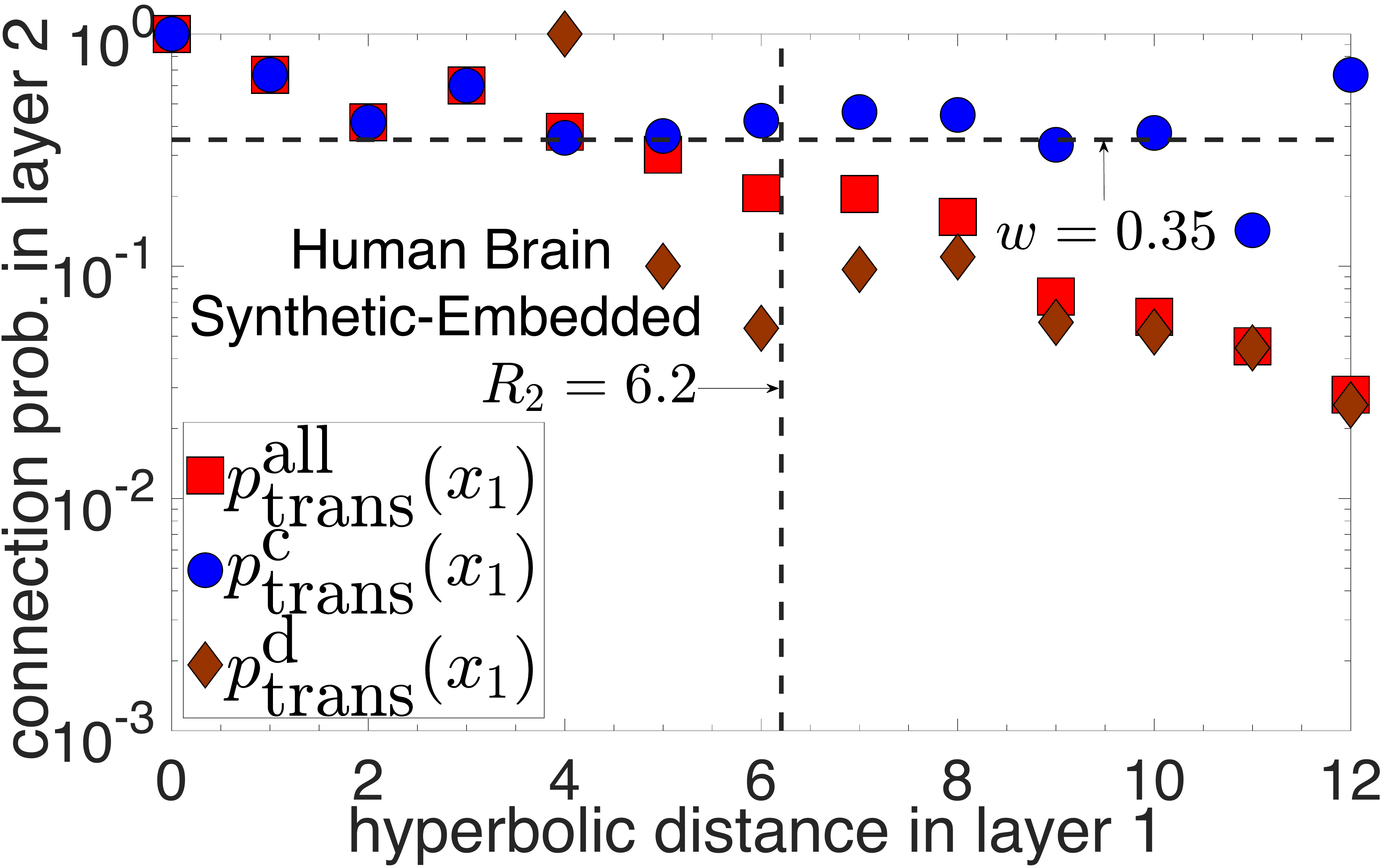}
\includegraphics[width=1.73in, height=1.1in]{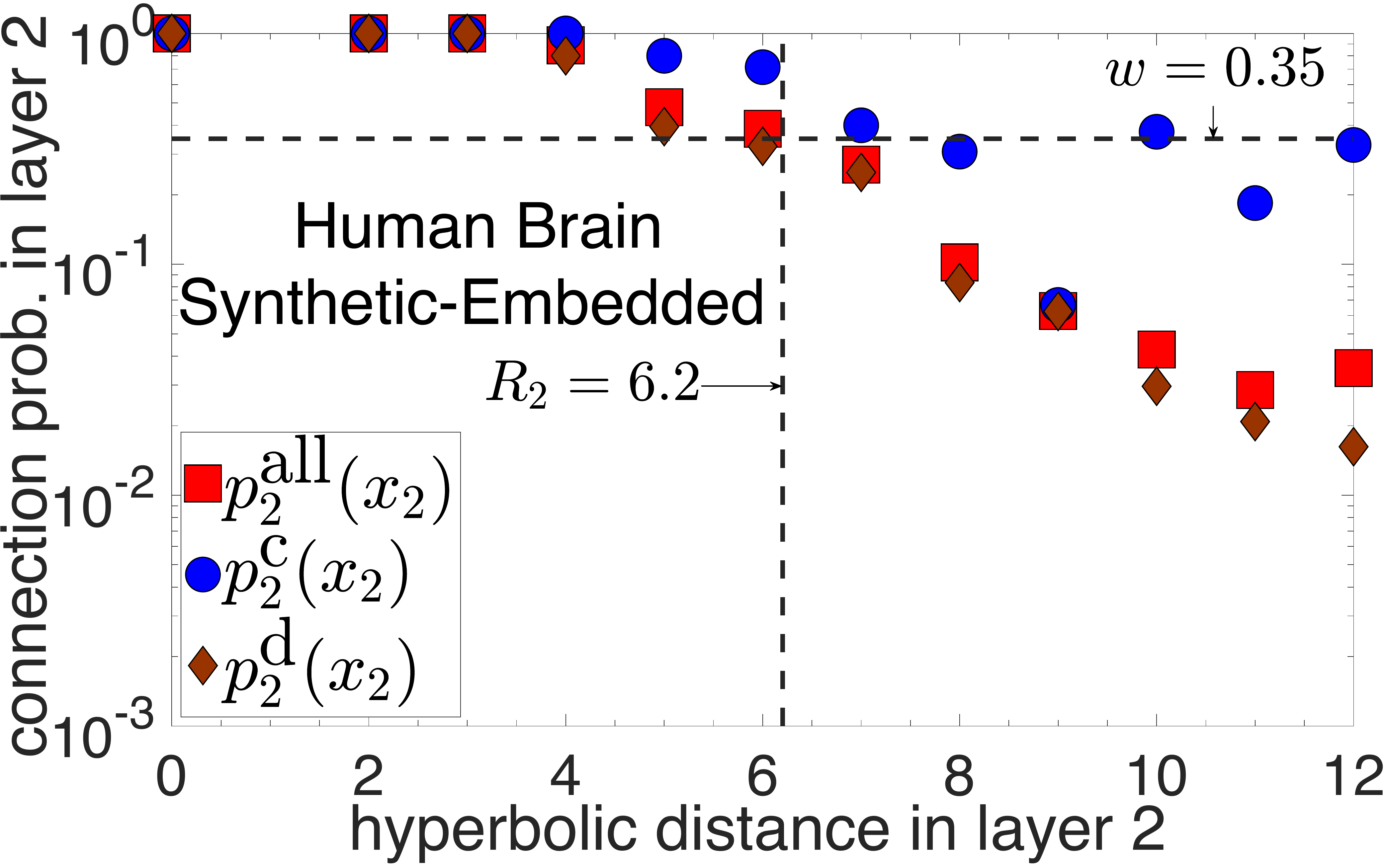}
}
\caption{Link persistence in synthetic versions of the multiplexes in Fig.~\ref{fig:empirical_results} constructed using the GMM-LP. The plots show the trans-layer connection probabilities between layers 1 and 2 and the connection probabilities in layer 2, as in Fig.~\ref{fig:empirical_results}. The first two columns are the results with the real node coordinates, while the next two columns are the results with the inferred node coordinates.    
\label{fig:sim_results_1}}
\end{figure*}

\begin{figure*}
\centering{
\includegraphics[width=1.73in, height=1.1in]{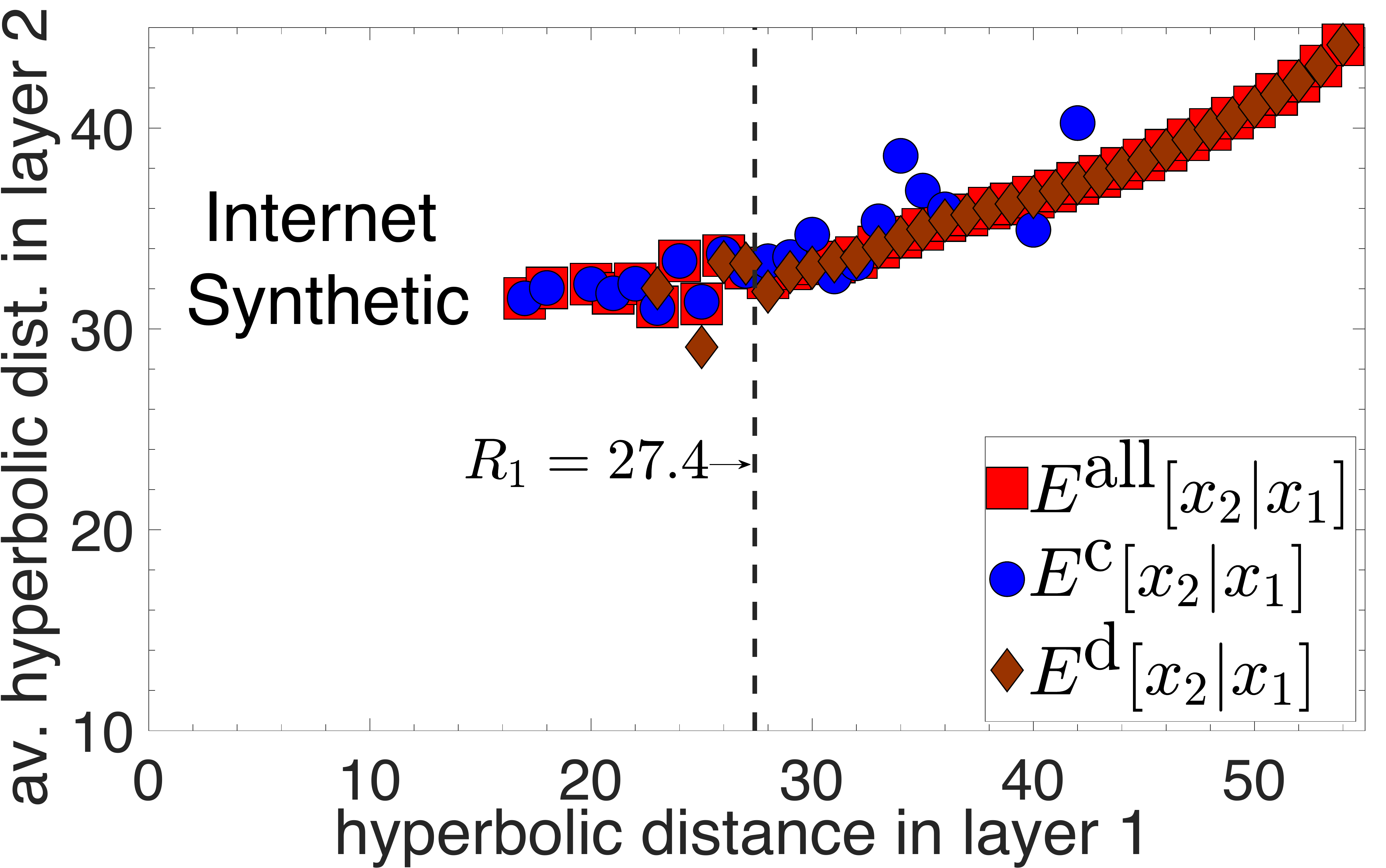}
\includegraphics[width=1.73in, height=1.1in]{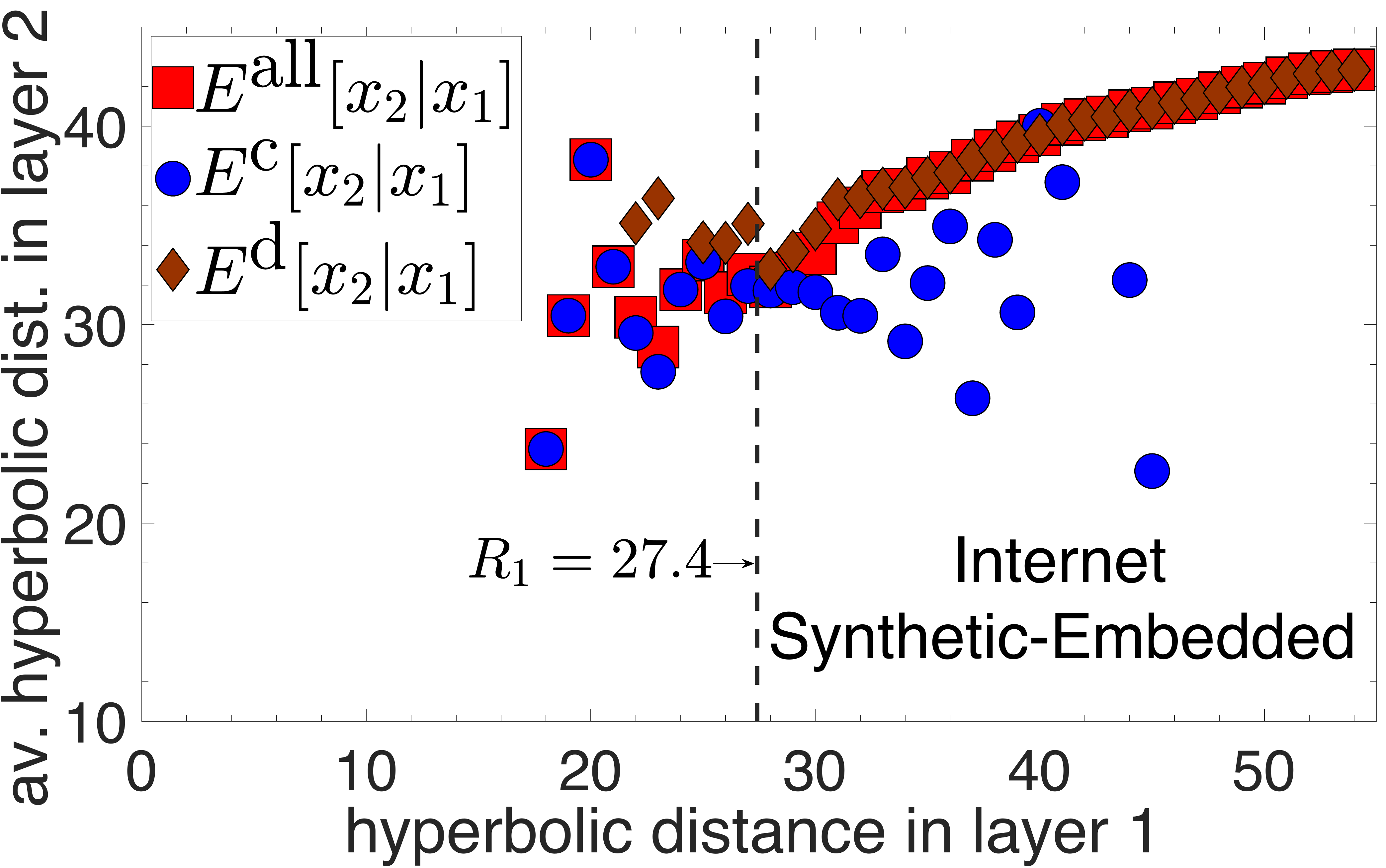}
\includegraphics[width=1.73in, height=1.1in]{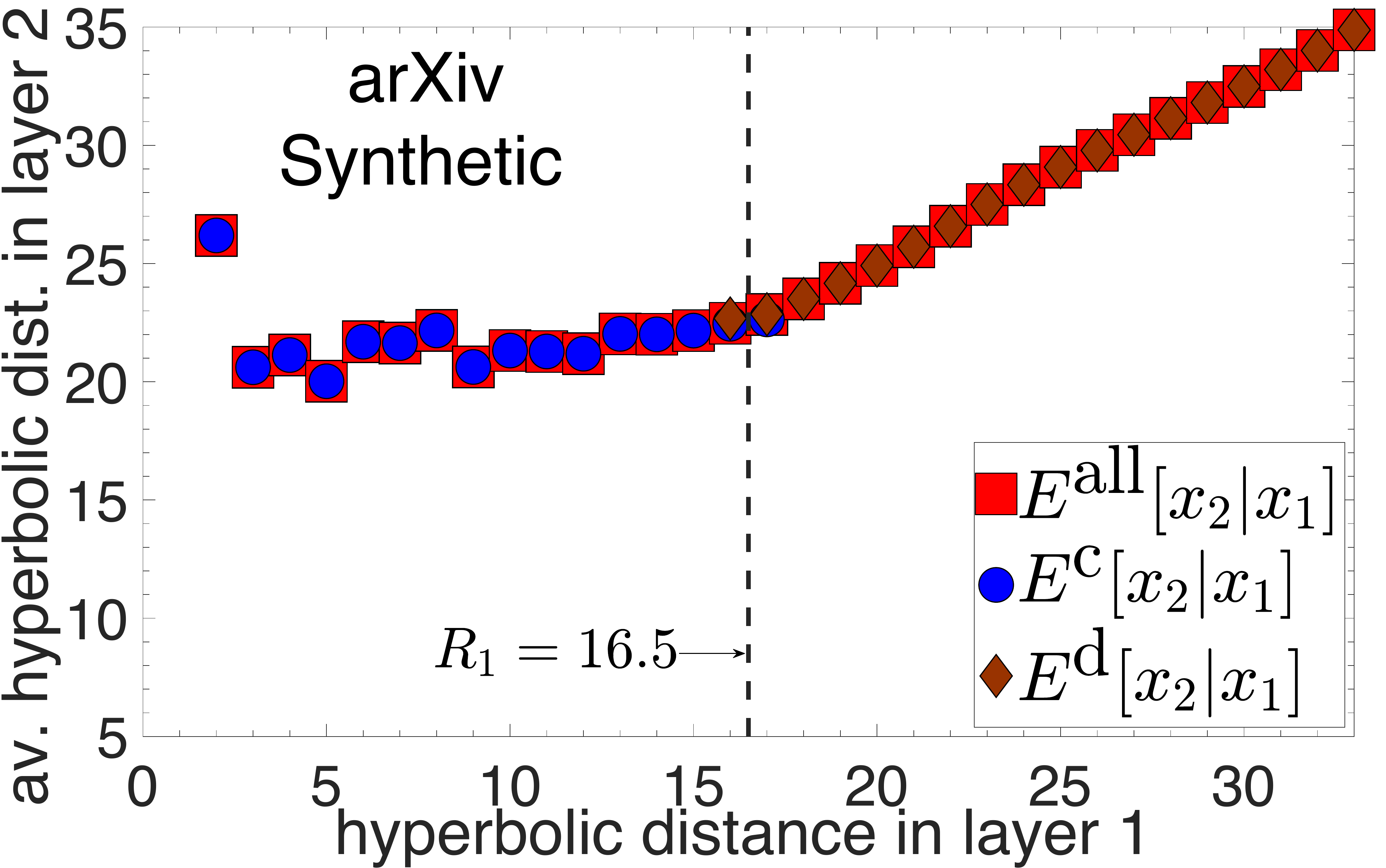}
\includegraphics[width=1.73in, height=1.1in]{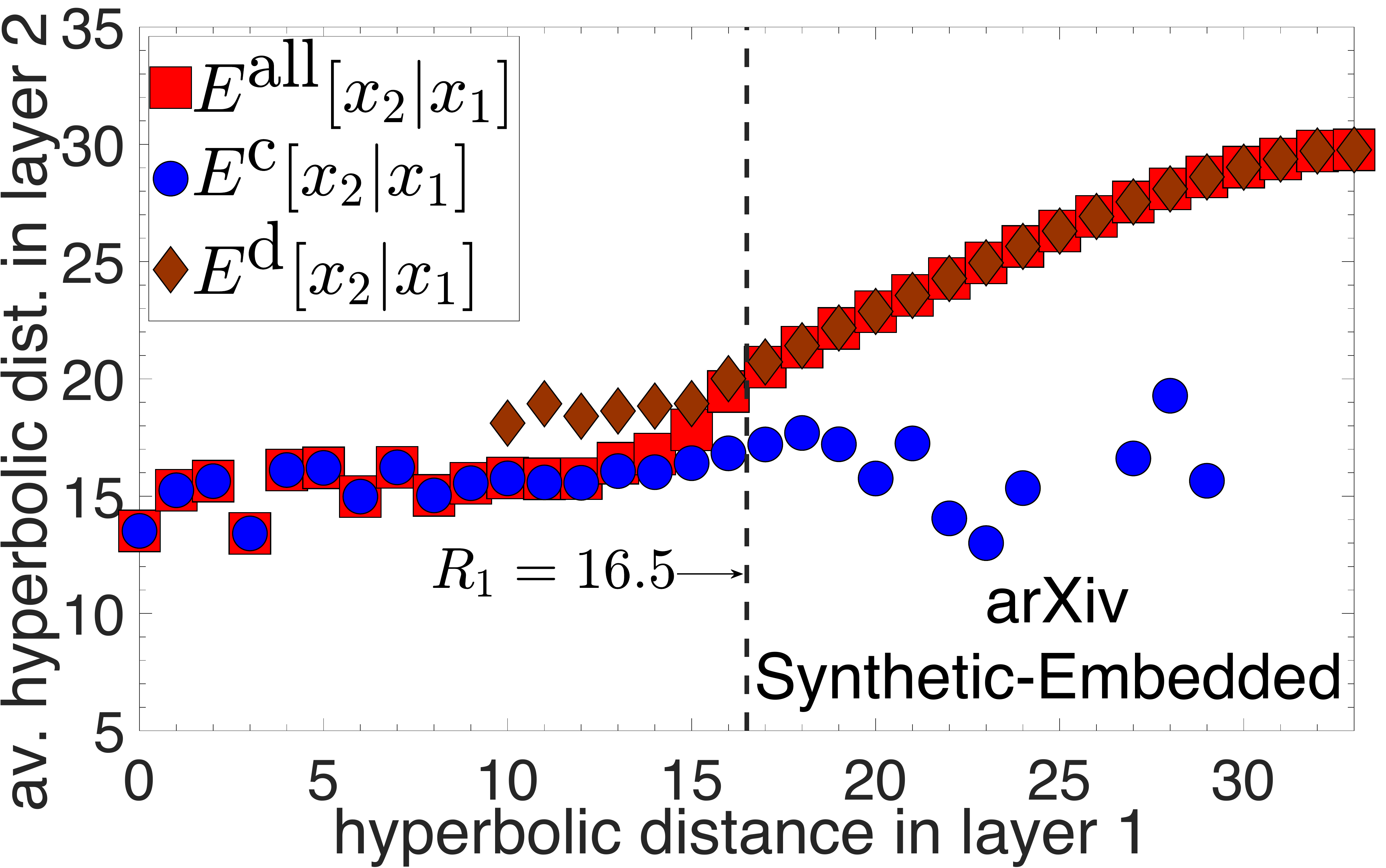}
\includegraphics[width=1.73in, height=1.1in]{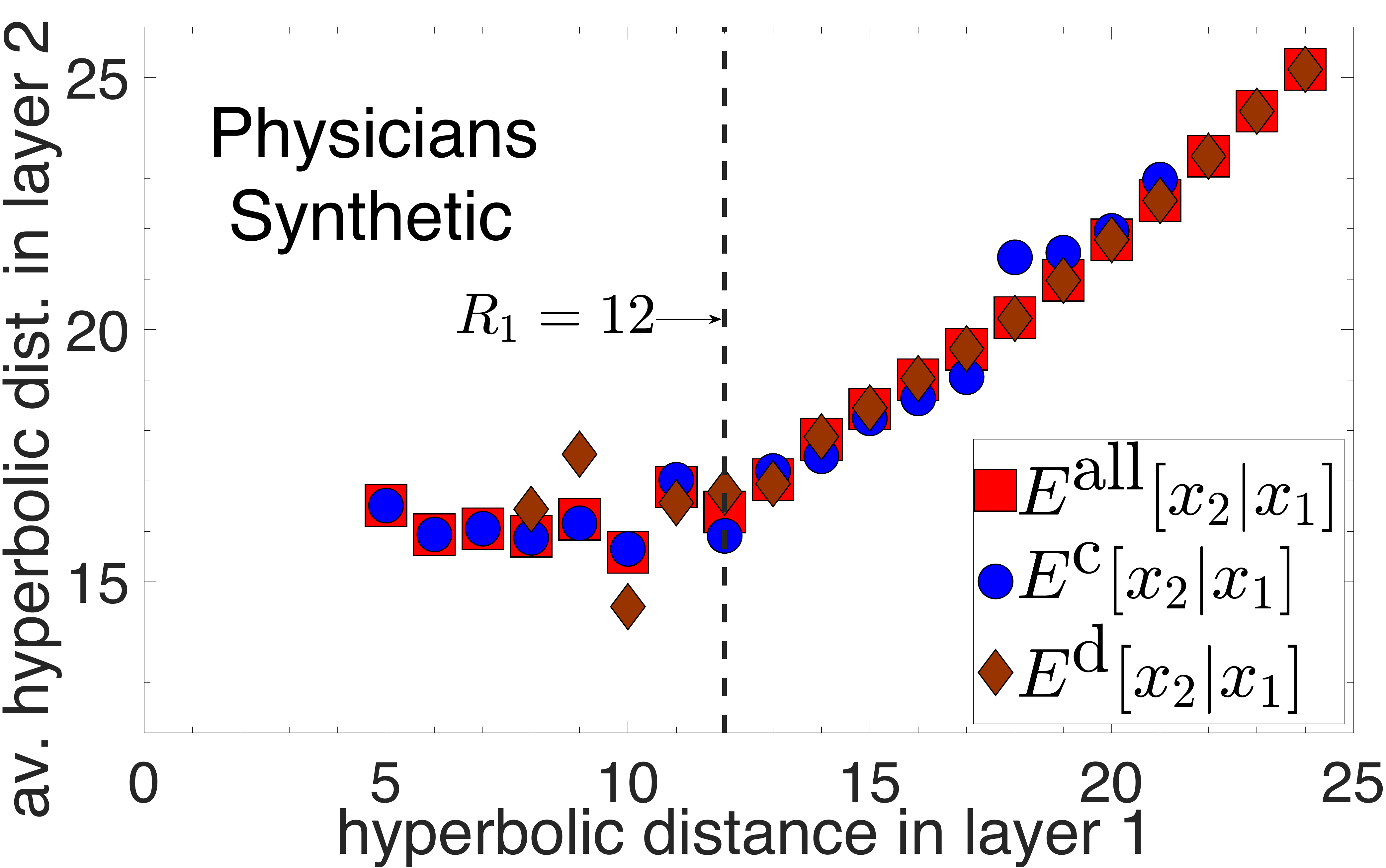}
\includegraphics[width=1.73in, height=1.1in]{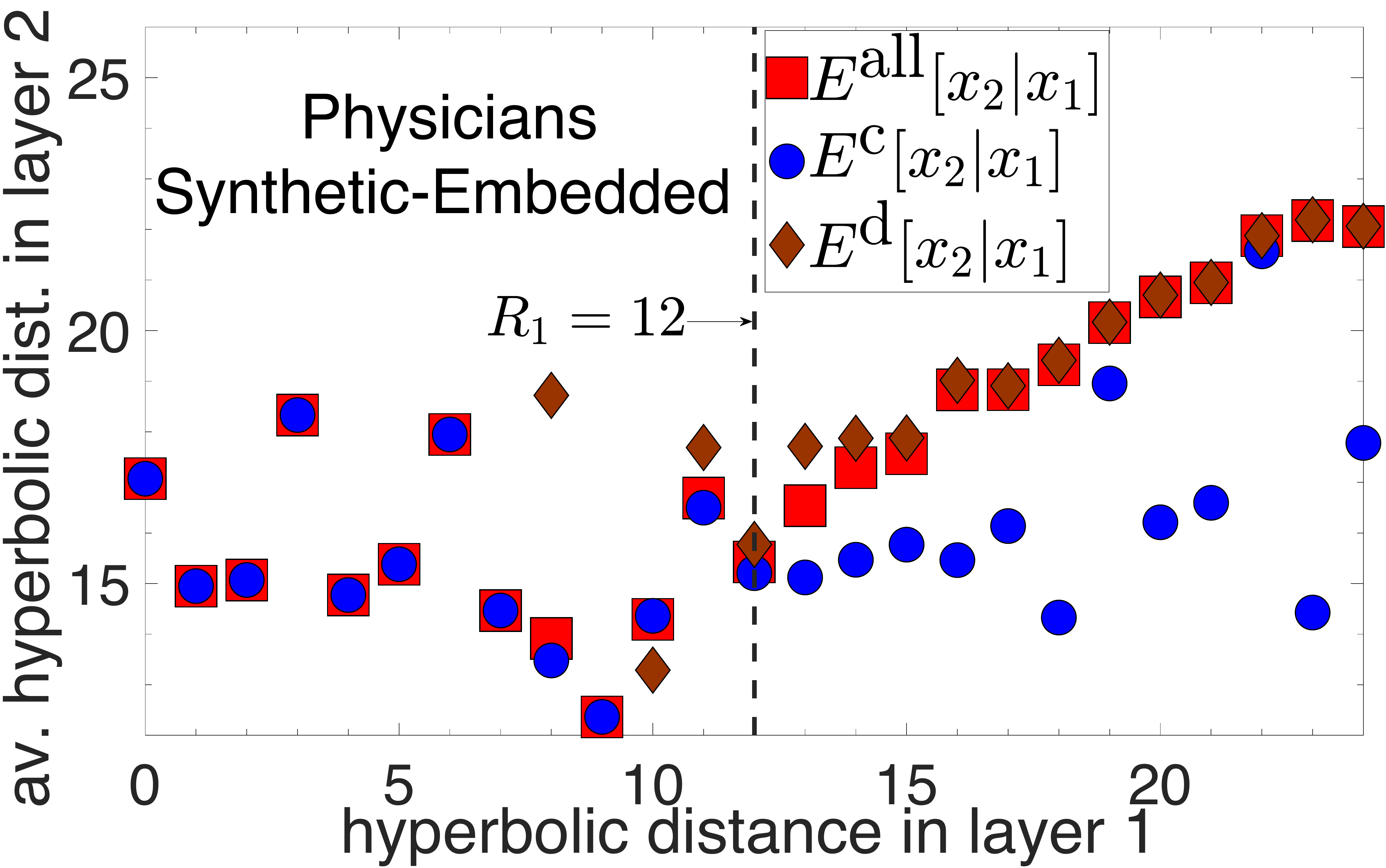}
\includegraphics[width=1.73in, height=1.1in]{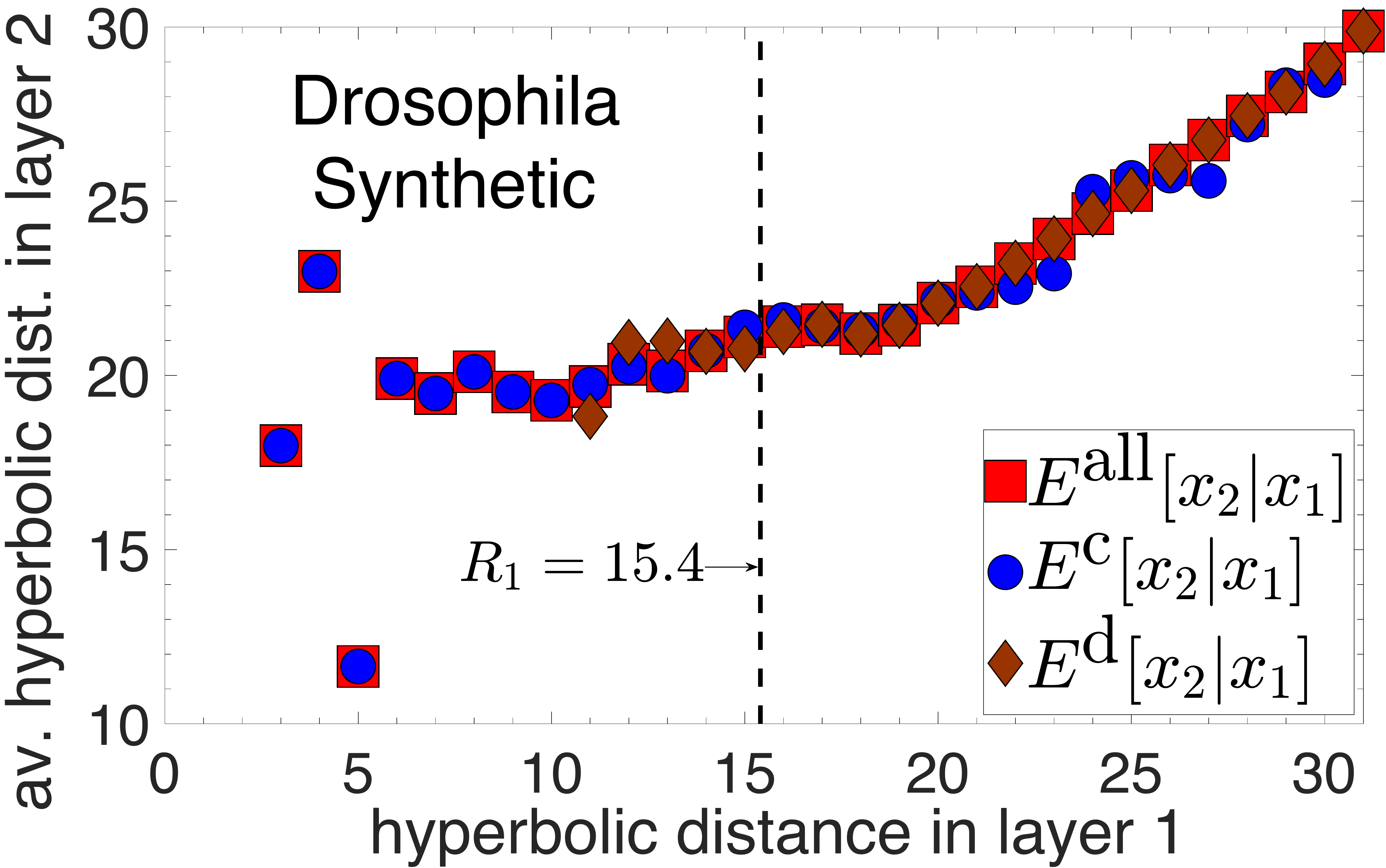}
\includegraphics[width=1.73in, height=1.1in]{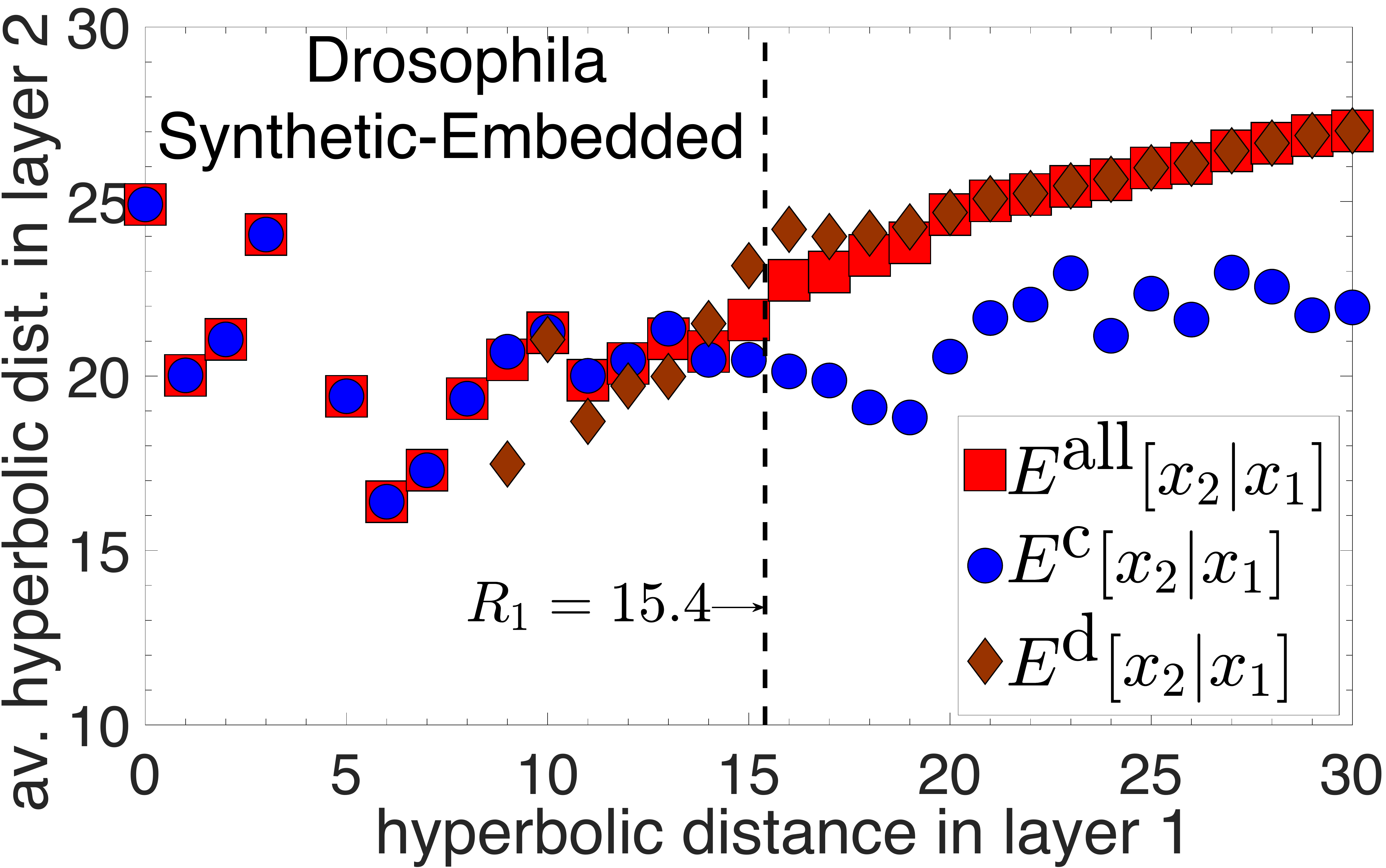}
\includegraphics[width=1.73in, height=1.1in]{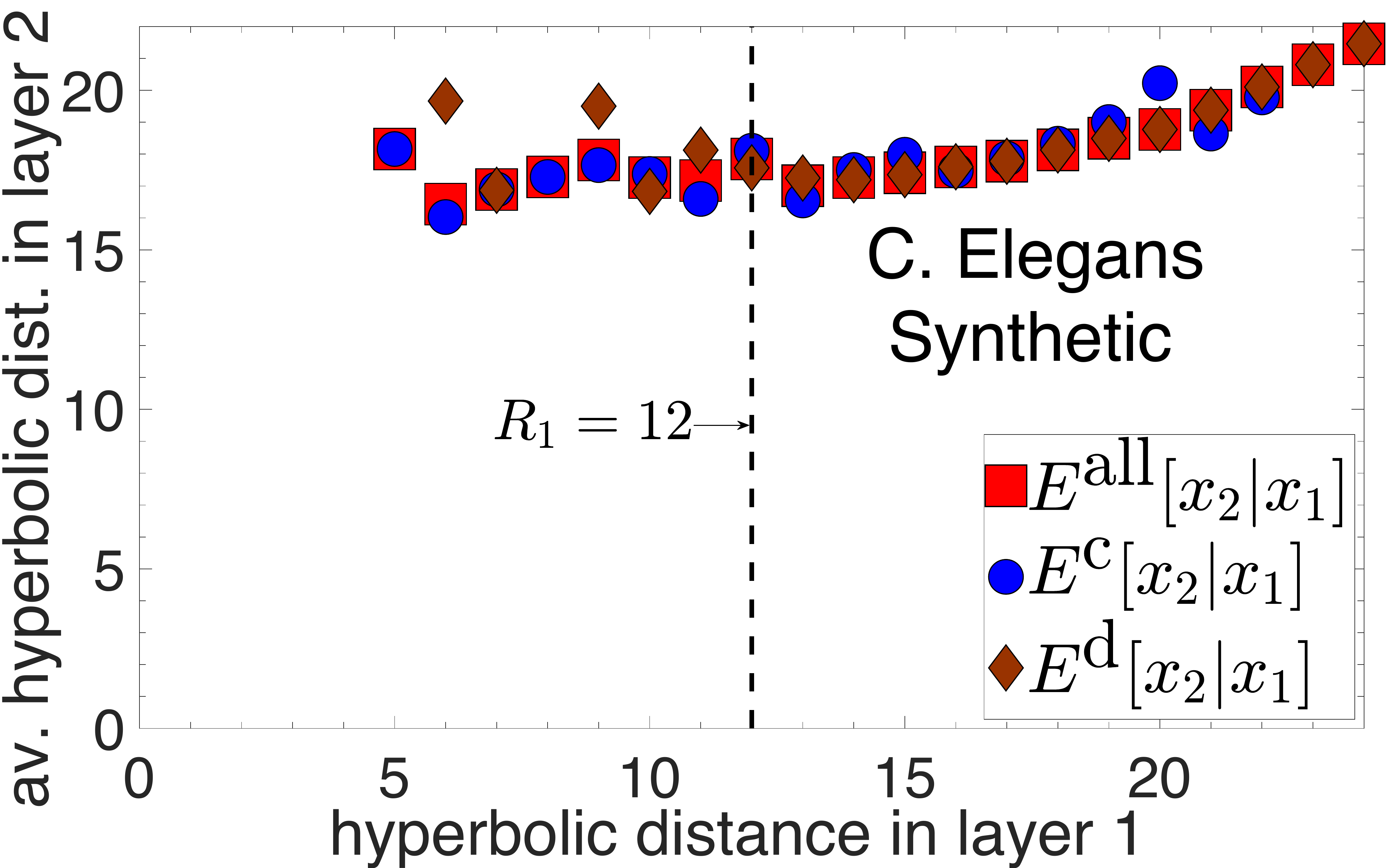}
\includegraphics[width=1.73in, height=1.1in]{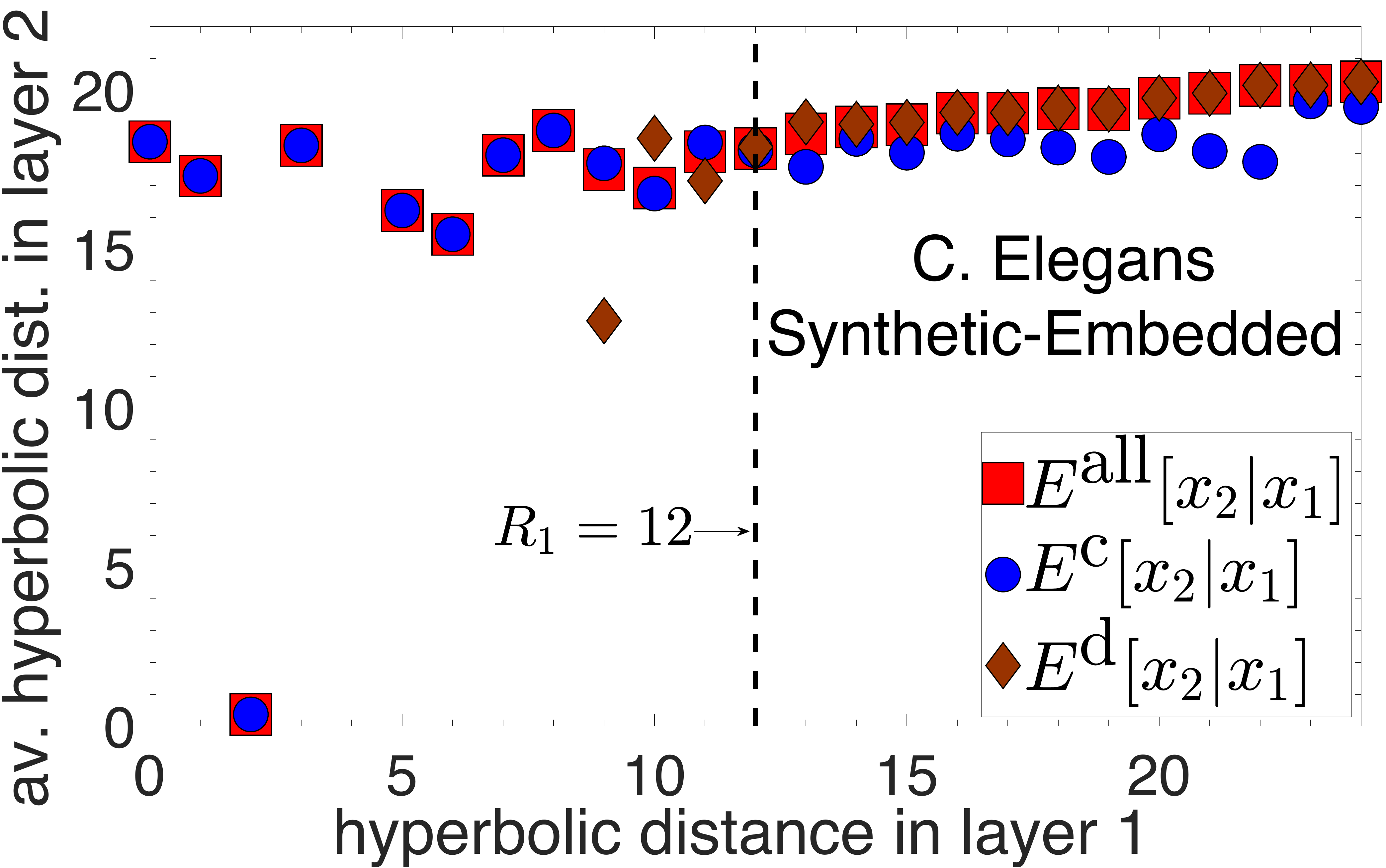}
\includegraphics[width=1.73in, height=1.1in]{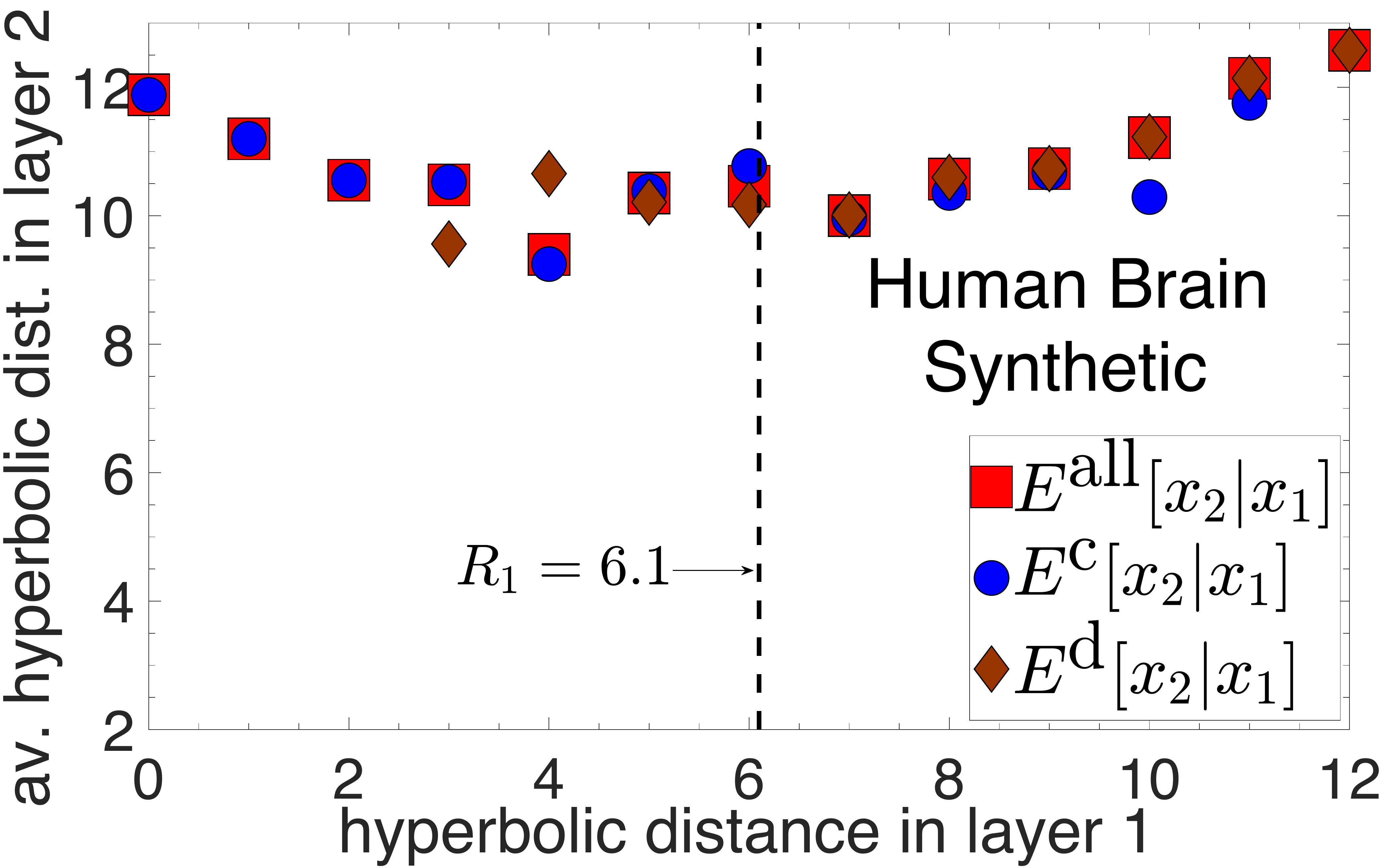}
\includegraphics[width=1.73in, height=1.1in]{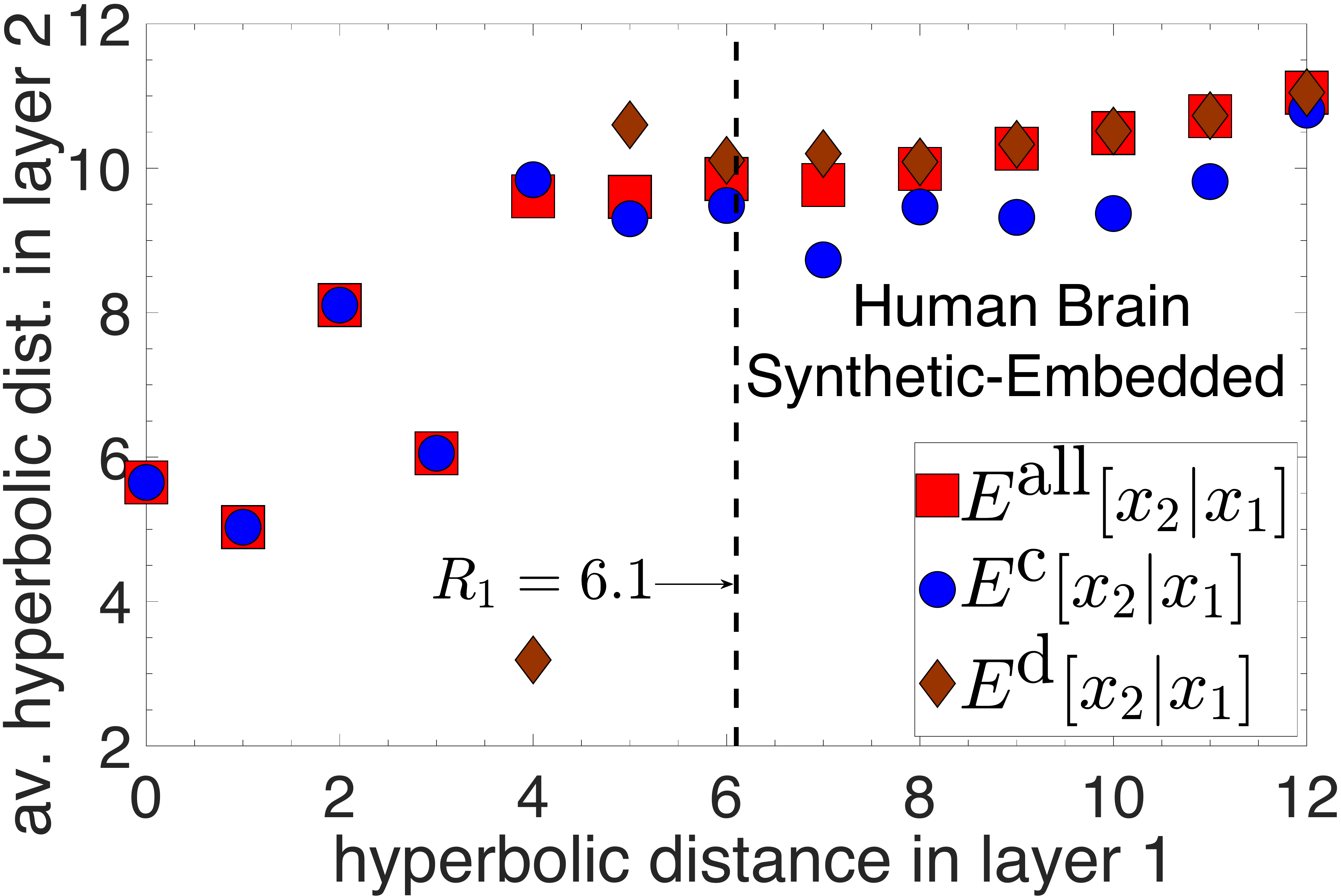}
}
\caption{Hyperbolic distance correlations in synthetic versions of the multiplexes in Fig.~\ref{fig:empirical_results} constructed using the GMM-LP. The plots show the conditional average hyperbolic distances as in Fig.~\ref{fig:empirical_results}. For each multiplex the first plot are the results with the real node coordinates while the second plot are the results with the inferred node coordinates.  
\label{fig:sim_results_2}}
\end{figure*}

Fig.~\ref{fig:sim_results_1} shows the trans-layer connection probabilities $p_{\text{trans}}^{\text{all}}(x_1)$, $p_{\text{trans}}^{\text{c}}(x_1)$, $p_{\text{trans}}^{\text{d}}(x_1)$ and the connection probabilities $p_2^{\text{all}}(x_2)$, $p_2^{\text{c}}(x_2)$, $p_2^{\text{d}}(x_2)$ in synthetic versions of the Internet, arXiv, Physicians,  Drosophila, C. Elegans and Human Brain multiplexes constructed using the GMM-LP. Fig.~\ref{fig:sim_results_2} also shows the corresponding average hyperbolic distances $E^{\text{all}}[x_2|x_1]$, $E^{\text{c}}[x_2|x_1]$, $E^{\text{d}}[x_2|x_1]$. Each synthetic layer $i=1,2$ has approximately the same number of nodes $N_i$, average degree, power law degree distribution exponent $\gamma_i$ and temperature $T_i$  as the corresponding real layer, see Appendix~\ref{sec:real_data} for the values of these parameters. For simplicity, in these synthetic systems all nodes in the smaller layer also exist in the larger, while the corresponding correlation strengths $\nu, g$ and link persistence probabilities $w$ are the ones shown in Table~\ref{tab_datasets} and Fig.~\ref{fig:empirical_results}.  

For each synthetic multiplex, Figs.~\ref{fig:sim_results_1}, \ref{fig:sim_results_2} show the results with the real node coordinates, as well as with the inferred node coordinates that are obtained after independently mapping each layer to its hyperbolic space using HyperMap. The reason of performing the latter is to facilitate a more direct comparison with the plots of Fig.~\ref{fig:empirical_results} where the layers are also embedded independently. We observe that the resemblance of Figs.~\ref{fig:sim_results_1}, \ref{fig:sim_results_2} with Fig.~\ref{fig:empirical_results} is remarkable, especially for the embedded synthetic layers. We note that by maximizing the likelihood in Eq.~(\ref{eq:likelihood}) connected nodes are attracted and placed closer to each other in the hyperbolic space while disconnected nodes repel. This explains why in general $E^{\text{c}}[x_2|x_1] < E^{\text{d}}[x_2|x_1]$ in the embeddings. This effect can also impose a decreasing trend in $p_2^{\text{c}}(x_2)$ at large $x_2 > R_2$ (also observed in Fig.~\ref{fig:empirical_results}), which does not exist if the real coordinates are used and where we can clearly see that $p_2^{\text{c}}(x_2) \approx w$ (Fig.~\ref{fig:sim_results_1}). $p_{\text{trans}}^{\text{c}}(x_1)$ appears less prone to embedding effects and we can see in Fig.~\ref{fig:sim_results_1} that  $p_{\text{trans}}^{\text{c}}(x_1) \approx w$ at $x_1 > R_2$, as in the real systems (Fig.~\ref{fig:empirical_results}).

Taken altogether, the GMM-LP can capture the behavior observed in real systems remarkably well. In Section~\ref{sec:analysis} we analyze the model and prove the behavior observed in Fig.~\ref{fig:sim_results_1}. Furthermore, we show that the main topological properties of layer~2 in GMM-LP are very similar to the ones in GMM. Below, we show that link persistence drives the high edge overlap observed in real systems.

\section{Link persistence and Edge Overlap}
\label{sec:overlap}

The edge overlap $O$ between two layers (layer~1, layer~2) is formally defined as the ratio of the number of overlapping (i.e., common) edges between the layers, to the maximum possible number of common edges~\cite{Battiston2014}:
\begin{equation}
O = \frac{\#(\text{overlapping edges)}}{\min[\#(\text{edges in layer 1}),\#(\text{edges in layer 2})]},
\label{eq:overlap}
\end{equation}
where the number of edges in each layer in the denominator is computed only among the common nodes.

\begin{figure}[!h]
\includegraphics[width=3.4in]{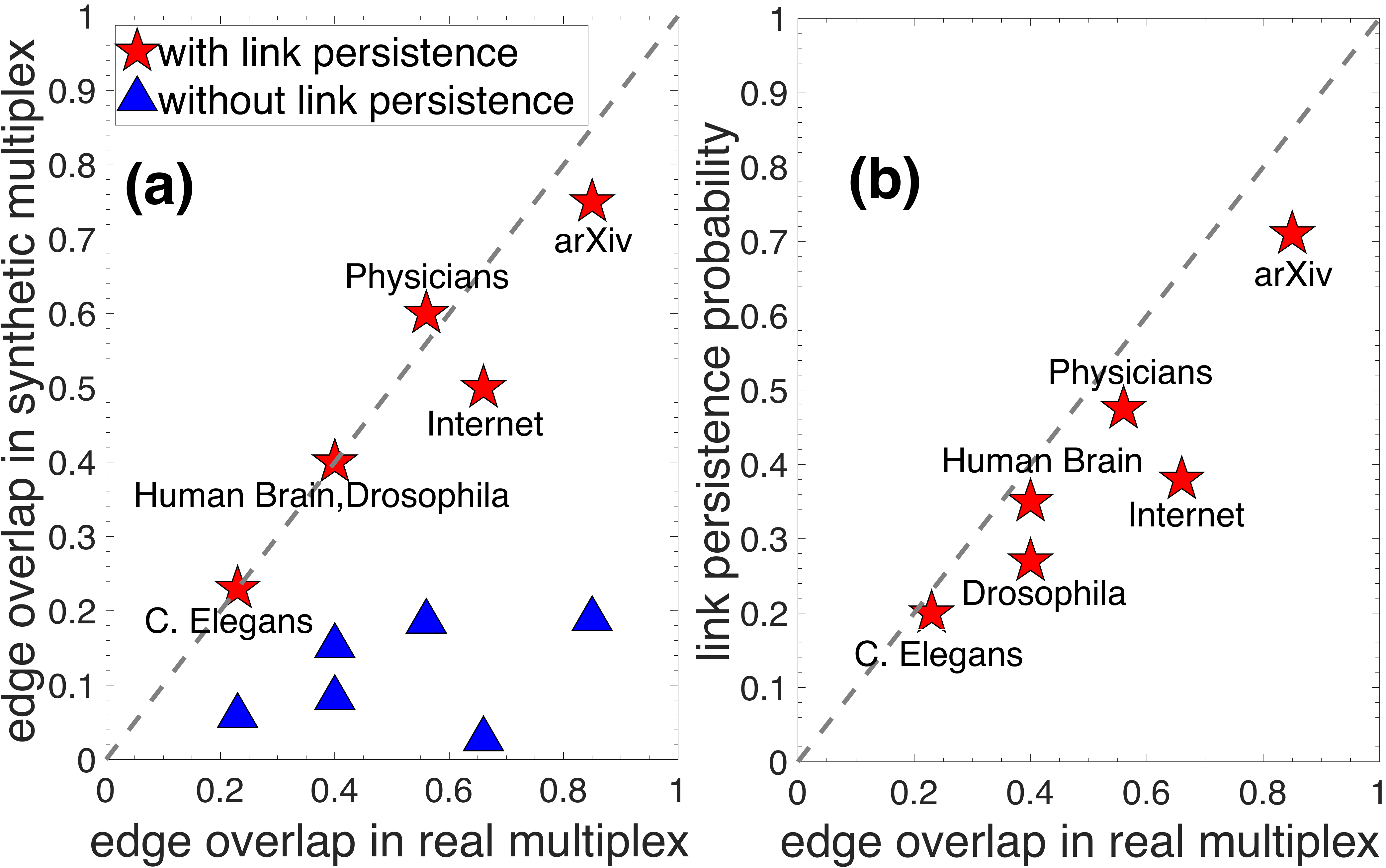}
\caption{\textbf{(a)} Edge overlap $O$ in real systems vs. synthetic counterparts with and without link persistence. \textbf{(b)} Edge overlap $O$ in real systems vs. link persistence probability $w$.
\label{fig:overlap}}
\end{figure}

Fig.~\ref{fig:overlap}(a) shows the edge overlap $O$ in the real systems of Fig.~\ref{fig:empirical_results} and in their synthetic counterparts of Figs.~\ref{fig:sim_results_1},~\ref{fig:sim_results_2}. The figure also shows the overlap in synthetic systems constructed with the same parameters and correlation strengths as the ones of Figs.~\ref{fig:sim_results_1},~\ref{fig:sim_results_2} but without link persistence ($w=0$). We see that the synthetic systems with link persistence exhibit a high edge overlap similar to the real systems, while the ones without link persistence have a significantly lower overlap. To achieve the same overlap in synthetic systems without link persistence we need significantly higher correlation strengths than those in Table~\ref{tab_datasets}. For example, in the synthetic versions of the Human Brain, C. Elegans and Internet multiplexes we would need $\nu=g\approx0.7, \nu=g \approx 0.8, \nu=g \approx 0.9$, respectively, while in Drosophila, Physicians and arXiv we cannot achieve the same overlap as in the real systems even with $\nu=g \approx1$. Fig.~\ref{fig:overlap}(b) shows that the edge overlap in the considered systems is very well correlated with their estimated link persistence probabilities $w$ (horizontal dashed lines in Fig.~\ref{fig:empirical_results}), further highlighting that link persistence is the driving force for their observed high edge overlap.

\section{Analysis of the GMM-LP}
\label{sec:analysis}

In this section we analyze the GMM-LP.  We first analyze the connection probabilities in layer 2,  $p_2^{\text{c}}(x_2)$, $p_2^{\text{d}}(x_2)$, $p_2^{\text{all}}(x_2)$, and investigate the layer's degree distribution and clustering. We then analyze the trans-layer connection probabilities $p_{\text{trans}}^{\text{c}}(x_1)$, $p_{\text{trans}}^{\text{d}}(x_1)$, $p_{\text{trans}}^{\text{all}}(x_1)$, which will inform us of how we can improve trans-layer link prediction by taking link persistence into account.

\subsection{Connection probabilities in layer 2}
\label{sec:con_prob}

The connection probabilities among connected and disconnected layer~1 pairs in layer 2 are given by Eqs.~(\ref{eq:p_c}),~(\ref{eq:p_d}). At $x_2 \gg R_2-2T_2\ln{w}$, $p_2^{\text{c}}(x_2) \approx (1-p_2(x_2))w \approx w$, while at $x_2 \ll R_2-2T_2\ln{w}$, $p_2^{\text{c}}(x_2) \approx p_2(x_2)=p_2^{\text{d}}(x_2)$, justifying the observed behavior in Fig.~\ref{fig:sim_results_1}. Below, we analyze $p_2^{\text{all}}(x_2)$. Since link persistence is relevant only among common pairs, to ease exposition we assume that the two layers consist of $N$ common nodes.

Let $f_1(x_1), f_2(x_2)$ be the PDFs of the hyperbolic distances $x_1, x_2$ among pairs in layers~1 and~2. 
Further, let $f(x_1|x_2)$ be the PDF of the hyperbolic distance $x_1$ of a pair in layer~1, conditioned on its distance $x_2$ in layer~2. Even though the PDFs admit closed-form expressions~\cite{PaPsKr12}, the conditional PDF depends on the radial and angular correlation strengths $\nu, g$ and does not have an analytic expression (see Appendix~\ref{sec:conditional_hyperbolic_pdf}). Nevertheless, we can still deduce the behavior of $p_2^{\text{all}}(x_2)$.

Let $\eta(x_2)$ be the probability that a pair of nodes at distance $x_2$ is connected in layer~1:
\begin{align}
\label{eq:eta}
\eta(x_2)=\int_0^{2 R_1}f(x_1|x_2)p_1(x_1)\mathrm{d}x_1,
\end{align}
where $p_1(x_1)$ is the connection probability in layer 1 (Eq.~(\ref{eq:c_prob_1})). If there are distance correlations, $\eta(x_2)$ decreases with $x_2$ since $f(x_1|x_2)$ concentrates over higher $x_1$ values and $p_1(x_1)$ decreases with $x_1$. The stronger the correlations, the faster $\eta(x_2)$ decreases with $x_2$ as $x_1$ is more narrowly distributed around $x_2$. If there are no correlations, $f(x_1|x_2)=f_1(x_1)$, and $\eta(x_2)$ is the constant:
\begin{equation}
\label{eq:eta_u}
\eta(x_2) \equiv \eta=\frac{\bar{k}_1}{N},
\end{equation}
where $\bar{k}_1$ is the average degree in layer 1.
On the other hand, in the maximally correlated case where nodes coordinates are identical in the two layers,\footnote{Angular coordinates are identical at $g=1$ (Eqs.~(\ref{eq:assign_theta_text}),~(\ref{eq:truncated_normal})), while radial coordinates are identical at $\nu \to 1$ if $\gamma_1=\gamma_2$ and $\bar{k}_1/\bar{k}_2= T_1 \sin{T_2 \pi}/(T_2 \sin{T_1 \pi})$ (Eqs.~(\ref{eq:r_cond_text}), (\ref{eq:not})).} $f(x_1|x_2)=\delta(x_1-x_2)$, where $\delta$ is the Dirac delta function, and:
\begin{equation}
\label{eq:eta_mc}
\eta(x_2)=p_1(x_2).
\end{equation}
Using Eqs.~(\ref{eq:p_c}),~(\ref{eq:p_d}),~(\ref{eq:eta}) we can write:
\begin{align}
\label{eq:p_all}
\nonumber p_2^{\text{all}}(x_2)&=\eta(x_2) p_2^{\text{c}}(x_2)+\left(1-\eta(x_2)\right) p_2^{\text{d}}(x_2)\\
&=p_2(x_2)+(1-p_2(x_2))w\eta(x_2).
\end{align}
Therefore, $p_2^{\text{all}}(x_2)$ differs from the connection probability in the $\mathbb{H}^{2}$ model, $p_2(x_2)$, by the term $(1-p_2(x_2))w\eta(x_2)$. The term $1-p_2(x_2)$ is the percentage of disconnected pairs at distance $x_2$ in the $\mathbb{H}^{2}$ model, while $w\eta(x_2)$ is the percentage of these pairs that are connected by a persistent link in GMM-LP.  At $x_2 \gg R_2$, $(1-p_2(x_2))w\eta(x_2) \approx w \eta(x_2)$ and $p_2^{\text{all}}(x_2) \approx p_2(x_2)+w\eta(x_2)$. At $x_2 \ll R_2$, $p_2^{\text{all}}(x_2) \approx p_2(x_2)$. 

\begin{figure}[!h]
\includegraphics[width=3.4in]{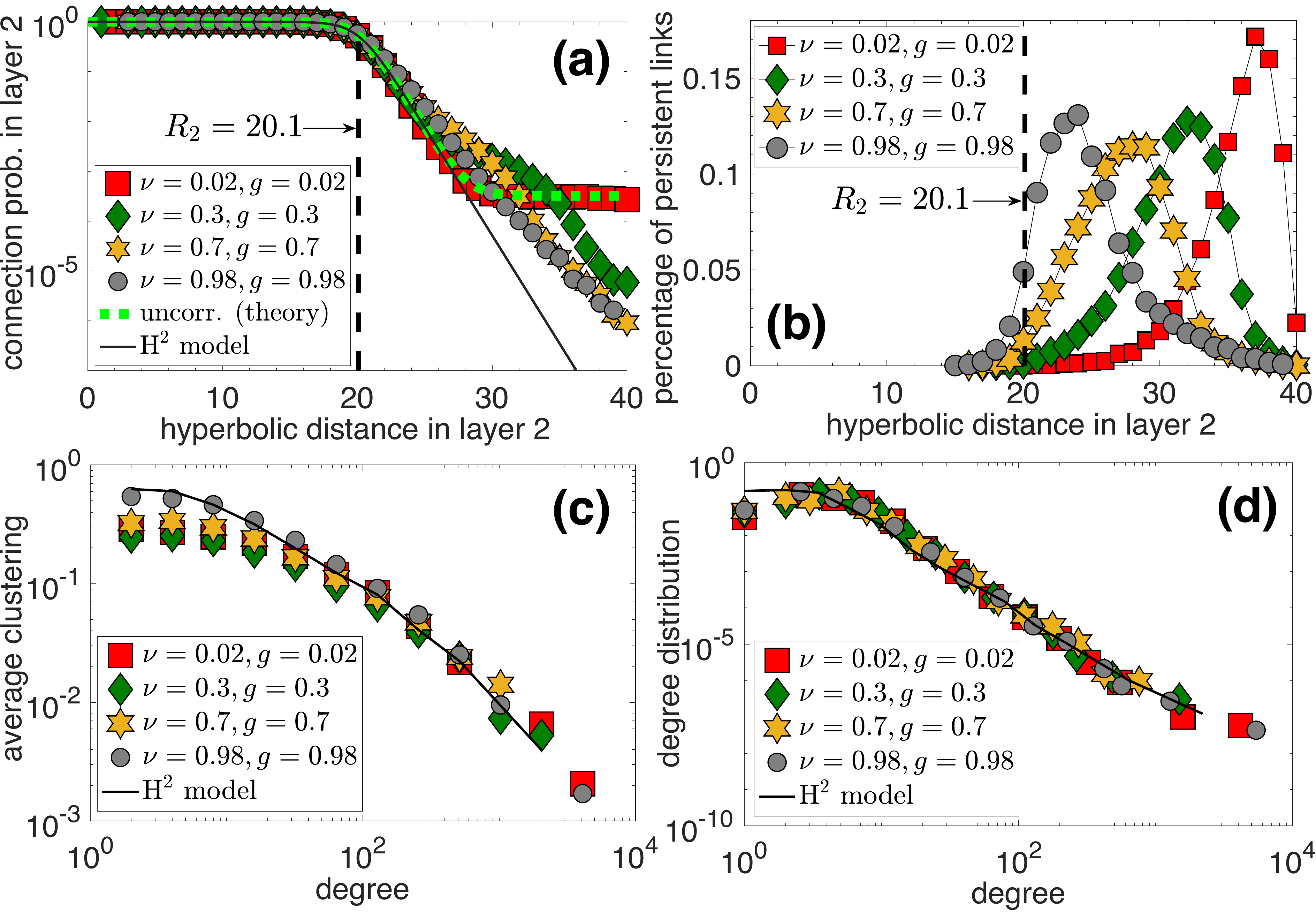}
\caption{\textbf{(a)} Connection probability $p_2^{\text{all}}(x_2)$.  \textbf{(b)} Distribution of persistent links over the hyperbolic distances in layer~2. \textbf{(c)} Degree-depended average clustering $\bar{c}(k)$. \textbf{(d)} Degree distribution $P(k)$.  The results correspond to layer~2 of a two-layer synthetic multiplex with $N=10^4$ nodes, $\gamma_1=2.8, \gamma_2=2.3, T_1=0.7, T_2=0.5, \bar{k}_1=\bar{k}_2=8$, link persistence probability $w=0.4$ and correlation strengths as shown in the legends. In~(a) the green dotted line is the theoretical prediction for the uncorrelated case ($\eta(x_2)$ given by Eq.~(\ref{eq:eta_u})), while the solid line is the connection probability in the $\mathbb{H}^{2}$ model (Eq.~(\ref{eq:c_prob_2})). In~(a),~(b) the vertical dashed lines indicate the hyperbolic disc radius $R_2$. In~(c),~(d) the solid lines are the corresponding results in a synthetic network constructed with the $\mathbb{H}^{2}$ model with the same parameters as layer~2.
\label{fig:analysis_validation_1}}
\end{figure}

We thus see that persistent links affect the connection probability mostly at large distances $x_2 \gg R_2$, increasing its tail. This is validated in Fig.~\ref{fig:analysis_validation_1}(a) where we also see that the stronger the correlations the faster the connection probability decreases at large distances. In other words, with stronger correlations pairs at smaller distances have higher chances of being connected  by a persistent link, as expected, and as also seen in Fig.~\ref{fig:analysis_validation_1}(b). Since persistent links increase the connection probability at large distances, they decrease the average clustering in the network, akin to increasing the temperature in the $\mathbb{H}^{2}$ model (Sec.~\ref{sec:H2}). This is seen in Fig.~\ref{fig:analysis_validation_1}(c), where we also see that the stronger the correlations the smaller the decrease is, as also expected, since shorter distance connections are preferred. Below, we show that the tail of the degree distribution remains the same as in the $\mathbb{H}^{2}$ model, irrespectively of the correlation strengths (Fig.~\ref{fig:analysis_validation_1}(d)).

\subsection{Degree distribution in layer 2}
\label{sec:degree_pdf}

We first recall that in the $\mathbb{H}^{2}$ model the average degree of a node decreases exponentially with its radial coordinate $r$, $\bar{k}(r) \propto e^{-\frac{1}{2}r}$ (Eq.~(\ref{eq:bar_k_r_H2})), while the density of radial coordinates increases exponentially, $\rho(r) \propto e^{\frac{1}{2\beta}r}$. The combination of these two exponentials gives a power law degree distribution, $P(k) \propto k^{-\gamma}$, $\gamma=1+1/\beta > 2$.

In our case the density of radial coordinates is the same as in the $\mathbb{H}^{2}$ model. Therefore, to investigate the degree distribution we analyze $\widetilde{\bar{k}}_2 (r)$, which is the average degree of a node with radial coordinate $r$ in layer~2, to see whether it still decreases as $e^{-\frac{1}{2}r}$. Since the hyperbolic distance $x$ between two nodes is a function of their radial coordinates $r, r'$ and angular distance $\Delta\theta$ (Eq.~(\ref{eq:hyperbolic_distance})), $p_2^{\text{all}}(x)$ is a function of $r, r', \Delta\theta$, and we can write:
\begin{equation}
\label{eq:bar_k_2_r}
\widetilde{\bar{k}}_2(r)= \frac{N}{\pi} \int_{0}^{R_2} \rho_2(r') \mathrm{d}r' \int_{0}^{\pi} p_2^{\text{all}}(r, r', \Delta\theta)  \mathrm{d}\Delta \theta. 
\end{equation}
The above relation does not have a closed-form expression except in the uncorrelated and maximally correlated cases (Eqs.~(\ref{eq:eta_u}),~(\ref{eq:eta_mc})), with the additional requirement in the latter that the temperatures of the two layers are the same, $T_1=T_2$. In the uncorrelated case we  have (see Appendix~\ref{sec:bar_k_corrs}):
\begin{align}
\label{eq:bar_k_2_r_no_corr}
\widetilde{\bar{k}}_2(r)& \approx (1-\frac{w\bar{k}_1}{N})\bar{k}_2(r)+w\bar{k}_1,
\end{align}
where:
\begin{align}
\label{eq:bar_k_2_r_H2_model}
\bar{k}_2(r) &\equiv \bar{k}_{0,2} e^{\frac{1}{2}(R_2-r)},~\bar{k}_{0, 2} \equiv \bar{k}_2\left(\frac{\gamma_2-2}{\gamma_2-1}\right).
\end{align}
We can see from the above relations that $\widetilde{\bar{k}}_2(r) \approx \bar{k}_2(r)$ for large $N$ and $r \ll R_2-2\ln{(w \bar{k}_1/\bar{k}_{0, 2})} \propto \ln{N}$.\footnote{In sparse networks $\bar{k}_1 \propto \bar{k}_{0,2} \propto \bar{k}_2 \ll N$ while $R_2 \propto \ln{N}$.}

In the maximally correlated case we can show (Appendix~\ref{sec:bar_k_corrs}) that: 
\begin{align}
\label{eq:bar_k_2_r_max_corr_bounds}
 A \bar{k}_2(r) \leq \widetilde{\bar{k}}_2(r)  \leq B \bar{k}_2(r),
\end{align}
where  $A\equiv (1+w C-w\sqrt{C(1-T_1)(1-T_2)})$, $B \equiv (1+w C)$,
$C \equiv T_1\sin{T_2\pi}/(T_2\sin{T_1\pi})$,
and $\bar{k}_2(r)$ in Eq.~(\ref{eq:bar_k_2_r_H2_model}). 
Therefore $\widetilde{\bar{k}}_2(r) \propto \bar{k}_2(r)$. If in addition $T_1=T_2$:
\begin{align}
\label{eq:bar_k_2_r_max_corr}
\widetilde{\bar{k}}_2(r) \approx (1+wT_2)\bar{k}_2(r).
\end{align}

\begin{figure}[!h]
\centering{
\includegraphics[width=3.4in]{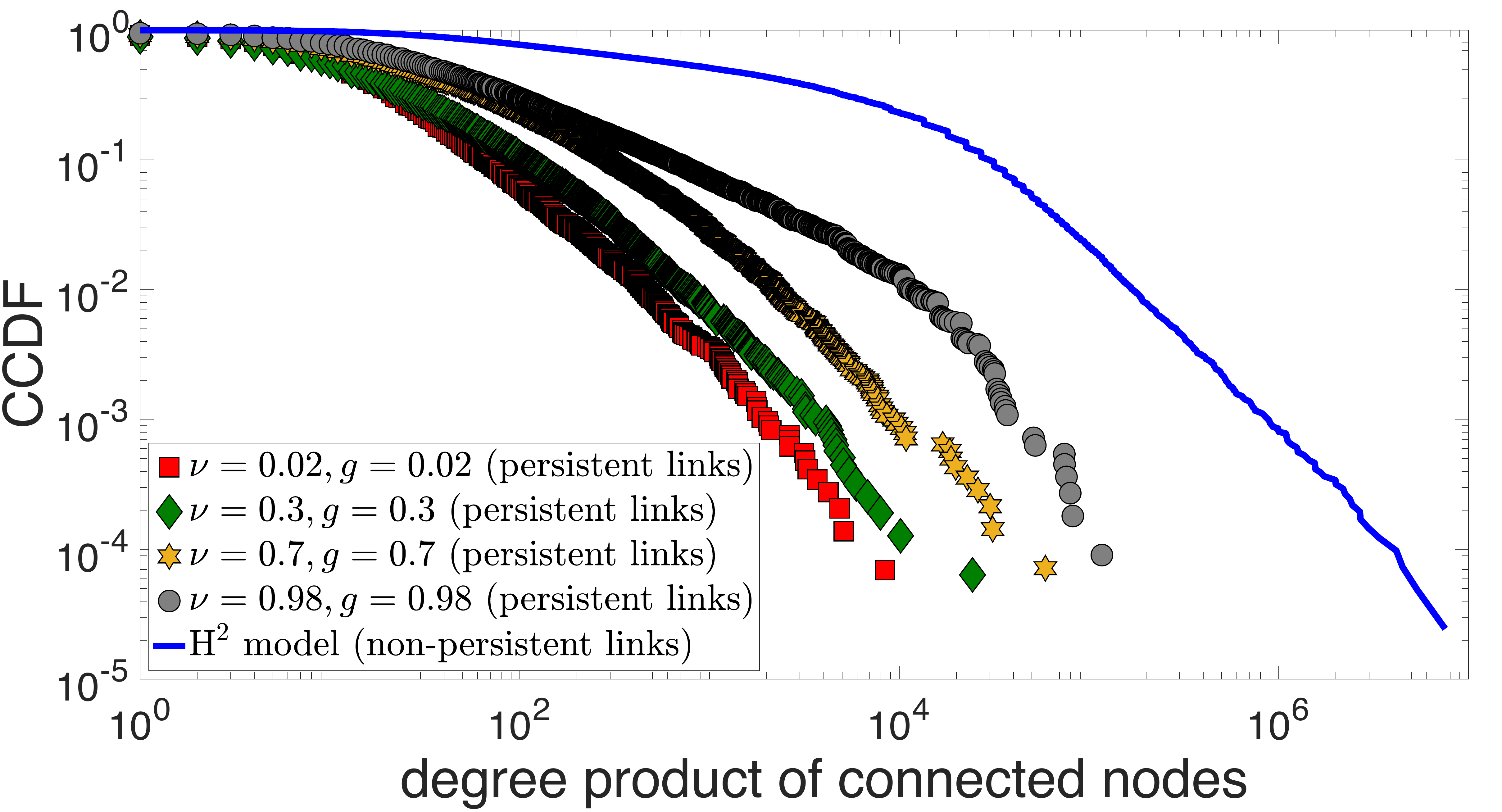}
\caption{Complementary cumulative distribution function (CCDF) of the degree product $k_1\times k_2$ of nodes connected by persistent and non-persistent links. The results correspond to the persistent links in layer~2 of the multiplexes of Fig.~\ref{fig:analysis_validation_1}, and to the links (non-persistent) in a network constructed with the $\mathbb{H}^{2}$ model with the same parameters as layer~2.
\label{fig:analysis_validation_2}}}
\end{figure}

Our analysis shows that if there are no distance correlations $\widetilde{\bar{k}}_2(r) \propto e^{-\frac{1}{2}r}$ for sufficiently small $r$. Small values of $r$ correspond to higher expected degrees and hence the degree distribution still scales as $P(k) \propto k^{-\gamma_2}$, $\gamma_2=1+1/\beta_2 > 2$. 
This result is expected since persistent links randomly connect pairs at larger distances (cf. Fig.~\ref{fig:analysis_validation_1}(a)), which correspond primarily to low degree nodes. As distance correlations increase, persistent links tend to connect pairs at smaller distances (Fig.~\ref{fig:analysis_validation_1}(b)), which results in more connections to higher degree nodes, see Fig.~\ref{fig:analysis_validation_2}. At strongest correlations, we see from Eq.~(\ref{eq:bar_k_2_r_max_corr_bounds}) that $\widetilde{\bar{k}}_2(r) \propto \bar{k}_2(r)~\forall r$, meaning again that $P(k) \propto k^{-\gamma_2}$.  We can thus conclude that no matter the correlation strengths, persistent links do not affect the tail of the degree distribution, which remains the same as in the $\mathbb{H}^{2}$ model. This result is validated in Fig.~\ref{fig:analysis_validation_1}(d). 

\subsection{Average degree in layer 2}
\label{sec:average_degree}

The resulting average degree in layer 2, $\widetilde{\bar{k}}_2$, is:
\begin{equation}
\label{new_bar_k_2}
\widetilde{\bar{k}}_2= N \int_{0}^{2 R_2}  p_2^{\text{all}}(x_2) f_2(x_2)\mathrm{d}x_2=\bar{k}_2+ w(\bar{k}_1-\bar{k}_{\textnormal{o}}),\\
\end{equation}
where $\bar{k}_2$ from Eq.~(\ref{eq:not}) and:
\begin{equation}
\nonumber \bar{k}_\textnormal{o}=N \int_{0}^{2 R_2} \int_0^{2 R_1}p_1(x_1)p_2(x_2)f(x_1|x_2)f_2(x_2) \mathrm{d}x_1\mathrm{d}x_2
\end{equation}
is the average local edge overlap between the two layers due to distance correlations. If there are no distance correlations, $\bar{k}_\textnormal{o}=(\bar{k}_1 \bar{k}_2)/N \to 0$ as $N\to \infty$ and $\widetilde{\bar{k}}_2=\bar{k}_2+ w\bar{k}_1$. On the other hand, if the topologies  of the two layers are identical,  $\bar{k}_{\textnormal{o}}=\bar{k}_1=\bar{k}_2$ and $\widetilde{\bar{k}}_2=\bar{k}_2$.  Therefore, $\bar{k}_2 < \widetilde{\bar{k}}_2 < \bar{k}_2+ w\bar{k}_1$, and we can tune $\widetilde{\bar{k}}_2$ by tuning $\bar{k}_2$. We note that for $w > 0$ we need a lower $\bar{k}_2$ to achieve the same $\widetilde{\bar{k}}_2$ as for $w=0$ where $\widetilde{\bar{k}}_2 = \bar{k}_2$. A lower $\bar{k}_2$ corresponds to a higher hyperbolic disc radius $R_2$ (Eq.~(\ref{eq:not})), which explains the small differences in the $R_2$s of the original and embedded synthetic multiplexes in Fig.~\ref{fig:sim_results_1}---in the embeddings $w=0$ as we do not account for link persistence.

\subsection{Trans-layer connection probabilities}
\label{sec:trans_layer}

Let's turn our attention now to the trans-layer connection probabilities. The trans-layer connection probability among connected layer~1 pairs, $p_{\text{trans}}^{\text{c}}(x_1)$, can be written as:
\begin{align}
\label{eq:p_trans_c}
p_{\text{trans}}^{\text{c}}(x_1)=w+(1-w) \int_0^{2 R_2}f(x_2|x_1)p_2(x_2)\mathrm{d}x_2,
\end{align}
where $f(x_2|x_1)$ is the PDF of the hyperbolic distance $x_2$ of a pair in layer~2 conditioned on its distance $x_1$ in layer~1, and $p_2(x_2)$ in Eq.~(\ref{eq:c_prob_2}). If there are distance correlations, the second term in Eq.~(\ref{eq:p_trans_c}) decreases with $x_1$ since $f(x_2|x_1)$ concentrates over higher $x_2$ values and $p_2(x_2)$ decreases with $x_2$. The stronger the correlations the faster the decrease. In the uncorrelated case, $f(x_2|x_1)=f_2(x_2)$, and:
\begin{equation}
\label{eq:p_c_uc}
p_{\text{trans}}^{\text{c}}(x_1)=w+(1-w)\frac{\bar{k}_2}{N} \approx w,
\end{equation}
in sparse networks ($\bar{k}_2 \ll N$). In the maximally correlated case, $f(x_2|x_1)=\delta(x_2-x_1)$, and:
\begin{align}
\label{eq:p_c_mc}
\nonumber p_{\text{trans}}^{\text{c}}(x_1)&=w+(1-w) p_2(x_1)\\
&=p_2(x_1)+(1-p_2(x_1))w.
\end{align}
In the above relation, $p_{\text{trans}}^{\text{c}}(x_1) \approx (1-p_2(x_1))w \approx w$ for $x_1 \gg R_2-2T_2\ln{w}$, while for $x_1 \ll R_2-2T_2\ln{w}$, $p_{\text{trans}}^{\text{c}}(x_1) \approx p_2(x_1)$. 

\begin{figure}[!h]
\includegraphics[width=3.4in, height=2.5in]{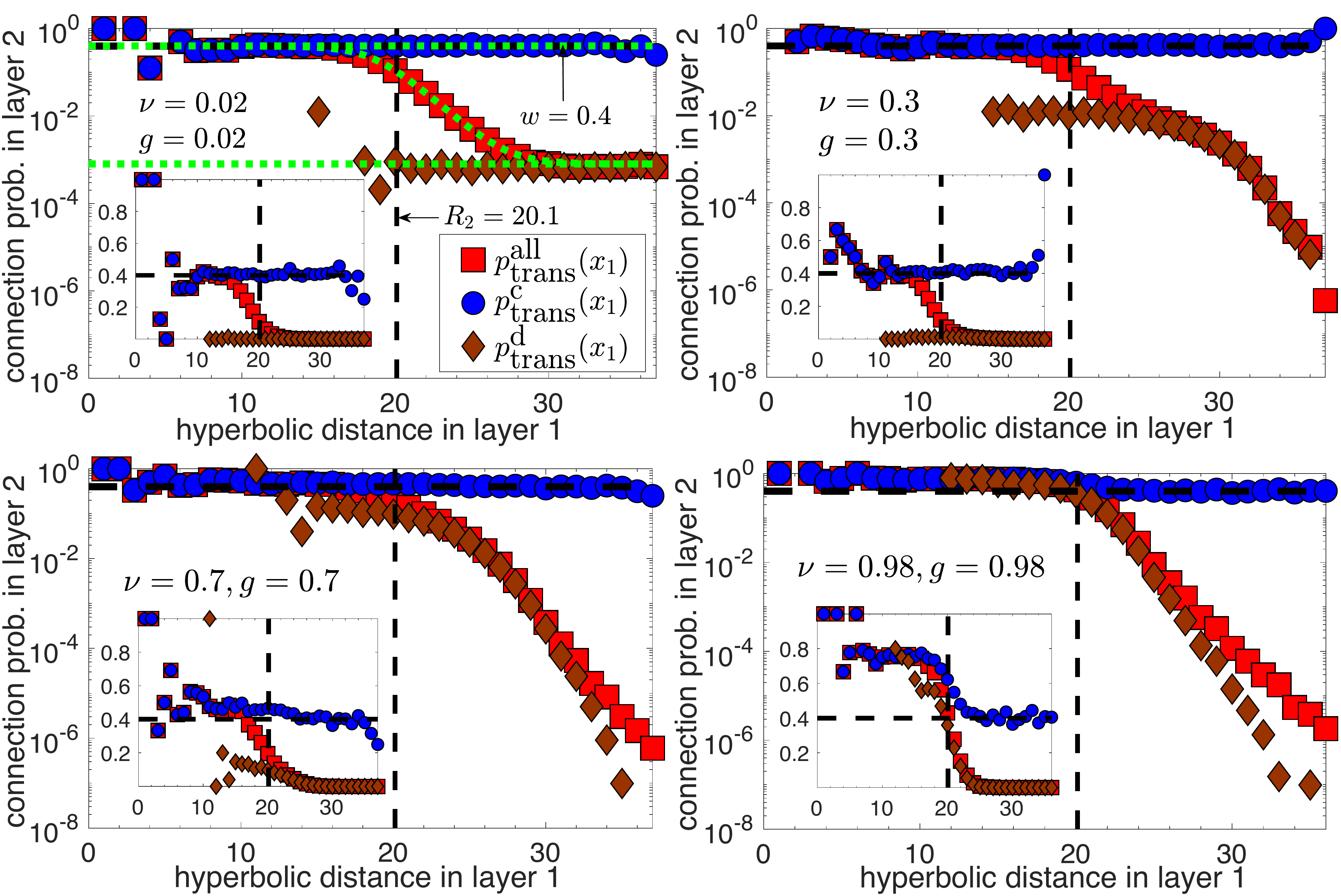}
\caption{ Trans-layer connection probabilities $p_{\text{trans}}^{\text{all}}(x_1)$, $p_{\text{trans}}^{\text{c}}(x_1)$, $p_{\text{trans}}^{\text{d}}(x_1)$ at different correlation strengths. The results correspond to the synthetic multiplexes of Fig.~\ref{fig:analysis_validation_1}. The $y$-axes in the plots are in log-scale, while the insets show the same results in linear scale.
The green dotted lines in the first plot ($\nu=g=0.02$) are the theoretical predictions for the uncorrelated case (Eqs.~(\ref{eq:p_c_uc}),~(\ref{eq:p_d_uc})~(\ref{eq:p_all_trans})). 
\label{fig:analysis_validation_3}}
\end{figure}

Taken altogether, from the above analysis we can conclude that: (i) no matter the correlation strengths, at large distances $x_1 \gg R_2$, $p_{\text{trans}}^{\text{c}}(x_1) \approx w$; and (ii) at smaller $x_1$, $p_{\text{trans}}^{\text{c}}(x_1) \geq w$ and has an increasing trend if there are distance correlations (due to the second term in Eq.~(\ref{eq:p_trans_c}))---the stronger the correlations the higher the increase. These conclusions are validated in Fig.~\ref{fig:analysis_validation_3} and are consistent with our empirical observations in Fig.~\ref{fig:empirical_results}.

The trans-layer connection probability among disconnected layer~1 pairs, $p_{\text{trans}}^{\text{d}}(x_1)$, is:
\begin{align}
\label{eq:p_trans_d}
p_{\text{trans}}^{\text{d}}(x_1)= \int_0^{2 R_2}f(x_2|x_1)p_2(x_2)\mathrm{d}x_2.
\end{align}
As with the second term in Eq.~(\ref{eq:p_trans_c}), $p_{\text{trans}}^{\text{d}}(x_1)$ decreases with $x_1$ if there are distance correlations---the stronger the correlations the faster the decrease. In the uncorrelated case: 
\begin{equation}
\label{eq:p_d_uc}
p_{\text{trans}}^{\text{d}}(x_1)=\frac{\bar{k}_2}{N}, 
\end{equation}
while in the maximally correlated case:
\begin{equation}
\label{eq:p_d_mc}
 p_{\text{trans}}^{\text{d}}(x_1)=p_2(x_1).
\end{equation}
This behavior is also validated in Fig.~\ref{fig:analysis_validation_3} and it is consistent with our empirical observations in Fig.~\ref{fig:empirical_results}. Finally, the trans-layer connection probability across all layer~1 pairs can be written as:
\begin{align}
\label{eq:p_all_trans}
 p_{\text{trans}}^{\text{all}}(x_1)&=p_1(x_1) p_{\text{trans}}^{\text{c}}(x_1)+\left(1-p_1(x_1)\right) p_{\text{trans}}^{\text{d}}(x_1).
\end{align}

\subsection{Estimating the link persistence probability $w$ in real multiplexes}
\label{sec:w_infer}

Our analysis shows that at large distances $x_1, x_2 \gg R_2$, $p_{\text{trans}}^{\text{c}}(x_1) \approx w$, $p_2^{\text{c}}(x_2) \approx w$. For each of the layer pairs in Fig.~\ref{fig:empirical_results} we have estimated $w$ as the average of the empirical $p_{\text{trans}}^{\text{c}}(x_1)$ at distances $x_1 > R_2$. The estimates of $w$ are shown by the horizontal dashed lines in Fig.~\ref{fig:empirical_results} and are also reported in Table~\ref{tab:ws} ($w_{12}$). We have chosen $p_{\text{trans}}^{\text{c}}(x_1)$ for the estimation instead of  $p_2^{\text{c}}(x_2)$ since as explained in Sec.~\ref{sec:model}, the former appears less prone to independent layer embedding effects. Table~\ref{tab:ws} also shows the standard deviation of $p_{\text{trans}}^{\text{c}}(x_1)$ at $x_1 > R_2$ ($\sigma_{w_{12}}$), as well as the estimated link persistence probability ($w_{21}$) and standard deviation ($\sigma_{w_{21}}$) in the direction from layer~2 to~1.

\begin{table}[!h]
\begin{tabular}{|c|c|c|c|c|c|}
\hline 
Multiplex & $w_{12}$ & $\sigma_{w_{12}}$ & $w_{21}$ & $\sigma_{w_{21}}$\\ 
\hline 
Internet & 0.38 & 0.08 & 0.71 &  0.11\\ 
\hline 
arXiv & 0.71 & 0.06 & 0.82 & 0.10 \\ 
\hline 
Physicians & 0.475 & 0.16 & 0.58 & 0.17 \\ 
\hline 
Drosophila & 0.27 & 0.12 & 0.31 & 0.11 \\ 
\hline 
C. Elegans & 0.20 & 0.13 & 0.09 & 0.06\\ 
\hline 
Human Brain & 0.35 & 0.18 & 0.42 & 0.13 \\ 
\hline 
\end{tabular}
\caption{Estimated link persistence probability $w_{ij}$ from layer $i$ to $j$, $i, j=1,2$, in the considered multiplexes. The estimated $w_{ij}$ is the average of the empirical $p_{\text{trans}}^{\text{c}}(x_i)$ at distances $x_i > R_j$, while $\sigma_{w_{ij}}$ is the standard deviation of $p_{\text{trans}}^{\text{c}}(x_i)$ at $x_i > R_j$. The empirical $p_{\text{trans}}^{\text{c}}(x_i)$, $i=1,2$, is computed as in Fig.~\ref{fig:empirical_results}.
\label{tab:ws}}
\end{table}

\section{Improving trans-layer link prediction}
\label{sec:link_prediction}

The results of Sec.~\ref{sec:trans_layer} suggest that we can improve \emph{trans-layer link prediction}, i.e., the prediction of whether two nodes are connected in one layer of a multiplex, if we know the hyperbolic distance between the same nodes in another independently embedded layer, and whether the two nodes are connected in that layer. Specifically, from Eqs.~(\ref{eq:p_trans_c}),~(\ref{eq:p_trans_d}) we see that for each pair of nodes $i, j$ in layer~1 we can assign a score:
\begin{align}
\label{eq:score}
s_{ij}=\psi(x_1^{ij})+[1-\psi(x_1^{ij})]w \alpha_{ij,1},
\end{align} 
where $\alpha_{ij,1}$ is the adjacency matrix of layer~1, while $0 \leq \psi(x_1^{ij}) \leq 1$ is a decreasing function of the hyperbolic distance between $i, j$ in layer~1, $x_1^{ij}$. The higher the $s_{ij}$ the higher is the likelihood that $i$ and $j$ are connected in layer~2.

Eq.~\eqref{eq:score} combines two mechanisms for link prediction. The first uses the fact that hyperbolically closer nodes in layer~1 have higher chances of being connected in layer~2. Eqs.~(\ref{eq:p_trans_c}),~(\ref{eq:p_trans_d}) suggest that $\psi(x_1)$ is given by Eq.~(\ref{eq:p_trans_d}), which requires knowledge of the conditional density $f(x_2|x_1)$ and the connection probability $p_2(x_2)$. However, since in link prediction it is only the relative score among pairs that matters, we can use any monotonously decreasing function $\psi(x_1)$ in Eq.~\eqref{eq:score}, which does not require the aforementioned knowledge. The second mechanism takes link persistence into account. In isolation, it predicts the existence of a link between two nodes in layer~2  with probability $w$ if the nodes are connected in layer~1, and with probability $0$ otherwise. 

The balance between the two mechanisms depends on the value of $w$ and the choice of $\psi(x_1)$. Below, we consider the function:
\begin{align} 
\label{eq:psi_x_1}
\psi(x_1) = e^{-x_1}.
\end{align}
From Eqs.~(\ref{eq:score}),~(\ref{eq:psi_x_1}) the score for the set of connected layer~1 pairs is given by $w+(1-w)e^{-x_1}$, while for the disconnected pairs by $e^{-x_1}$. For each set of pairs the score decreases with the hyperbolic distance $x_1$. At $x_1 \geq \ln{(1/w)}$ the score among disconnected pairs is always smaller than the score among connected pairs. At smaller $x_1$, disconnected pairs can have a higher score than connected pairs at larger $x_1$.
Therefore, $w$ in Eq.~(\ref{eq:score}) can be seen as a ``mixing" parameter than mixes the ordering of the scores of  the two sets of pairs. At $w=0$ link persistence is ignored, while at $w=1$  all connected pairs are assigned the maximum score $1$. The case where distance correlations are ignored is equivalent to choosing $\psi(x_1) = C$, where $C$ is a constant, $0 \leq C < 1$. 

\begin{figure}[!h]
\includegraphics[width=3.45in]{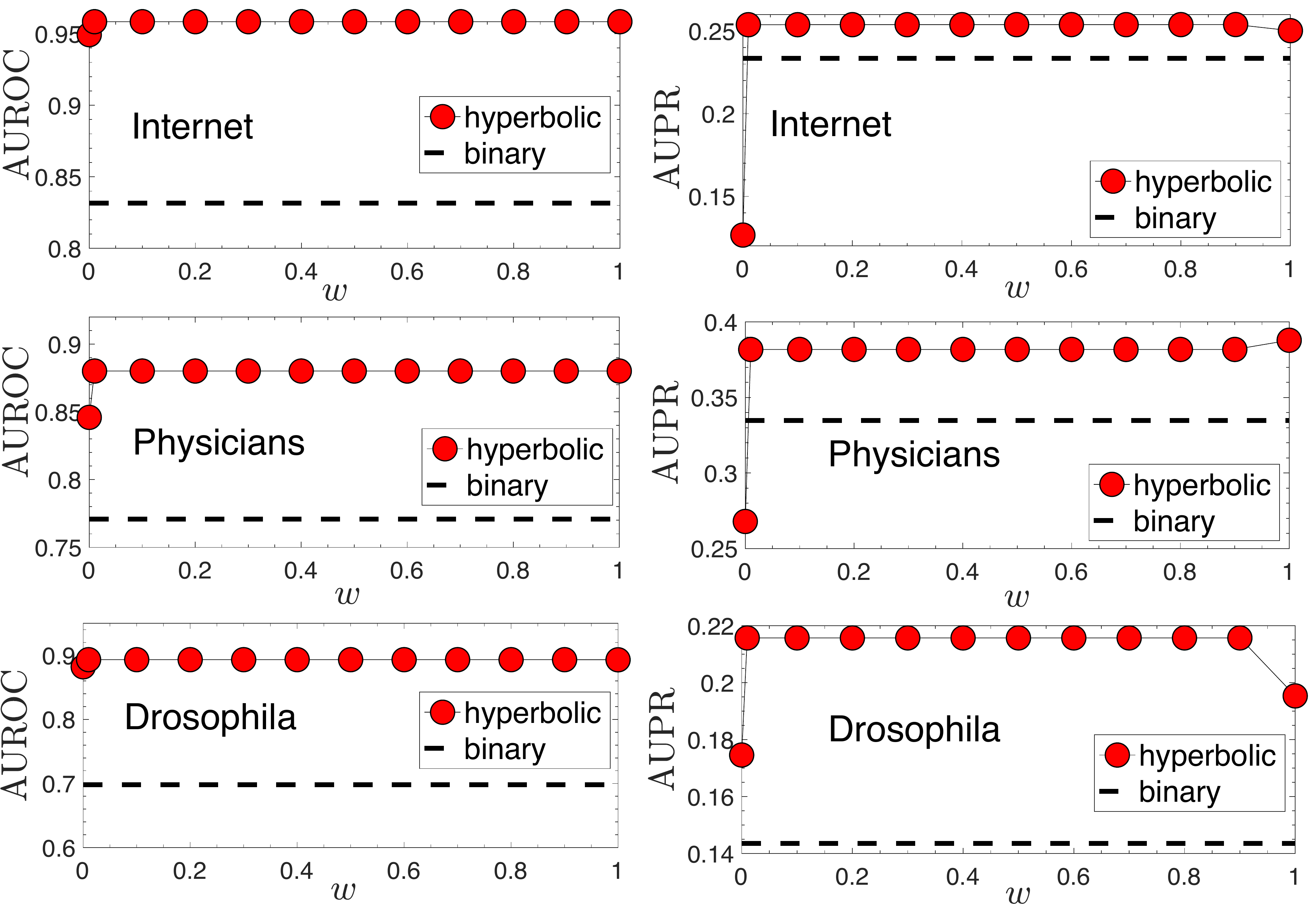}
\includegraphics[width=3.45in]{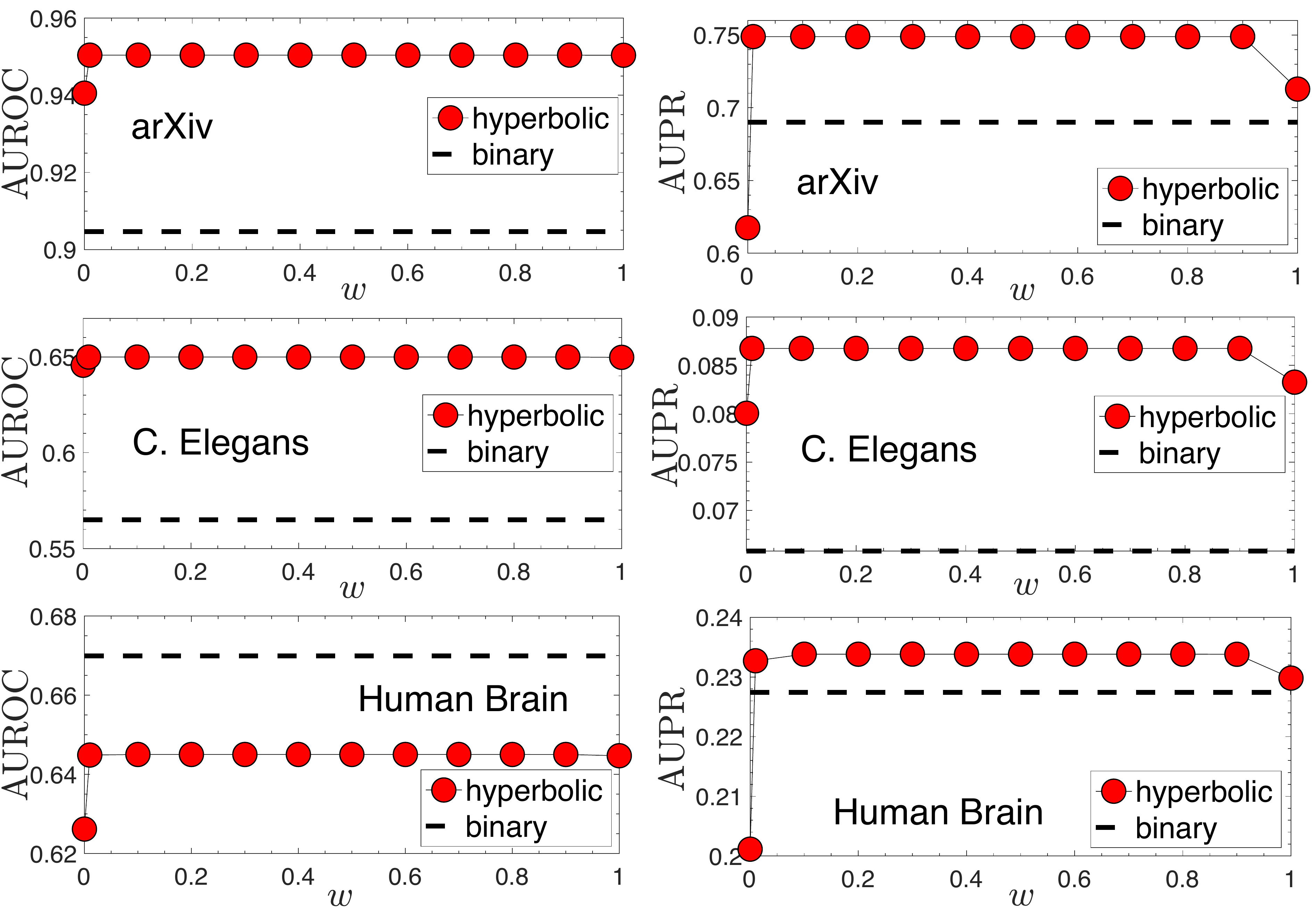}
\caption{Trans-layer link prediction (layer~1 to layer~2) in the multiplexes of Fig.~\ref{fig:empirical_results}.  \textbf{Left column:} Area Under the Receiver Operating Characteristic Curve (AUROC). \textbf{Right column:}  Area Under the Precision-Recall Curve (AUPR). 
\label{fig:trans_layer_link_prediction}}
\end{figure}

In Fig.~\ref{fig:trans_layer_link_prediction} we quantify the quality of trans-layer link prediction in the real multiplexes of Fig.~\ref{fig:empirical_results}. To this end, we use the \emph{Area Under the Receiver Operating Characteristic Curve} (AUROC) and the \emph{Area Under the Precision-Recall Curve} (AUPR)~\cite{precrec}, computed over the set of common nodes in the two layers. The AUROC represents the probability that a randomly selected link from the set of links among the common nodes in layer $2$ is given a higher score than a randomly selected nonexistent link, where the ``non-existent" links are the disconnected common node pairs in layer $2$. The degree to which the AUROC exceeds $0.5$ indicates how much better the method performs than pure chance, while $\textnormal{AUROC}=1$ is the best possible AUROC. The AUPR represents how accurately one can classify layer~2 pairs as connected and disconnected based on their scores. It is a standard metric used when classes are imbalanced, i.e., when the number of negatives (disconnected pairs) is significantly larger than the number of positives (connected pairs), as in our case. The higher the AUPR the better the model is, while a perfect classifier has $\textnormal{AUPR}=1$. See~\cite{precrec} for further details. 

Fig.~\ref{fig:trans_layer_link_prediction} shows the AUROC and AUPR as a function of $w$ if we use the scores prescribed by Eqs.~(\ref{eq:score}),~(\ref{eq:psi_x_1}) (hyperbolic). The same figure also shows the results if instead of Eq.~(\ref{eq:psi_x_1}) we use $\psi(x_1)=0$, which results in a simple binary link predictor, where if a link exists between two common nodes in layer~1, we predict that this link will also exist in layer~2 (binary). We see in Fig.~\ref{fig:trans_layer_link_prediction} that by taking both hyperbolic distance correlations and link persistence into account we can improve trans-layer link prediction, especially with respect to AUPR. Specifically, we see that for $w > 0$ the hyperbolic AUPR improves and it is higher than the binary AUPR in all cases; in Drosophila and C. Elegans it is higher even at $w=0$. The hyperbolic AUROC is significantly higher than the binary AUROC in all cases (even at $w=0$), except from the Human Brain, and it improves at $w > 0$. 

We also see that in all cases the performance of prediction is virtually the same for the considered values of $w \in (0, 1)$. This is expected since as explained, for $x_1 \geq \ln{(1/w)}$ connected layer~1 pairs are always ranked higher (have a better score) than disconnected pairs.  As the majority of disconnected pairs are separated by large $x_1$ distances, this condition is expected to hold even for small values of $w > 0$ (e.g., $w=0.01$). Furthermore, $w$ does not affect the relative ranking within each set of pairs. In other words, the considered values of $w \in (0, 1)$ are not  expected to significantly affect the overall ranking across node pairs, which justifies the observed behavior. As also explained, when $w=1$ we ignore the distances among connected layer~1 pairs, which are all assigned the same score $1$. We see that in this case the AUPR decreases in all cases, except from the Physicians. In general, one would require learning the optimal value(s) of $w$ for a specific option of $\psi(x_1)$ in Eq.~(\ref{eq:psi_x_1}) using a training set of layer~2 links. Exploring other options for $\psi(x_1)$ is beyond the scope of this paper.

\section{Discussion and Conclusion}
\label{sec:conclusion} 

We have seen that GMM-LP can reproduce the link-persistence behavior observed in real multiplexes remarkably well. In addition to geometric correlations, in GMM-LP the explicit link formation process in layer~1 impacts the formation of links in layer~2. As a result, the connection probability in layer~2 differs from the one in the $\mathbb{H}^2$ model (Eq.~(\ref{eq:p_all}) vs. Eq.~(\ref{eq:c_prob_2})). Specifically, in GMM-LP one can view link persistence as ``noise" imposed to the tail of the connection probability of the $\mathbb{H}^2$ model, allowing connections at larger distances with higher probability. The stronger the distance correlations among the layers the weaker the effect of link persistence
is, i.e., the lower the probability for longer distance connections (Figs.~\ref{fig:analysis_validation_1}(a),(b)). In this sense, link persistence has a similar effect as temperature in the $\mathbb{H}^{2}$ model, and does not significantly affect the main topological properties of layer~2, which are similar to those in the $\mathbb{H}^2$ model (Figs.~\ref{fig:analysis_validation_1}(c),(d)). An outstanding question is whether there exists a model that can capture the link-persistence behavior of real systems while preserving the connection probability of the $\mathbb{H}^2$ model in the layers. 

We have also seen that link persistence can explain the high edge overlap in real systems, which cannot be explained by coordinate correlations alone (Fig.~\ref{fig:overlap}). Specifically, to achieve the same overlap in synthetic systems without link persistence, we need significantly higher correlation strengths than those in Table~\ref{tab_datasets}. Furthermore, in some cases we cannot achieve the same overlap even with maximal correlations. We note that this result does not contradict the work in~\cite{Halu}, which shows that high edge overlap naturally emerges among layers embedded in Euclidean spaces, where nodes have the same coordinates in each layer. In fact, our results also suggest that high overlap without link persistence can emerge in cases where the node coordinates are very strongly correlated across layers and the layers have sufficiently low temperatures. However, this is not the case with the considered real systems~(Fig.~\ref{fig:overlap}). In addition, we have seen that by taking both link persistence and hyperbolic distance correlations into account we can improve trans-layer link prediction (Fig.~\ref{fig:trans_layer_link_prediction}). 

All considered models ($\mathbb{H}^2$, GMM, GMM-LP) use for simplicity a uniform distribution for the angular similarity coordinates. This leads to generated topologies without community structure~\cite{Zuev2015, npso, npso2}. On the other hand, the considered real layers (Table~\ref{tab_datasets}) exhibit community structure and trans-layer community correlations, which are manifested in their embeddings as groups of nodes that are similar---close along the angular similarity direction---in both layers simultaneously~\cite{kleineberg2016np, saeed_prl}. It would be interesting to modify the assignment of angular coordinates in the GMM-LP along the lines of~\cite{npso} so that the model can also generate synthetic layers with community correlations. Community correlations are expected to promote the overlap among intra-community links, without however explaining the observed link persistence that occurs irrespectively of the hyperbolic distances that the connections span (cf. Fig.~\ref{fig:empirical_results}).  

The work in~\cite{link_persistence_paper} considers link persistence (also called stability) in dynamic networks, in conjunction with node hidden variables (or fitnesses) that determine the nodes' capability of forming links. For each network snapshot in~\cite{link_persistence_paper} both connections and disconnections can be copied from the previous snapshot with a certain probability, or formed with a probability that depends on the nodes' fitnesses. Similarly to our work, this work also attempts to disentangle the importance of the two mechanisms (link persistence vs. node hidden variables) in link formation. However, differently from our work, it does not consider multiplex networks, nor networks embedded into hyperbolic spaces, i.e., networks where the node hidden variables are their coordinates in their underlying hyperbolic space. Furthermore, it does not analyze the effect of link persistence to the resulting topological properties of the network. Finally, in~\cite{link_persistence_paper} both connections (links) and disconnections (non-links) can persist (copied) from one snapshot to another with possibly different probabilities. In GMM-LP only connections can persist from one layer to another, with the same probability $w$. It appears that this simple mechanism is sufficient for reproducing the link-persistence behavior observed in real systems. In future work, it would be interesting to investigate the reasons behind the variability of $w$ across different real systems and layers (cf. Table~\ref{tab:ws}).

Our results guide the development of multiplex embedding methods, in which the layers of a multiplex are simultaneously and not independently embedded into hyperbolic spaces, suggesting that such methods should be accounting for both coordinate correlations and link persistence across layers. An important aspect of multiplex embeddings is that they can improve link prediction in the individual layers compared to the case where the layers are embedded independently, cf.~\cite{mell}. Further, they could potentially lead to improved multidimensional community detection on a geometric basis and yield more realistic multilayer greedy routing success rates~\cite{kleineberg2016np}, as they are expected to better capture the relation between the layers. To infer the hyperbolic node coordinates $\{r_{2, i},\theta_{2, i}\}$ for each node $i=1, \dots, N$ in layer~2 of GMM-LP, along with the correlation strengths $\nu, g$ and the link persistence probability $w$, one needs to maximize a conditional likelihood of the form:
\begin{align}	
\label{eq:penalized_likelihood}
\nonumber \mathcal{L_\textnormal{cond}} &\propto \text{Prob}(\{r_{2,i},\theta_{2,i}\} | \{r_{1,i},\theta_{1,i}\}, \nu, g)\\
&\times \mathcal{L}(\alpha_{ij, 2} | \{r_{i, 2},\theta_{i, 2}\},\alpha_{ij, 1}, w).
\end{align}
The first term in the above relation is the PDF of the node coordinates in layer~2 conditioned on their values in layer~1 and the correlation strengths:
\begin{align}
\nonumber \text{Prob}(\{r_{2, i},\theta_{2,i}\} | \{r_{1,i},\theta_{1, i}\}, \nu, g) =& \prod_{1\leq i \leq N}  \rho_2(r_{2, i} | r_{1,i}, \nu)\\
\nonumber \times& \prod_{1 \leq i \leq N} f(\theta_{2,i} | \theta_{1, i}, g), 
\end{align}
where $\rho_2(r_{2, i} | r_{1,i}, \nu)$ in Eq.~(\ref{eq:r_cond_text}) and:
\begin{equation}
\nonumber f(\theta_{2,i} | \theta_{1, i}, g) = \frac{N}{2 \pi} f_g\left(\frac{N}{2\pi}(\pi-|\pi-|\theta_{2, i}-\theta_{1, i}||)\right),
\end{equation}
where $f_g(l)$ in Eq.~(\ref{eq:truncated_normal}). The second term in Eq.~(\ref{eq:penalized_likelihood}), $\mathcal{L}(\alpha_{ij, 2} | \{r_{i, 2},\theta_{i, 2}\},\alpha_{ij, 1}, w) \equiv \mathcal{L}_2$, is the likelihood to have the network adjacency matrix $\alpha_{ij, 2}$ in layer~2 if the node coordinates have the values $\{r_{i, 2},\theta_{i, 2}\}$, the adjacency matrix of layer~1 is $\alpha_{ij, 1}$, and the link persistence probability is $w$:
\begin{align}
\label{eq:Lml}
\nonumber \mathcal{L}_2 &= \prod_{1 \leq j < i \leq N} [w \alpha_{ij, 1}\alpha_{ij, 2}+(1-w\alpha_{ij, 1})p_2(x_2^{ij})^{\alpha_{ij, 2}}\\
&\times (1-p_2(x_2^{ij}))^{1-\alpha_{ij, 2}}].
\end{align}
The above product goes over all node pairs in layer~2, while $p_2(x_2^{ij})$ is given in Eq.~(\ref{eq:c_prob_2}). An initial estimate of $w$ can be obtained using the procedure in Sec.~\ref{sec:w_infer}. We leave this maximization problem open for future work.

\begin{acknowledgments}
F. P. acknowledges support by the EU H2020 NOTRE project (grant 692058).
K.-K. K. acknowledges support by the EU H2020 Program under the funding scheme FET-PROACT-1-2014: Global Systems Science (GSS), grant agreement 641191 CIMPLEX: Bringing CItizens, Models and Data together in Participatory, Interactive SociaL EXploratories.
\end{acknowledgments}

\appendix
\section{Real-world multiplex network data}
\label{sec:real_data}
Here we provide details on the considered real-world multiplex network data. For further details see~\cite{kleineberg2016np}.

\textbf{IPv4/IPv6 Internet.} The IPv4 and IPv6 Autonomous Systems (AS) Internet topologies were extracted from the data collected by CAIDA~\cite{ark2009, as_topo}. The connections in each topology are not physical but logical, representing AS relationships. The IPv4 dataset consists of ASs that can route Internet packets with IPv4 destination addresses, while the IPv6 dataset consists of ASs that can route packets with IPv6 destination addresses. The considered topologies correspond to January 2015. The IPv4 topology (layer~1) consists of $N_1=37563$ nodes (ASs), and has a power law degree distribution with exponent $\gamma_1=2.1$, average node degree $\bar{k}_1=5.06$, and average clustering
$\bar{c}_1=0.63$ ($T_1=0.5$).\footnote{The average clustering is calculated excluding nodes of degree $1$.} The IPv6 topology (layer~2) consists of $N_2=5162$
nodes, has a power law degree distribution with exponent $\gamma_2=2.1$, average node degree $\bar{k}_2=5.21$, and average clustering $\bar{c}_2=0.55$ ($T_2=0.5$). There are $4819$ common nodes in the two topologies, i.e., ASs that can route both IPv4 and IPv6 packets. 

\textbf{Drosophila Melanogaster.} The Drosophila Melanogaster dataset is taken from~\cite{biogrid, manlio2015}. In this dataset, the networks represent protein--protein interactions and the layers correspond to interactions of different nature. Layer 1 corresponds to suppressive genetic interaction, while layer 2 corresponds to additive genetic interaction. Layer~1 has 
$N_1=839$ nodes, average degree $\bar{k}_1=4.43$, and average clustering $\bar{c}_1=0.28$ ($T_1=0.68$). Its degree distribution can be approximated by a power law with exponent $\gamma_1=2.6$. Layer~2 has $N_2=755$ nodes, average degree $\bar{k}_2=3.77$, and average clustering $\bar{c}_2=0.29$ ($T_2=0.65$). Its degree distribution can be approximated by a power law with exponent $\gamma_2=2.8$. There are $557$ common nodes in the two layers. 

\textbf{C. Elegans Connectome.} The C. Elegans dataset is taken from~\cite{pnas:celegans, muxviz}. It corresponds to the neuronal network of the nematode Caenorhabditis Elegans. The nodes are neurons and each layer corresponds to a different type of synaptic connection: Electric (layer~1) and Chemical Monadic (layer~2). Layer 1 has $N_1=253$ nodes, average degree $\bar{k}_1=4.06$, and average clustering $\bar{c}_1=0.24$ ($T_1=0.65$). Layer 2 has $N_2=260$ nodes, average degree $\bar{k}_2=6.83$, and average clustering $\bar{c}_2=0.21$ ($T_2=0.7$).  The degree distribution in both layers can be approximated by a power law with exponent $\gamma_1=\gamma_2=2.9$. There are $238$ common nodes in the two layers. 

\textbf{Human Brain.} The human brain data is taken from~\cite{Simas2015}. The data consists of a structural (anatomical) network, as well as a functional network. In both networks, nodes are brain regions. The structural network (layer~1) consists of $85$ nodes, with average degree $\bar{k}_1=5.41$, maximum degree $k_1^{\textnormal{max}}=12$, and average clustering $\bar{c}_1=0.49$ ($T_1=0.4$). The functional network (layer~2) has 
$80$ nodes, average degree $\bar{k}_2=5.48$, maximum degree $\bar{k}_2^{\textnormal{max}}=14$, and average clustering $\bar{c}_2=0.40$ ($T_2=0.5$). The two layers have $77$ nodes in common and $\gamma_1=\gamma_2=6$. 

\textbf{arXiv.} The arXiv data is taken from~\cite{prx:modular} and contains co-authorship networks. The nodes are authors that are connected if they have co-authored a paper. In arXiv, each paper is assigned to one or more relevant categories. The data considers only papers with the word ``networks" in the title or abstract from different categories up to May 2014. In the considered data, layer 1 corresponds to the network formed by the authors of papers in the `Disordered Systems and Neural Networks" (cond-mat.dis-nn) category and layer~2 corresponds to ``Biological Physics" (physics.bio-ph). The corresponding size, average degree, average clustering, and power law exponent for each layer are $N_1=3506, N_2=2956$, $\bar{k}_1=4.19, \bar{k}_2=4.13$, $\bar{c}_1=0.81, \bar{c}_2=0.83~(T_1=T_2=0.05)$, and $\gamma_1=\gamma_2=2.6$. There are $1514$ common nodes in the two layers. 
 
\textbf{Physicians.} The Physicians dataset is taken from~\cite{physicians:data}. The network layers correspond to different types of relations among physicians in four US towns. In the considered data, layers 1, 2 correspond respectively to discussion and advice relations among the physicians. The corresponding size, average degree, average clustering, and power law exponent for each layer are $N_1=231, N_2=215$, $\bar{k}_1=4.31, \bar{k}_2=4.18$, $\bar{c}_1=\bar{c}_2=0.28~(T_1=T_2=0.65)$, and $\gamma_1=2.8, \gamma_2=2.7$. There are $212$ common nodes in the two layers. 

In all hyperbolic embeddings of both real and synthetic layers, the angular node coordinates are obtained using HyperMap~\cite{frag:hypermap_cn, hypermap_code}. The radial coordinate $r_i$ of each node $i$ is obtained from Eq.~(\ref{eq:bar_k_r_H2}) after setting $\bar{k}(r_i)=\kappa_i \equiv \max\{\bar{k}_0, k_i-\gamma T\}$, where $k_i$ is the observed degree of the node in the layer~\cite{Boguna2010}. In other words, we use the inferred radial coordinates prescribed by the static $\mathbb{H}^{2}$ model~\cite{Boguna2010} that we work with, instead of the ones obtained by HyperMap that are akin to the growing popularity$\times$similarity model~\cite{Papadopoulos2012}.

\section{Conditional PDF of radial coordinates}
\label{sec:conditional_radial_pdf}

Here we derive Eq.~(\ref{eq:r_cond_text}). We note that the GMM~\cite{kleineberg2016np} uses the $\mathbb{S}^{1}$ model~\cite{Serrano2008}, where instead of a radial coordinate $r_i$ each node $i$ has a hidden degree variable $\kappa_i$. The degree variables are then transformed to radial coordinates in the $\mathbb{H}^{2}$ model, which is isomorphic to the $\mathbb{S}^{1}$, via Eq.~(\ref{eq:bar_k_r_H2}) after setting $\bar{k}(r_i)=\kappa_i$. Here we work directly with radial coordinates. 

By integrating the PDFs of the radial coordinates in layers~1 and~2, $\rho_1(r_1), \rho_2(r_2)$ (Eqs.~(\ref{eq:eq_rho_1}),~(\ref{eq:eq_rho_2})), we get the corresponding CDFs:
\begin{align}
F_1(r_1) \approx e^{-\phi_1}, \phi_1 \equiv \frac{R_1-r_1}{2\beta_1},\\
F_2(r_2) \approx e^{-\phi_2}, \phi_2 \equiv \frac{R_2-r_2}{2\beta_2}.
\end{align}
To derive the joint PDF of the radial coordinates we use the bivariate Gumbel-Hougaard copula as in~\cite{kleineberg2016np}, defined as:
\begin{align}
\label{eq:copula}
\nonumber C_\eta(r_1, r_2) &=  e^{-[(-\ln{F_1(r_1)})^\eta+(-\ln{F_2(r_2)})^\eta]^\frac{1}{\eta}}\\
&= e^{-(\phi_1^\eta+\phi_2^\eta)^\frac{1}{\eta}}, \eta \equiv \frac{1}{1-\nu} \in [1, \infty),
\end{align}
where $\nu \in [0, 1)$ is the radial correlation strength parameter. The joint PDF of $r_1$ and $r_2$, $\rho_\eta(r_1, r_2)$, can be obtained by differentiating $C_\eta(r_1, r_2)$ with respect to $r_1$ and $r_2$. The conditional PDF in Eq.~(\ref{eq:r_cond_text}) is obtained as $\rho_{2} (r_2 | r_1, \eta)=\rho_{\eta} (r_1, r_2)/\rho_1(r_1)$.

\section{$\widetilde{\bar{k}}_2(r)$ in the uncorrelated and maximally correlated cases}
\label{sec:bar_k_corrs}

The last approximation for the hyperbolic distance in Eq.~(\ref{eq:hyperbolic_distance}) allows us to write the connection probability in layer~$i=1, 2$ as:
\begin{align}
p_i(r,r',\Delta \theta) \approx \frac{1}{1+\left(\frac{\Delta \theta}{2} e^{\frac{1}{2}(r+r'-R_i)}\right)^{1/T_i}}.
\end{align}
Using the fact that $\int_{0}^\infty 1/(1+\chi^\frac{1}{T}) \mathrm{d}\chi = T \pi/\sin{T \pi}$ for $T < 1$, and that the angular distance $\Delta \theta$ is uniformly distributed on $[0, \pi]$, we can write:
\begin{align}
\label{eq:fact_0}
\frac{1}{\pi} \int_{0}^{\pi} p_i (r, r', \Delta\theta)  \mathrm{d}\Delta \theta \approx \frac{2T_i}{\sin{T_i \pi}} e^{-\frac{1}{2}(r+r'-R_i)}.
\end{align}
Using Eqs.~(\ref{eq:fact_0}) and (\ref{eq:eq_rho_2}) yields:
\begin{align}
\label{eq:fact_1}
\frac{N}{\pi} \int_{0}^{R_2} \rho_2(r') \mathrm{d}r' \int_{0}^{\pi} p_2 (r, r', \Delta\theta)  \mathrm{d}\Delta \theta \approx \bar{k}_2(r),\\
\label{eq:fact_2}
\frac{N}{\pi} \int_{0}^{R_2} \rho_2(r') \mathrm{d}r' \int_{0}^{\pi} p_1 (r, r', \Delta\theta)  \mathrm{d}\Delta \theta \approx C \bar{k}_2(r),
\end{align}
where $\bar{k}_2(r)$ in Eq.~(\ref{eq:bar_k_2_r_H2_model}), $C=c_2 T_1\sin{T_2\pi}/(c_1T_2\sin{T_1\pi})$ and $c_i, i=1,2$ in Eq.~(\ref{eq:not}). If the radial coordinates are identical in the two layers, $R_1=R_2$, and $c_1=c_2$. 

Eq.~(\ref{eq:bar_k_2_r_no_corr}) follows from Eqs.~(\ref{eq:eta_u}),~(\ref{eq:p_all}),~(\ref{eq:bar_k_2_r}) and (\ref{eq:fact_1}). 
To derive Eq.~(\ref{eq:bar_k_2_r_max_corr_bounds}) we also use the additional fact that $\int_{0}^\infty [1/(1+\chi^\frac{1}{T})]^2 \mathrm{d}\chi = (1-T)T \pi/\sin{T \pi}$ for $T < 1$, which allows us to write:
\begin{align} 
\label{eq:fact_3}
\nonumber \frac{N}{\pi} \int_{0}^{R_2} \rho_2(r') \mathrm{d}r' \int_{0}^{\pi} p_2 (r, r', \Delta\theta)^2  \mathrm{d}\Delta \theta\\
\approx (1-T_2)\bar{k}_2(r),\\
\label{eq:fact_4}
\nonumber \frac{N}{\pi} \int_{0}^{R_2} \rho_2(r') \mathrm{d}r' \int_{0}^{\pi} p_1 (r, r', \Delta\theta)^2  \mathrm{d}\Delta \theta\\
\approx (1-T_1) C \bar{k}_2(r).
\end{align}
Using the above two relations and the Cauchy-Schwarz inequality we have:
\begin{align}
\label{eq:fact_5}
\nonumber \frac{N}{\pi} \int_{0}^{R_2} \rho_2(r') \mathrm{d}r' \int_{0}^{\pi} p_1 (r, r', \Delta\theta) p_2 (r, r', \Delta\theta) \mathrm{d}\Delta \theta\\
\leq \bar{k}_2(r) \sqrt{C (1-T_1) (1-T_2)}.
\end{align}
Eq.~(\ref{eq:bar_k_2_r_max_corr_bounds}) follows from Eqs.~(\ref{eq:eta_mc}),~(\ref{eq:p_all}),~(\ref{eq:bar_k_2_r}),~(\ref{eq:fact_1}),~(\ref{eq:fact_2}) and (\ref{eq:fact_5}). If in addition $T_1=T_2$, $p_1 (r, r', \Delta\theta)=p_2 (r, r', \Delta\theta)$, $C=1$, and Eq.~(\ref{eq:bar_k_2_r_max_corr}) follows from Eqs.~(\ref{eq:eta_mc}),~(\ref{eq:p_all}),~(\ref{eq:bar_k_2_r}),~(\ref{eq:fact_1}),~(\ref{eq:fact_2}) and (\ref{eq:fact_3}). 

\section{Conditional hyperbolic distance PDF}
\label{sec:conditional_hyperbolic_pdf}

Finally, in this section we show how to compute $f(x_2|r_1, r_1', \Delta \theta_1)$, which is the PDF of the hyperbolic distance $x_2$ between two nodes in layer~2 conditioned on the nodes' radial coordinates $r_1, r_1'$ and angular distance $\Delta \theta_1$ in layer~1. This conditional PDF does not have an analytic expression and we show here how to compute it using numerical integration. To this end, we first need to derive $P(\Delta \theta_2 \leq \Delta \theta | \Delta \theta_1)$, which is the CDF of the angular distance $\Delta \theta_2$ of a pair in layer 2 conditioned on its angular distance $\Delta \theta_1$ in layer 1. This CDF admits an analytic expression. As we explain (Appendix~\ref{sec:f_x_2_x_1}), the conditional PDF $f(x_2|x_1)$, where $x_1$ is the hyperbolic distance between the pair in layer 1, can be obtained by integrating $f(x_2|r_1, r_1', \Delta \theta_1)$ over $r_1, r_1'$.

\subsection{Conditional CDF $P(\Delta \theta_2 \leq \Delta \theta | \Delta \theta_1)$}

Let $\Delta \theta_1$ and $\Delta \theta_2$ be random variables denoting respectively the angular distance between the same pair of nodes in layers 1 and 2, whose angles are $\theta_1, \theta_1'$ (layer~1) and $\theta_2, \theta_2'$ (layer~2).  From Eq.~(\ref{eq:assign_theta_text}) and the fact that $\Delta \theta_2=\pi-|\pi-|\theta_2-\theta_2'||$, we can see that $\Delta \theta_2$ is obtained by first moving the points at $\theta_1$ and $\theta_1'$ on the circle by $2 \pi l/N$ and $2 \pi l'/N$, respectively, and then computing the angular distance between the new points. Equivalently, we can view $\Delta \theta_2$ as being obtained by first computing the angular distance between the points at $\theta_1$ and $\theta_1'$, $\Delta \theta_1=\pi-|\pi-|\theta_1-\theta_1'||$, and then adding to this distance the term $2 \pi \tilde{l}/N$, where $\tilde{l} = l-l'$. The PDF of $\tilde{l}$ can be obtained from the PDFs of $l, l'$ (Eq.~(\ref{eq:truncated_normal})):
\begin{align}
\label{eq:tilde_pdf}
\nonumber \tilde{f}_g(\tilde{l})=& \frac{\mathrm{d}}{\mathrm{d}\tilde{l}} \left(\int_{-\frac{N}{2}}^{\frac{N}{2}} \int_{-\frac{N}{2}}^{\tilde{l}+l'} f_g(l) f_g(l')  \mathrm{d}l \mathrm{d}l'\right)\\
\nonumber =& \frac{\phi\left(\frac{\tilde{l}}{\sqrt{2}\sigma}\right) }{2 \sqrt{2} \sigma}\left[\frac{\erf(\frac{N-\tilde{l}}{2\sigma})+\erf(\frac{N+\tilde{l}}{2\sigma})}{\erf(\frac{N}{ \sqrt{2} \sigma})^2}\right]\\
\approx& \frac{\phi\left(\frac{\tilde{l}}{\sqrt{2}\sigma}\right) }{\sqrt{2} \sigma \erf(\frac{N}{4\sigma})}.
\end{align} 
In other words, the PDF of $\tilde{l} \in [-N/2, N/2]$ is approximately the same as Eq.~(\ref{eq:truncated_normal}) except that its variance is $2 \sigma^2$. Therefore, we can write:
\begin{align}
\label{eq:dtheta_dis}
\Delta\theta_2 \overset{d}{=} \pi-|\pi-\textnormal{mod}\left[\Delta \theta_1+ \frac{2 \pi \tilde{l}}{N}, 2\pi\right]|,
\end{align} 
where $\tilde{l}$ is sampled from Eq.~(\ref{eq:tilde_pdf}) and the symbol $\overset{d}{=}$ means \emph{equal in distribution}.

\begin{figure}[!h]
\centering{
\includegraphics[width=3.0in]{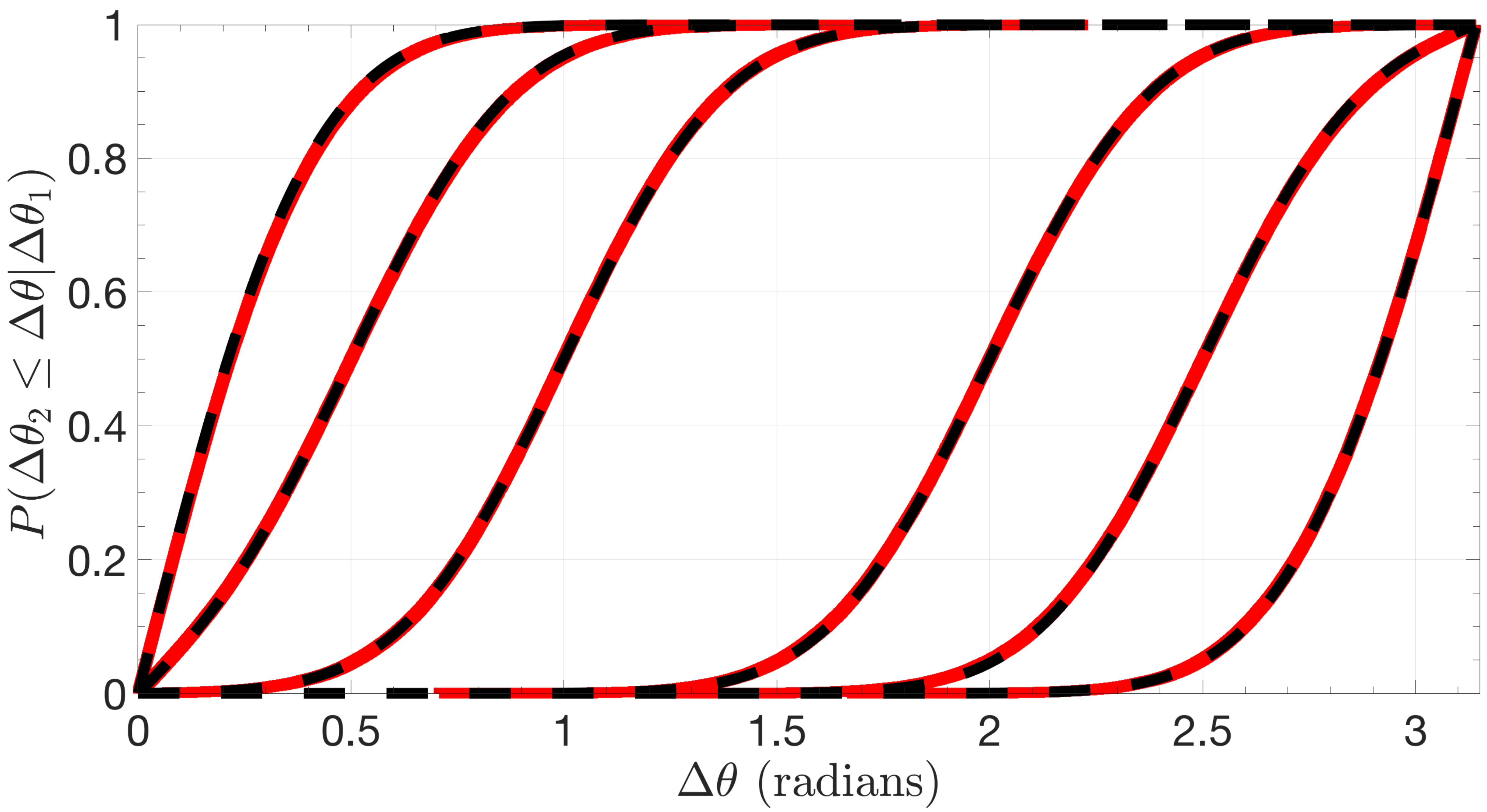}
\caption{Conditional angular distance CDF.  The results are from a two layer synthetic multiplex with $N_1=5000$ and $N_2=3000$ nodes, and angular correlation strength $g=0.5$. All nodes in the smaller layer also exist in the larger. The conditional CDFs from left to right correspond to $\Delta \theta_1=0.1, 0.5, 1.0, 2.0, 2.5, 3.0$ radians. The red solid lines are the empirical distributions while the dashed black lines are the corresponding theoretical predictions given by Eqs.~(\ref{eq:cond_angular_cdf_1}),~(\ref{eq:cond_angular_cdf_2}) with $N=N_2$.
\label{fig:conditional_angular_cdf}}}
\end{figure}

Eq.~(\ref{eq:dtheta_dis}) suggests that for a given $\Delta \theta_1$, $\Delta\theta_2$ is Gaussian with mean $\Delta \theta_1$ and variance $\tilde{\sigma}^2 = (2 \pi/N)^2 2\sigma^2$. In fact, it is a \emph{folded Gaussian}~\cite{folded_gaussian} since probability mass is ``folded" at $0$. Specifically, when $\Delta \theta_1+2 \pi \tilde{l}/N \in [-\pi, 0)$, $\Delta\theta_2=|\Delta \theta_1+2 \pi \tilde{l}/N| \in (0, \pi]$. Further, probability mass is also folded at $\pi$, since when $\Delta \theta_1+2 \pi \tilde{l}/N \in (\pi, 2\pi]$, $\Delta\theta_2=2 \pi - \Delta \theta_1-2 \pi \tilde{l}/N \in [0, \pi)$. Using the PDF of a folded Gaussian~\cite{folded_gaussian} we can write:
\begin{align}
\label{eq:cond_angular_pdf}
f(\Delta \theta_2 | \Delta \theta_1)=&\frac{e^{-\frac{(\Delta \theta_2-\Delta \theta_1)^2}{2\tilde{\sigma}^2}}+e^{-\frac{(\pi-|\pi-\Delta \theta_2-\Delta \theta_1|)^2}{2\tilde{\sigma}^2}}}{K \sqrt{2\pi}\tilde{\sigma}},\\
\nonumber \tilde{\sigma} =& \frac{2 \sqrt{2} \pi \sigma}{N}, K= \erf(\frac{N}{4\sigma}),
\end{align}
where $K$ is the normalizing constant such that $\int_{0}^{\pi} f(\Delta \theta_2 | \Delta \theta_1) \mathrm{d} \Delta \theta_2=1, \forall  \Delta \theta_1 \in [0, \pi]$. The second term in the numerator of Eq.~(\ref{eq:cond_angular_pdf}) accounts for the folding at $0$ when $-\pi \leq \Delta \theta_1-\Delta \theta_2 < 0$ and at $\pi$ when $ \pi < \Delta \theta_1+\Delta\theta_2 \leq 2\pi$. By integrating Eq.~(\ref{eq:cond_angular_pdf}) we get the CDF $P(\Delta \theta_2 \leq \Delta \theta | \Delta \theta_1)$:
\begin{align}
\label{eq:cond_angular_cdf_1}
\nonumber P(\Delta \theta_2 \leq \Delta \theta | \Delta \theta_1) &=\frac{\erf\left(\frac{N(\Delta\theta-\Delta\theta_1)}{4\pi \sigma}\right)+\erf\left(\frac{N(\Delta \theta+\Delta\theta_1)}{4\pi \sigma}\right)}{2\erf(\frac{N}{4 \sigma})}\\
&\equiv  P_1^g(\Delta \theta | \Delta\theta_1),
\end{align}
if $\Delta \theta \leq \pi - \Delta \theta_1$, or, otherwise,
\begin{align}
\label{eq:cond_angular_cdf_2}
\nonumber P(\Delta \theta_2 \leq \Delta \theta | \Delta \theta_1) =1-\frac{\erf\left(\frac{N(2\pi-\Delta\theta-\Delta\theta_1)}{4\pi \sigma}\right)}{2\erf(\frac{N}{4 \sigma})}\\
+\frac{\erf\left(\frac{N(\Delta \theta-\Delta\theta_1)}{4\pi \sigma}\right)}{2\erf(\frac{N}{4 \sigma})} \equiv P_2^g(\Delta \theta | \Delta \theta_1).
\end{align}
Eqs.~(\ref{eq:cond_angular_cdf_1}) and (\ref{eq:cond_angular_cdf_2}) are validated in Fig.~\ref{fig:conditional_angular_cdf}, perfectly matching simulations.

\subsection{Conditional PDF $f(x_2  | r_1, r_1', \Delta \theta_1)$}
\label{sec:f_x_r_r_dtheta}

Now, consider two nodes with radial coordinates $r_1 \in [0, R_1]$, $r_1' \in [0, R_1]$ and angular distance $\Delta \theta_1 \in [0, \pi]$ in layer~1. Further, let $X_2$ be a random variable denoting the hyperbolic distance $x_2$ between these nodes in layer~2, where their radial coordinates are $r_2 \in [0, R_2]$, $r_2' \in [0, R_2]$, and their angular distance is $\Delta \theta_2 \in [0, \pi]$. Since $x_2 \approx  r_2+r_2'+2\ln{\sin{(\Delta\theta_2/2)}}$ (Eq.~(\ref{eq:hyperbolic_distance})), we can write the CDF of $X_2$ conditioned on $r_1, r_1', \Delta \theta_1$ as:
\begin{align}
\label{eq:general_1}
\nonumber P(X_2 \leq x_2 | r_1, r_1', \Delta \theta_1)=\\
\nonumber \int \int P\left (\Delta \theta_2 \leq 2 \arcsin{e^{\frac{1}{2}(x_2-r_2-r_2')}} | \Delta \theta_1 \right)\\
\times \rho_2 (r_2' | r_1', \nu)  \rho_2 (r_2 | r_1, \nu) \mathrm{d}r_2' \mathrm{d}r_2,
\end{align}
where $\rho_2 (r_2 | r_1, \nu)$ in Eq.~(\ref{eq:r_cond_text}). The above integral can be evaluated numerically using Eqs.~(\ref{eq:cond_angular_cdf_1}) and (\ref{eq:cond_angular_cdf_2}). To this end, we need to identify the different limits of integration and the corresponding integrands. To ease notation, let:
\begin{align}
\label{eq:not1}
\mathrm{d}\rho_{\nu} \equiv  \rho_2 (r_2' | r_1', \nu)  \rho_2 (r_2 | r_1, \nu) \mathrm{d}r_2' \mathrm{d}r_2,\\
\label{eq:not2}
\widetilde{\Delta \theta}_2 \equiv 2 \arcsin{e^{\frac{1}{2}(x_2-r_2-r_2')}},\\
\label{eq:not3}
\tilde{x}_2 \equiv \min\left[x_2-2\ln{\sin{\left(\frac{\pi-\Delta\theta_1}{2}\right)}}, 2 R_2\right].
\end{align}
We observe that $x_2 \leq \tilde{x}_2$. Furthermore, Eq.~(\ref{eq:cond_angular_cdf_1}) holds if $\tilde{x}_2 \leq r_2+r_2'$, while Eq.~(\ref{eq:cond_angular_cdf_2}) holds if $r_2+r_2' \leq \tilde{x}_2$. Finally, $P(X_2 \leq x_2 | r_1, r_1', \Delta \theta_1)=1$ if $r_2+r_2' \leq x_2$.  

Given the above observations we distinguish three cases: (i) $\tilde{x}_2 \leq R_2$; (ii) $x_2 \leq R_2 \leq \tilde{x}_2$; and (iii) $R_2 \leq x_2$. Using Eqs.~(\ref{eq:cond_angular_cdf_1}), (\ref{eq:cond_angular_cdf_2}) and the notation in (\ref{eq:not1})-(\ref{eq:not3}) we can write:\\

\begin{figure}[!h]
\centering{
\includegraphics[width=3.4in]{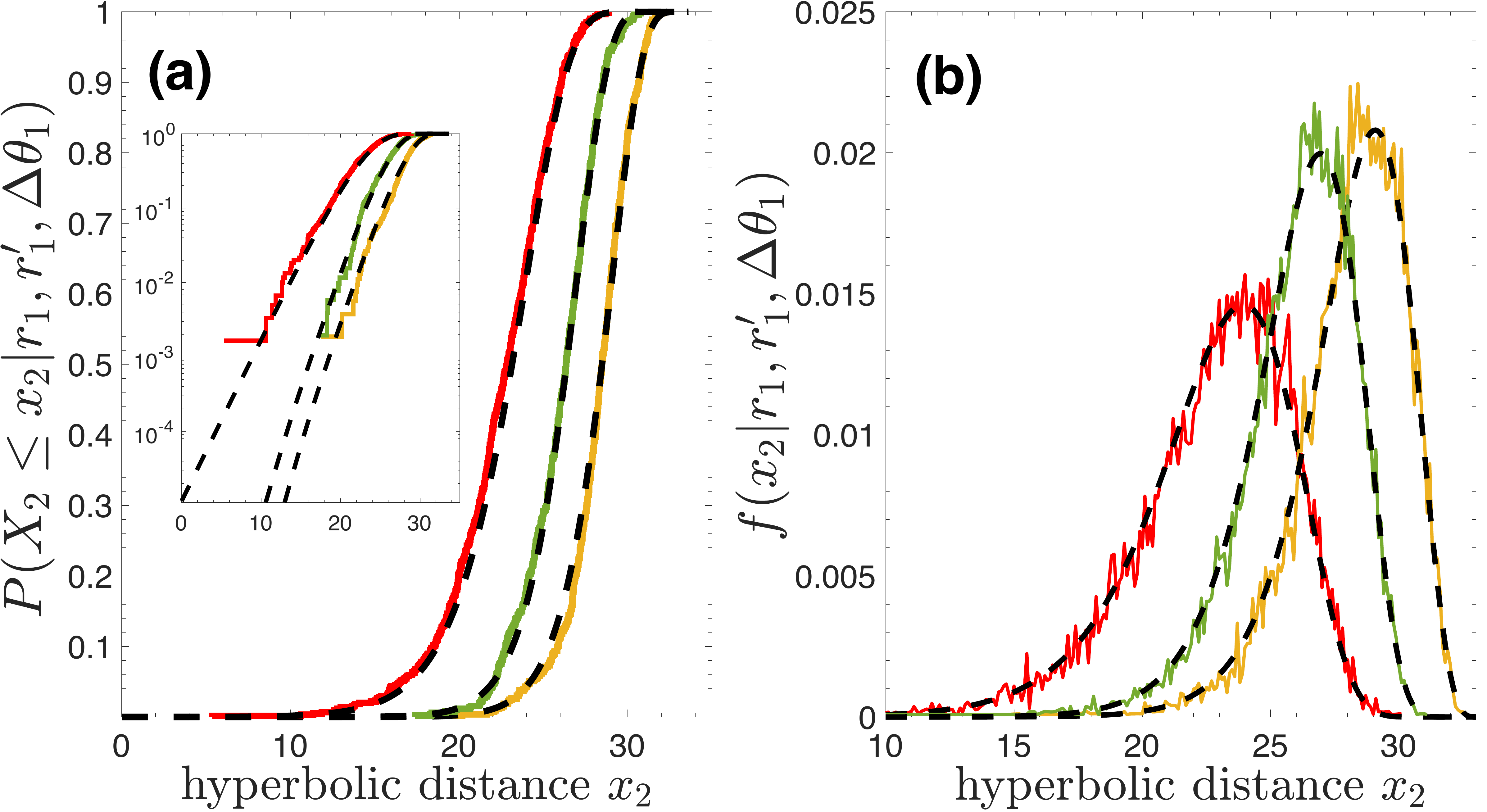}
\caption{The results are from a two layer synthetic multiplex with $N_1=5000$ and $N_2=3000$ nodes, $\gamma_1=2.1, \gamma_2=2.5, T_1=T_2=0.5$ and $\bar{k}_1=\bar{k}_2=6$ ($R_1=23, R_2=16.8$). The correlation strengths are $\nu=0.5, g=0.7$, and all nodes in the smaller layer also exist in the larger. The plots show conditional hyperbolic distance CDFs and PDFs (histograms) for $r_1=18$, $r_1'=20$, and $\Delta \theta_1=0.1, 0.5, 1.5$ radians (from left to right, corresponding respectively to hyperbolic distances $x_1= 32, 35.2, 37.2$). The $y$-axis in the inset of~(a) is in log-scale. The $x$-axis in~(b) is binned into bins of size $0.1$. The solid lines in the plots are the empirical distributions while the dashed black lines are the corresponding theoretical predictions. The empirical distributions are computed over all node pairs with $r_1=18 \pm 0.5, r_1'=20 \pm 0.5$ at the corresponding angular distance $\Delta \theta_1 \pm 0.05$. The empirical distributions in~(b) are average distributions over 20 simulation runs.
\label{fig:conditional_hyperbolic_cdf}}}
\end{figure}

If \underline{$\tilde{x}_2 \leq R_2$}:
\small
\begin{eqnarray}
\nonumber P(X_2 \leq x_2 | r_1, r_1', \Delta \theta_1) = \int\limits_{0}^{\tilde{x}_2} \int\limits_{\tilde{x}_2-r_2}^{R_2} P_1^g(\widetilde{\Delta \theta}_2 |\Delta\theta_1) \mathrm{d} \rho_{\nu}\\
\nonumber + \int\limits_{\tilde{x}_2}^{R_2} \int\limits_{0}^{R_2} P_1^g(\widetilde{\Delta \theta}_2 | \Delta\theta_1)\mathrm{d} \rho_{\nu}+\int\limits_{0}^{x_2} \int\limits_{x_2-r_2}^{\tilde{x}_2-r_2} P_2^g(\widetilde{\Delta \theta}_2 | \Delta \theta_1) \mathrm{d} \rho_{\nu}\\
\nonumber +\int\limits_{x_2}^{\tilde{x}_2} \int\limits_{0}^{\tilde{x}_2-r_2} P_2^g(\widetilde{\Delta \theta}_2 | \Delta \theta_1)\mathrm{d} \rho_{\nu}+\int\limits_{0}^{x_2}\int\limits_{0}^{x_2-r_2} \mathrm{d} \rho_{\nu}\\
\nonumber \equiv I_1 (x_2 | r_1, r_1', \Delta \theta_1);
\label{eq:I1}
\end{eqnarray}
\normalsize
if \underline{$x_2 \leq R_2 \leq \tilde{x}_2$}:
\small
\begin{eqnarray}
\nonumber P(X_2 \leq x_2 | r_1, r_1', \Delta \theta_1)=\int\limits_{\tilde{x}_2-R_2}^{R_2} \int\limits_{\tilde{x}_2-r_2}^{R_2} P_1^g(\widetilde{\Delta \theta}_2 | \Delta \theta_1) \mathrm{d} \rho_{\nu}\\
\nonumber +\int\limits_{0}^{x_2} \int\limits_{x_2-r_2}^{R_2-r_2} P_2^g(\widetilde{\Delta \theta_2} | \Delta \theta_1)\mathrm{d} \rho_{\nu}+\int\limits_{x_2}^{R_2} \int\limits_{0}^{R_2-r_2} P_2^g(\widetilde{\Delta \theta_2} | \Delta \theta_1)\mathrm{d} \rho_{\nu}\\
\nonumber+\int\limits_{0}^{\tilde{x}_2-R_2} \int\limits_{R_2-r_2}^{R_2} P_2^g(\widetilde{\Delta \theta_2} | \Delta\theta_1)\mathrm{d} \rho_{\nu}+\int\limits_{\tilde{x}_2-R_2}^{R_2} \int\limits_{R_2-r_2}^{\tilde{x}_2-r_2} P_2^g(\widetilde{\Delta \theta_2} | \Delta\theta_1)\mathrm{d} \rho_{\nu}\\
\nonumber + \int\limits_{0}^{x_2} \int\limits_{0}^{x_2-r_2} \mathrm{d} \rho_{\nu} \equiv I_2 (x_2 | r_1, r_1', \Delta \theta_1);
\label{eq:I2}
\end{eqnarray}
\normalsize
and if \underline{$R_2 \leq x_2$}:
\small
\begin{eqnarray}
\nonumber P(X_2 \leq x_2 | r_1, r_1', \Delta \theta_1)=\int\limits_{\tilde{x}_2-R_2}^{R_2} \int\limits_{\tilde{x}_2-r_2}^{R_2} P_1^g(\widetilde{\Delta \theta}_2 | \Delta \theta_1) \mathrm{d} \rho_{\nu}\\
\nonumber +\int\limits_{x_2-R_2}^{\tilde{x}_2-R_2} \int\limits_{x_2-r_2}^{R_2} P_2^g(\widetilde{\Delta \theta}_2 | \Delta \theta_1)\mathrm{d} \rho_{\nu} +\int\limits_{\tilde{x}_2-R_2}^{R_2} \int\limits_{x_2-r_2}^{\tilde{x}_2-r_2} P_2^g(\widetilde{\Delta \theta}_2 | \Delta \theta_1)\mathrm{d} \rho_{\nu}\\
+ \int\limits_{0}^{x_2-R_2} \int\limits_{0}^{R_2} \mathrm{d} \rho_{\nu}+\int\limits_{x_2-R_2}^{R_2} \int\limits_{0}^{x_2-r_2} \mathrm{d} \rho_{\nu} \nonumber \equiv I_3 (x_2 | r_1, r_1', \Delta \theta_1).
\label{eq:I3}
\end{eqnarray}
\normalsize
By differentiating the above relations with respect to $x_2$ we get the corresponding PDFs:
\begin{align}
\label{eq:pdf1}
f_1(x_2 | r_1, r_1', \Delta \theta_1) = \frac{\mathrm{d} I_1 (x_2 | r_1, r_1', \Delta \theta_1)}{\mathrm{d}x_2},
\end{align}
\begin{align}
\label{eq:pdf2}
f_2(x_2 | r_1, r_1', \Delta \theta_1) = \frac{\mathrm{d} I_2 (x_2 | r_1, r_1', \Delta \theta_1)}{\mathrm{d}x_2},
\end{align}
\begin{align}
\label{eq:pdf3}
f_3(x_2 | r_1, r_1', \Delta \theta_1) = \frac{\mathrm{d} I_3 (x_2 | r_1, r_1', \Delta \theta_1)}{\mathrm{d}x_2}.
\end{align}
The above analysis is validated in Fig.~\ref{fig:conditional_hyperbolic_cdf}. To evaluate our integrals we use the Cuba library for multidimensional numerical integration~\cite{cuba1, cuba2}. Our code computing both conditional CDFs and PDFs is available at~\cite{gmm_lp}. 

\subsection{Conditional PDF $f(x_2 | x_1)$}
\label{sec:f_x_2_x_1}

Finally, let's consider $f(x_2 | x_1)$.  Since $x_1 \approx  r_1+r_1'+2\ln{\sin{(\Delta\theta_1/2)}}$, we have:
\begin{align}
\Delta \theta_1 \approx 2 \arcsin{e^{\frac{1}{2}(x_1-r_1-r_1')}}.
\label{eq:not4}
\end{align}
Now, since we know how to compute $f(x_2 | r_1, r_1', \Delta \theta_1)$ (Eqs.~(\ref{eq:pdf1})-(\ref{eq:pdf3})), we can write:
\begin{align}
\label{eq:general_2}
f(x_2 | x_1) =& \int \int f(x_2 | r_1, r_1', \Delta \theta_1)
\mathrm{d} \rho_1,\\
\nonumber \mathrm{d} \rho_1 \equiv& \rho_1(r_1') \rho_1(r_1) \mathrm{d}r_1' \mathrm{d}r_1,
\end{align}
where $\Delta \theta_1$  is given in Eq.~(\ref{eq:not4}) and $\rho_1(r_1)$ in Eq.~(\ref{eq:eq_rho_1}). To be able to numerically evaluate Eq.~(\ref{eq:general_2}) we need to identify the limits of integration for the $r_1, r_1'$ variables and the corresponding integrands. Since Eqs.~(\ref{eq:pdf1})-(\ref{eq:pdf3}) consist of two-dimensional integrals, Eq.~(\ref{eq:general_2}) consists of four-dimensional integrals. This analysis is beyond the scope of this paper.

\input{paper.bbl}

\end{document}

%% file: paper.bbl
%